\documentclass[twocolumn,twocolappendix]{aastex63}
\usepackage{fp}
\usepackage{amsmath}
\usepackage[normalem]{ulem}
\usepackage{gensymb}

\newcommand{\mZEROtrgbLMC}{14.439}
\newcommand{\mZEROtrgbSMC}{14.927}
\newcommand{\distmodLMC}{18.477} 
\newcommand{\distmodSMC}{18.977}
\newcommand{\calSTATerrLMC}{0.012}
\newcommand{\calSYSerrLMC}{0.032}
\newcommand{\calSTATerrSMC}{0.030}
\newcommand{\calSYSerrSMC}{0.039}
\FPeval{\MtrgbLMC}{\mZEROtrgbLMC-\distmodLMC}
\FPeval{\MtrgbSMC}{\mZEROtrgbSMC-\distmodSMC}

\FPeval{\MtrgbLMC}{round(\MtrgbLMC,3)}
\FPeval{\MtrgbSMC}{round(\MtrgbSMC,3)}

\shorttitle{TRGB Calibration in the MCs}
\shortauthors{Hoyt}

\graphicspath{{./}{figs/}}

\begin{document}

\title{On Zero Point Calibration of the Red Giant Branch Tip in the Magellanic Clouds}

\correspondingauthor{Taylor Hoyt}
\email{tjhoyt@uchicago.edu, taylorjhoyt@gmail.com}

\author[0000-0001-9664-0560]{Taylor J. Hoyt}
\affiliation{University of Chicago \\
5640 S. Ellis Ave \\
Chicago, IL 60615, USA}

\begin{abstract}
  A zero point calibration of the Red Giant Branch Tip (TRGB) in the $I$-band is determined from OGLE photometry of the Magellanic Clouds (MCs).
  It is shown that TRGB measurements made in star-forming regions, with concomitantly high quantities of gas and dust, are less precise and biased to fainter magnitudes, as compared to the same measurements made in quiescent regions.
  Once these low accuracy fields are excluded from consideration, the TRGB can be used for the first time to constrain the three-dimensional plane geometry of the LMC.
  Composite CMDs are constructed for the SMC and LMC from only those fields with well-defined TRGB features, and the highest accuracy TRGB zero point calibration to date is presented. The $I$-band TRGB magnitude is measured to be flat over the color range  $ 1.45 < (V-I)_0 < 1.95\mathrm{~mag}$, with a modest slope introduced when including metal-rich (up to $(V-I)_0 = 2.2$~mag) Tip stars into the fit.
  Both the flat, blue zero point and the shallow slope calibration are consistent with the canonical value of $-4.05$~mag for the old, metal-poor TRGB, and would appear to resolve a recent debate in the literature over the method's absolute calibration.
\end{abstract}

\section{Introduction}

The Tip of the Red Giant Branch (TRGB) is arguably the most ``standard'' candle in astrophysics, with well-understood stellar core physics regulating the observed luminosity of low mass red giants at the time of the Helium Flash \citep{Salaris_2005_book}. The ubiquity of TRGB stars in galaxies of all types has enabled measurement of TRGB distances to over 500 galaxies \citep{Anand_2021}.
The TRGB has also been used recently to anchor the Hubble Diagram \citep{Jang_2017_h0, Freedman_2019} via HST imaging of the stellar halos of nearby SN~Ia Host galaxies. 

There have since emerged a number of proposed recalibrations of the \citeauthor[][Carnegie Chicago Hubble Program (CCHP)]{Freedman_2019} distance scale, leading to an apparent debate in the literature regarding the TRGB zero point \citep[see][for a review]{Capozzi_2020}. In this paper, it will be shown that, after a careful and comprehensive re-examination of all recent measurements and data taken of the Magellanic Clouds (MCs), this dispute does not hold, and is being driven by systematic biases in literature measurements and subsequently underestimated uncertainties.

At the nearby distance of $50$~kpc, the Large Magellanic Cloud (LMC) has long been a focal point of the extragalactic distance scale. In the HST Key Project's resolution of the Factor of Two debate, the distance to the LMC was the largest single uncertainty ($5\%$) in the classical distance ladder measurement of $H_0$ \citep{Freedman_2001}. And in the last decade, high-accuracy distances to the Clouds have been determined through the use of detached eclipsing binaries \citep[DEBs,][]{Graczyk_2012, Pietrzynski_2013, Graczyk_2014}. Specifically, the move to late-type DEBs (for which the surface brightness-color relation can be constrained to better than 0.02 mag) has made possible a pair of recent 1.1\% and 1.7\% geometric distances measured to the LMC \citep{Pietrzynski_2019} and the SMC \citep{Graczyk_2020}, respectively. 

This marked improvement in absolute precision and accuracy has brought with it a flurry of TRGB calibrations based in the MCs \citep{Jang_2017_color, Hoyt_2018, Gorski_2018, Groenewegen_2019, Freedman_2020}.
However, concerns have been raised regarding calibration of the TRGB in the LMC \citep{Yuan_2019, Nataf_2021}, specifically regarding the effects of dust extinction and star-formation (i.e., contamination from intermediate-aged stellar populations), leading to recent pursuits of a TRGB calibration via other geometric anchors, e.g., NGC~4258 \citep{Reid_2019, Jang_2021}, or Galactic globular clusters \citep{Soltis_2021, Cerny_2020}.

Still, the Clouds, in particular the LMC, remain prime targets for zero point calibration of the TRGB thanks to the high accuracy of the DEB distances, the presence of a predominantly old, metal-poor stellar component in both satellite galaxies, and the abnormally high integrated mass contained in such populations ($M_* \sim 10^8/10^9 M_{\odot}$ for SMC and LMC, respectively) providing an extremely well-populated metal-poor RGB luminosity function, all while being our nearest Galactic neighbors and making possible very high signal-to-noise photometry of upper RGB stars.

In this paper, it will be shown that the systematics of both dust reddening and mixed-populations are concomitantly minimized by considering only those regions in the LMC that exhibit unambiguous, well-defined RGB and TRGB features. Together with a TRGB measurement made in the more pristine (but more extended) SMC, the most accurate zero point calibration, and comprehensive error budget, of the TRGB to date will be presented.

There are three complete and independent studies of the TRGB in the MCs in the recent literature
\citep{Jang_2017_color, Gorski_2018, Freedman_2020}. I will explore in great detail the degree to which each was affected by systematics due to dust content and star formation (that were successfully controlled for in the present analysis). Each study's results will then be renormalized onto modern reddening and distance zero points. In doing so, their TRGB calibrations will be shown to be consistent with each other at the $\sim0.05$~mag level (albeit with larger uncertainties than the new calibration presented here). Furthermore, once regions identified as having low-quality TRGB features are removed from these calibrations, agreement is then found to better than 0.015~mag, at once resolving the debate over the TRGB's zero point calibration in the MCs, and in general.\footnote{Other recently proposed calibrations based in the LMC \citep{Yuan_2019, Soltis_2021} presented post-processed modifications of one or a combination of the three revisited studies, so the updates to be presented in this study are also direct updates and corrections to those studies.}

This paper is organized as follows: the data are described in \autoref{sect:data}. Details of the TRGB measurements made in the LMC and SMC are covered in \autoref{sect:trgb_measure}, with the final calibration presented in \autoref{sect:calib}. Implications of the TRGB results are discussed in \autoref{sect:discussion}, and the aforementioned literature calibrations are revisited in \autoref{sect:lmc_discuss}. Summary and conclusions are presented in \autoref{sect:conclusion}.

\section{Data} \label{sect:data}

The OGLE collaboration has been monitoring the Magellanic Clouds since the late 1990s \citep{Udalski_2003, Udalski_1997,Udalski_2000,Udalski_2008,Udalski_2015}, and with each iteration our understanding of the Clouds has advanced considerably. In light of the comprehensive Red Clump (RC) results presented in \citet[][hereafter S21]{Skowron_2021}, that trend looks likely to continue as the OGLE-IV data and results continue to roll out.

Adopted for the present analysis are: the publicly available OGLE-III \citep{Udalski_2008_maps} photometric maps, \citep[recalibrated using OGLE-IV observations of the bulge,][]{Szymanski_2011}, a set of recent DEB distances to the LMC \citep[][hereafter P19]{Pietrzynski_2019} and SMC \citep[][hereafter G20]{Graczyk_2020}, and the recent S21 reddening map based on OGLE-IV observations of RC stars.

S21 used the observed colors of RC stars to measure line of sight reddening values across the entirety of each MC, reaching resolutions of $1.7' \times 1.7'$ in the inner-most regions.
The large area covered by the OGLE-IV survey enabled a direct calibration of the RC intrinsic color using the \citet[][hereafter SF11]{SF11} re-normalization of the \citet[][hereafter SFD98]{SFD98} maps, a concept identical to that implemented by \citet{Pawlak_2016}. Thus, the S21 reddening map and its zero point(s) rely purely on homogeneous empirical measurements made in each Cloud, which sets it apart from previous reddening maps.

The RC has been used frequently in quantifying dust attenuation along MC sight lines \citep{Haschke_2011,Pawlak_2016, Choi_2018,Gorski_2020,Skowron_2021} and has produced consistent results in a relative sense. However, a debate has arisen over calibration of the RC intrinsic color in the MCs \citep{Gorski_2020, Skowron_2021, Nataf_2021}. This value of the RC intrinsic color is fundamentally tied to an eventual zero point calibration of the TRGB that uses RC-determined reddening corrections (for reference, bluer RC intrinsic color calibrations result in larger measured reddenings).
The reader is referred to Section 5 of S21 for a discussion on prior determinations of the RC intrinsic color and why their result is the state-of-the-art, and preferred for this analysis.

\begin{figure}
    \centering
    \includegraphics[width=\columnwidth]{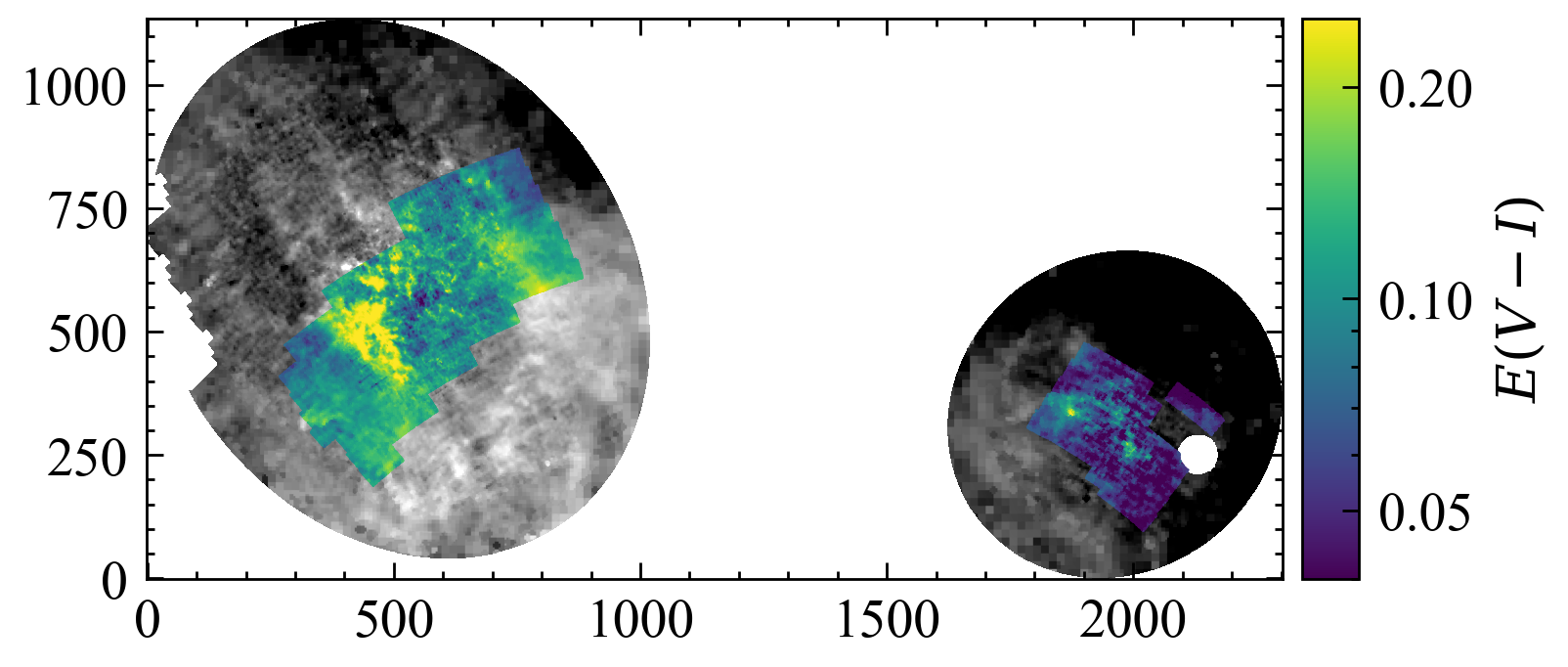}
    \caption{The \citet[][S21]{Skowron_2021} Red Clump (RC) OGLE-IV reddening map of the Magellanic Clouds is plotted on their native pixel grid (a Hammer projection). Overlap with the OGLE-III survey is hued, while the rest of the maps are in grayscale. Note the stretch is on a log scale.}
    \label{fig:skowron_ogle}
\end{figure}

In summary, the S21 results are comprehensive, fully empirical, and most applicable to the OGLE photometry adopted here. Each calibration step required to use the RC as a standard crayon was undertaken using only homogeneous measurements made directly in each Cloud (rather than invoking theoretical predictions or external observations). These steps include: (1) a direct color zero point calibration made in the outer regions of each MC via SFD98/SF11 foreground maps, (2) empirical determination of the metallicity gradients in each Cloud using recent abundance measurements from \citet{Nidever_2020}, and (3) an empirical calibration of the RC color-metallicity relation using homogeneous photometry of clusters in the MCs. For these many reasons, the S21 OGLE-IV reddening maps are preferred for this OGLE-based calibration of the TRGB. This view is shared with \citet{Soltis_2021}, who stated the S21 maps are a ``significant improvement'' over the \citet{Haschke_2011} maps. Furthermore, S21 found agreement with \citet{Nataf_2021} on the metallicity dependence of the intrinsic RC color.

Throughout this article, the following reddening law coefficients are adopted: $E(V-I) = 1.4E(B-V)$, and $A_I = 1.219 E(V-I)$. The former is computed directly from the SF11 Landolt coefficients for $R_V = 3.1$, while the latter is computed by taking the \citet{Fitzpatrick_1999} curve (built from the spline points provided in their Table 3) and convolving it with the OGLE-IV flux transmission curve (\url{http://ogle.astrouw.edu.pl/main/OGLEIV/mosaic.html}).\footnote{The difference between this exact calculation of the selective to total absorption and the Landolt values provided in SF11 is 0.19\%, and thus used solely for the sake of completeness, having no measurable effect on the absolute photometry used here.} Throughout this manuscript, dereddened magnitudes will always be identified with \textit{Naught}$_0$ subscripts. 

The OGLE-III photometric maps are trimmed to include only sources approximately classified as upper-RGB ($I < 17$~mag, $(V-I) > 0.5$~mag).
Duplicate sources defined as within 0.2'' of another source (about a factor of five smaller than the minimum seeing at the Warsaw telescope) in the OGLE maps are merged via weighted average of their respective mean fluxes, using the number of good observations underlying each mean flux as the weights. All duplicate candidates are located on the individual chip and field boundaries in the OGLE footprint. Also, magnitude differences between duplicate candidates are far below the standard deviation of each source's photometry. Both indicate that all candidates are true duplicates and not fluke matches.

High amplitude LPVs (those classified as Miras or semi-regular variables) in either Cloud \citep{Soszynski_2009, Soszynski_2011} are masked from the adopted science catalogs. Also, high probability foreground sources are removed from the photometry using Gaia proper motions. The details of the proper motion cleaning are described in the Appendix.

\section{TRGB Measurements in the Magellanic Clouds} \label{sect:trgb_measure}

In what follows, the TRGB measurements to each of the Clouds will be presented. The LMC will be given a more sophisticated treatment than that afforded the SMC for two reasons: (1) its complex distribution (and larger cumulative amount) of gas and dust, as well as higher rates of recent star formation, and (2) the potential for a much higher accuracy calibration, a result of the LMC's back-to-front geometry being much simpler to model than the SMC, which is significantly extended along the line of sight (for comparison, the LMC DEB distance is five times more precise than that measured to the SMC with comparable numbers of DEB systems). Indeed, it will be shown that the geometry of the LMC can be constrained remarkably well via the observed magnitudes of TRGB stars alone.

\subsection{LMC} \label{subsect:lmc}
For the TRGB calibration in the LMC, both the OGLE-Shallow \citep{Ulaczyk_2012} and OGLE-III photometric maps (\url{http://ogle.astrouw.edu.pl/cont/4_main/map/map.html}) are considered. The two catalogs produce slightly ($\lesssim 0.01$~mag) different TRGB measurements, which is due to small differences in the stellar magnitudes. This effect is taken into account in the final error budget. As mentioned, the S21 reddening map, determined from RC colors, is used to de-redden the photometry. 

\subsubsection{Field-by-field TRGB Measurements} \label{subsect:trgb_measure}

\begin{deluxetable*}{ccccccccc}

\tabletypesize{\small}

\tablecaption{LMC Field Definitions and Characteristics}

\tablehead{
\colhead{Field} &
\colhead{R.A. (J2000)} & 
\colhead{Dec. (J2000)} &
\colhead{Area} &
\colhead{$r^{HZ09}_{1}$} & 
\colhead{$r^{HZ09}_{2}$} &
\colhead{$E(B-V)$\tablenotemark{a}} & 
\colhead{$E(B-V)$\tablenotemark{b}} & 
\colhead{Rank}\\
\colhead{} &
\colhead{(deg)} & 
\colhead{(deg)} &
\colhead{(deg$^2$)} & 
\colhead{} &
\colhead{} &
\colhead{(mag)} & 
\colhead{(mag)} & 
\colhead{[1-4]}
} 
\startdata
 1 & 82.185812 & -68.095806 & 1.3 & 0.89 & 0.91 & 0.101 & 0.520 & 4 \\
 5 & 77.790750 & -67.777458 & 2.8 & 1.16 & 0.69 & 0.076 & 0.292 & 2 \\
 6 & 74.543542 & -67.385139 & 3.6 & 1.77 & 2.04 & 0.065 & 0.185 & 3 \\
 7 & 85.722708 & -68.515889 & 2.9 & 1.65 & 1.67 & 0.104 & 0.335 & 4 \\
 8 & 79.672396 & -68.788694 & 1.3 & 1.39 & 0.60 & 0.081 & 0.571 & 4 \\
10 & 82.432646 & -69.046653 & 1.4 & 1.37 & 0.96 & 0.082 & 0.509 & 3 \\
11 & 90.007729 & -69.788069 & 7.1 & 0.12 & 0.90 & 0.070 & 0.089 & 1 \\
13 & 69.572708 & -68.265278 & 7.1 & 0.33 & 0.90 & 0.064 & 0.073 & 1 \\
14 & 78.535375 & -69.345000 & 0.7 & 0.49 & 1.34 & 0.077 & 0.828 & 3 \\
15 & 76.943917 & -68.761236 & 1.2 & 1.62 & 1.56 & 0.073 & 0.514 & 4 \\
16 & 73.944479 & -68.695528 & 2.5 & 3.31 & 2.49 & 0.090 & 0.322 & 3 \\
17 & 82.406583 & -69.780778 & 0.5 & 0.37 & 1.37 & 0.058 & 0.363 & 3 \\
18 & 80.376125 & -69.496944 & 0.6 & 0.31 & 1.27 & 0.050 & 0.924 & 3 \\
19 & 78.776000 & -70.097417 & 1.1 & 1.04 & 0.57 & 0.070 & 0.287 & 1 \\
22 & 85.103813 & -69.699486 & 1.7 & 1.51 & 0.95 & 0.161 & 1.329 & 4 \\
23 & 80.986250 & -69.983083 & 0.5 & 0.37 & 1.53 & 0.071 & 0.518 & 3 \\
25 & 76.136208 & -69.599056 & 1.5 & 1.21 & 0.73 & 0.075 & 0.198 & 2 \\
26 & 86.236479 & -70.626833 & 1.5 & 2.00 & 0.98 & 0.126 & 0.449 & 4 \\
27 & 81.526312 & -70.576833 & 1.0 & 0.98 & 0.43 & 0.069 & 0.212 & 3 \\
28 & 83.518729 & -70.289056 & 0.8 & 0.79 & 0.97 & 0.072 & 0.409 & 4 \\
29 & 79.584458 & -71.179000 & 2.0 & 0.64 & 0.48 & 0.070 & 0.156 & 3 \\
30 & 87.534083 & -71.650167 & 5.4 & 0.61 & 0.42 & 0.096 & 0.094 & 3 \\
31 & 83.681437 & -71.307958 & 1.6 & 0.85 & 0.78 & 0.086 & 0.262 & 2 \\
32 & 75.580479 & -70.645722 & 2.3 & 1.12 & 1.39 & 0.069 & 0.138 & 2 \\
33 & 71.684167 & -69.876556 & 4.0 & 0.70 & 2.16 & 0.104 & 0.132 & 2 \\
\enddata

\tablenotetext{a}{Median reddening values as per \citet{Skowron_2021}.}
\tablenotetext{b}{Median reddening values from \citet{SF11} recalibration of \citet{SFD98} dust maps.}
\label{tab:fields}
\end{deluxetable*}

\begin{figure*}
    \centering
    \includegraphics[width=\textwidth]{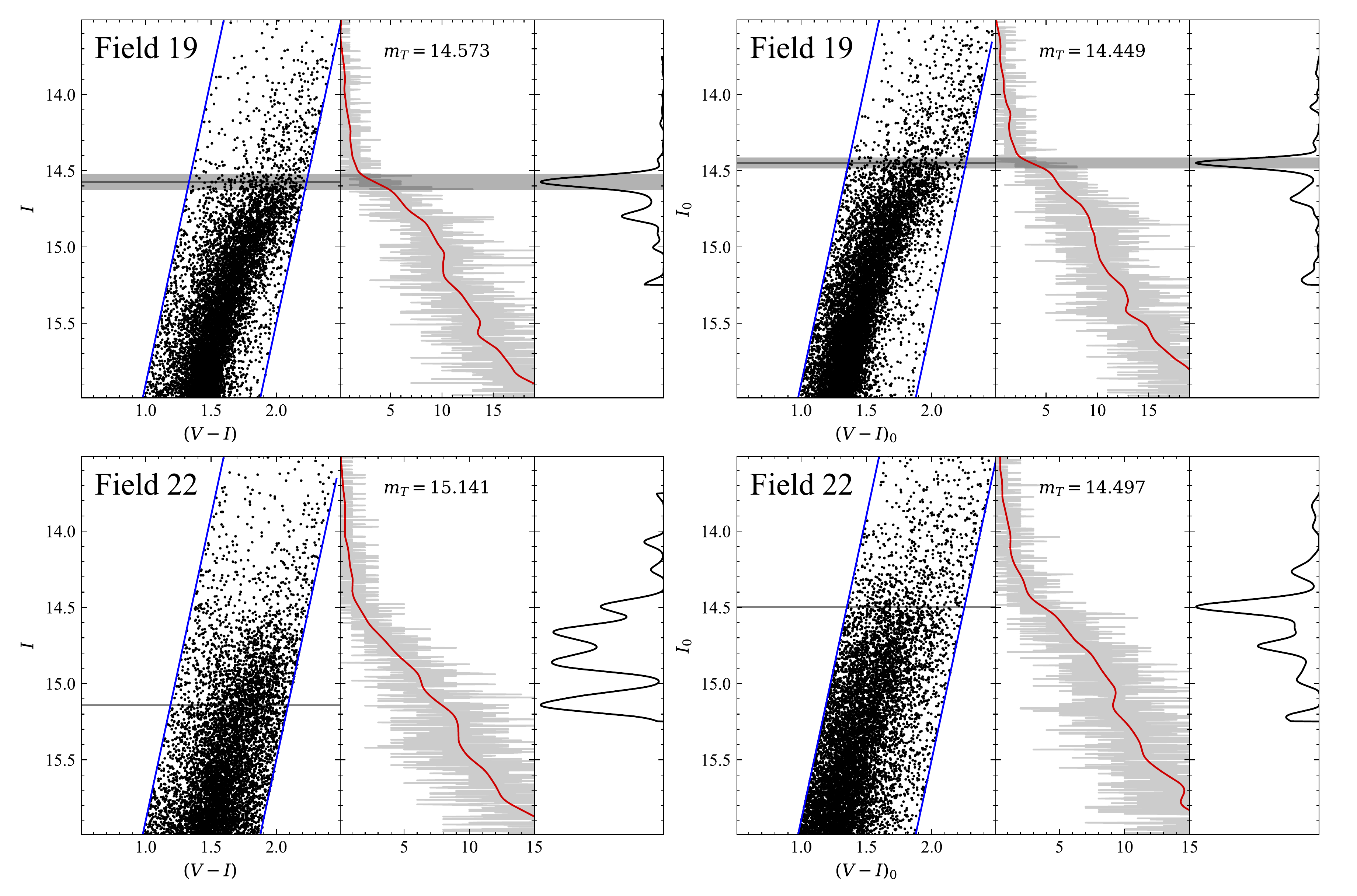}
    \caption{Example TRGB detections for Voronoi Fields 19 (Rank 1) and 22 (Rank 4), without reddening corrections (left column) and after applying the S21 reddening corrections (right column). The RGB luminosity function (LF) in 0.002~mag is isolated from the color-magnitude diagram (CMD) via color selection $1.35 < (V-I) < 2.25$~mag and slope $-4$~mag mag$^{-1}$ (dark blue lines).
    The edge detector response (EDR) function, the peak of which defines the measured TRGB magnitude, is shown in the right-most subplot of each tripanel. Note the increased precision (plotted as a gray band for Field 19) -- visible to the eye in the CMDs, LFs, and EDRs -- after application of the S21 reddening corrections, indicating that their dust reddening measurements, determined from the colors of Red Clump stars, accurately probe the same sight lines through which the Tip stars are observed. An accurate uncertainty could not be estimated for Field 22 and likely exceeds 0.1~mag.}
    \label{fig:lmc_example}
\end{figure*}

The distribution of dust and young stellar populations in the inner LMC is complex and not easily described by a smooth distribution (see \autoref{fig:skowron_ogle}). It is also expected that the TRGB will be both less accurate and less precise in those regions with high levels of dust, gas, and recent star formation. Therefore, it is necessary to divide the LMC into spatial bins. For this analysis two sets of spatial bins are considered: a set of 25 Voronoi bins (described in \autoref{tab:fields}) that were created in \citet{Hoyt_2018} to trace the isodensity contours of RGB stars, and the 116 native LMC OGLE-III fields (numbered 100-215). 
The TRGB is then measured in each of these fields and classified into four ``Ranks'' based on the observed quality of the TRGB detection. 

To measure the TRGB magnitude, a methodology used by the Carnegie Chicago Hubble Program (CCHP) is adopted. The details are presented in \citet{Hatt_2017} and a summary is provided here. The RGB luminosity function (RGB LF), isolated with a color selection box, is binned in this case at 0.002~mag intervals and smoothed via Gaussian-weighted Local Regression \citep[GLOESS,][]{Persson_2004, Monson_2017}. To produce the edge detector response (EDR), the discrete first derivative of the smoothed RGB LF is multiplied by the Poisson Signal-to-noise contributing to each bin of the first derivative, i.e., $(N_{i+1} - N_{i-1})/\sqrt{N_{i+1}+N_{i-1}}$ \citep{Madore_2009}. The peak of the EDR then signals the measured location of the TRGB magnitude. 

In \autoref{fig:lmc_example}, example TRGB detections are shown for Voronoi Fields 19 (Rank 1) and 22 (Rank 4) using either apparent or de-reddened photometry. The measurement uncertainty for Field 19 is estimated by first finding the smallest smoothing scale $\sigma_{min}$ at which the measured peak located at magnitude $m_{min}$ remains twice the amplitude of other peaks (if possible). The smoothing window size is then increased until noisy peaks in the EDR reach a stable minimum in amplitude, with the peak location at that smoothing scale denoted $m_{max}$. The $\sigma_{min}$ quantity is then added in quadrature with the quantity $\Delta m_T =| m_{min} - m_{max} |$ to capture the full uncertainty associated with the edge detector Tip determination, $\sqrt{\sigma_{min}^2 + \Delta m_T^2}$. The former term is best interpreted as finding the characteristic length scale of the edge feature that defines the final measurement. The latter folds in the non-uniqueness associated with choosing a single smoothing size to perform the edge detection. This approach will fail in cases where there is no single dominant edge (such as Field 22 in \autoref{fig:lmc_example}), and these cases will be classified into lower quality rankings as a result. 

In the Field 19 plots of \autoref{fig:lmc_example}, the $2\sigma$ intervals on the TRGB measurements are shown as gray bands, representing $I^{TRGB} = 14.578 \pm 0.026$~mag before reddening corrections, and $I_0^{TRGB} = 14.456 \pm 0.018$~mag after. From visual inspection of the CMDS, LFs, and EDRs, the error estimation is confirmed to be reasonable and an accurate representation of the data. Note the increased precision after applying the S21 reddening corrections, which can be visually verified through the increased contrast in the CMD, the sharper step in the RGB LF, and the narrower peak in the EDR, indicating the RC is a good tracer of the dust attenuation along sightlines to TRGB stars contained in the field.

When ranking the TRGB quality, the following metrics are considered: (1) The visual contrast and clarity of the TRGB as seen in the CMD, (2) the steepness of the LF at the TRGB transition, and (3) the width and structure of the EDR (e.g., one or multiple peaks, a long tail, extent of substructure in the dominant peak, etc.).
A detection of Rank 1 displayed a single, sharply-peaked TRGB edge in both the apparent and de-reddened CMDs. Rank 2 indicates the existence of one dominant edge location (with some substructure), that is slightly broader than Rank 1, and is sharpened after application of the S21 reddening corrections, indicating a well-behaved distribution of dust that is accurately traced by RC stars. Rank 3 indicates a CMD that contained two equally probable nearby peaks, a peak location in the EDR that contradicts visual inspection of the CMD, or a response function that worsened after applying the reddening correction (indicating significant differential reddening not accurately traced by the RC map). A Rank 4 detection showed no perceptible TRGB edge in the apparent CMD, indicating very high absolute and differential reddening that would significantly reduce the accuracy in applying the S21 corrections.
The Rank assigned to each spatial bin is tabulated in the last column of \autoref{tab:fields}. Further details of the TRGB ranking procedure are detailed in \autoref{app:fields}, with representative examples of each rank shown in \autoref{fig:example_fields} and summary statistics in \autoref{tab:detect_summary}.

\subsubsection{Validation of TRGB Rankings} \label{subsect:lmc_regions}
\begin{figure*}
    \centering
    \includegraphics[width=\textwidth]{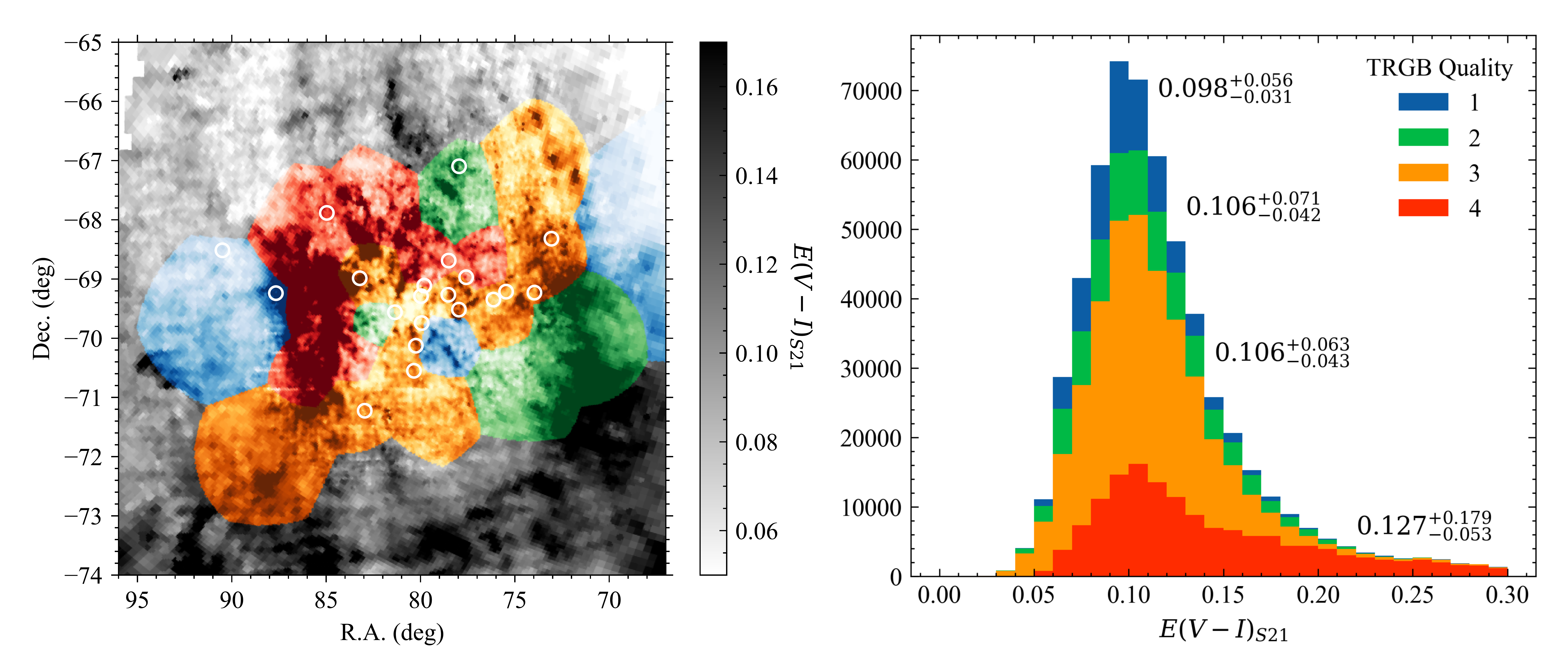}
    \caption{Results of the TRGB selection procedure for the LMC. \textit{Left:} Spatial Voronoi bins (red, orange, green, blue) projected onto the \citet{Skowron_2021} reddening map (grayscale). The spatial bins are color-coded according to their TRGB Rank, as defined in \autoref{subsect:lmc_regions}. The high-ranking (1 and 2) fields used in our TRGB calibration (green and blue shading) are plotted along with the low-ranking (3 and 4) regions (orange and red shading). The 20 DEBs of \citet{Pietrzynski_2019} are marked (white circles). \textit{Right:} Distributions of reddening values for the four sets of regions, matched in color to their hues in the left panel. The 5th and 95th percentiles are printed adjacent to their associated distributions. This quantitatively confirms what is visually apparent in the left panel, that low quality TRGB detections were predominantly found in regions of the LMC with high levels of internal reddening, and that the high quality regions are well-behaved, their reddening distributions being more symmetric and having smaller dispersions.}
    \label{fig:fields}
\end{figure*}

\begin{figure*}
    \centering
    \includegraphics[width=\textwidth]{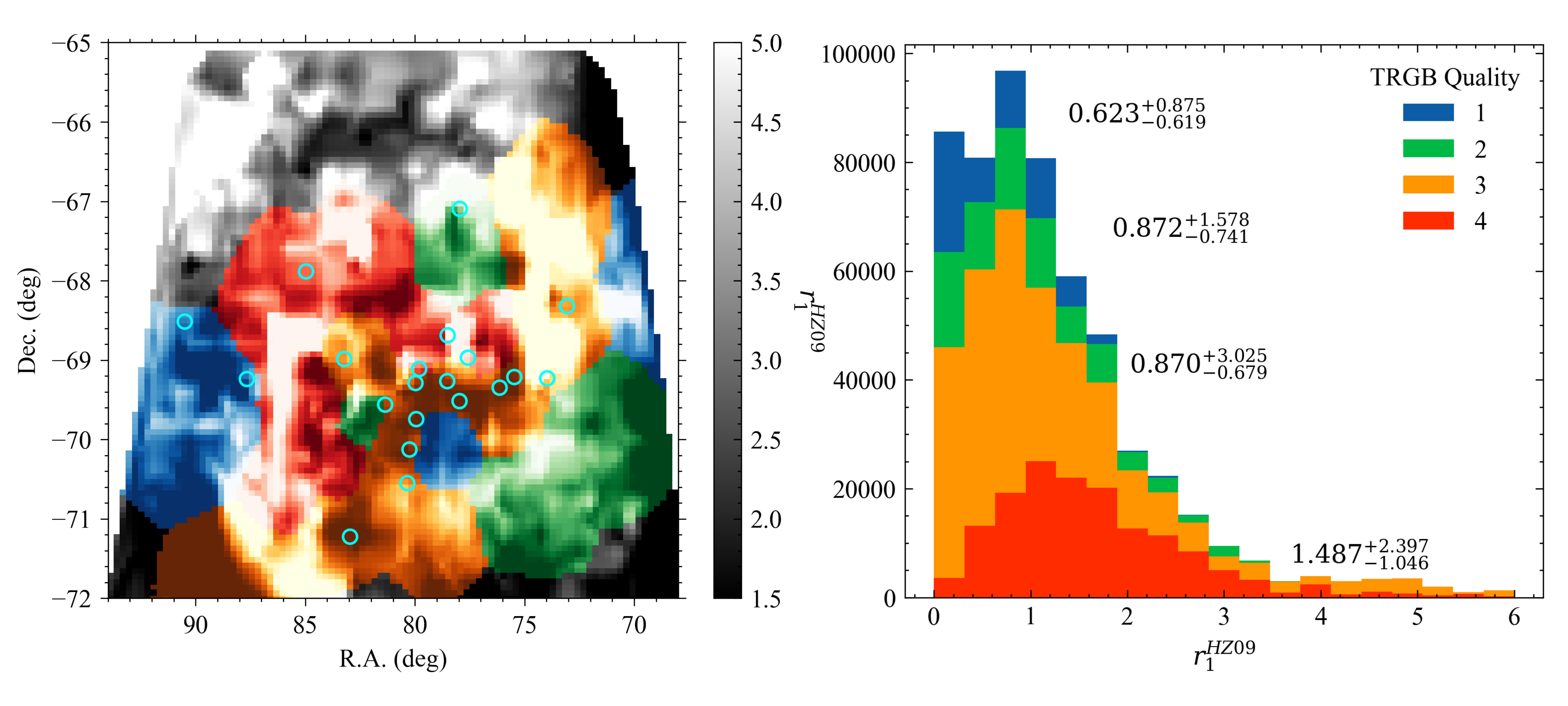}
    \caption{
    Same as \autoref{fig:fields} but for the $r^{HZ09}_1$ parameter defined in \autoref{eq:sfr_norm}. Note the smaller RA/DEC limits compared to \autoref{fig:fields}.
    A trend similar to that seen in \autoref{fig:fields} is apparent, with regions that contain a larger fraction of intermediate-aged stars exhibiting lower quality TRGB detections.} 
    \label{fig:lmc_fields2}
\end{figure*}

To externally validate the TRGB Ranking system, a comparison will now be made with a set of physically-motivated quantities expected to (anti)correlate with TRGB measurement accuracy: (1) the S21 $E(V-I)$ reddenings, and (2) maps of photometrically determined star formation histories \citep[][HZ09]{Harris_2009}. For the latter, it is useful to define the following quantities to estimate the contribution from intermediate-aged ($200$~Myr$<t<2$~Gyr) and young ($t<200$~Myr) populations relative to the LMC's old ($t>4$~Gyr) stellar population,
\begin{align}
  f^{HZ09}_1 &= \frac{ \int^{t<2Gyr}_{t>200Myr} SFH_{HZ09}(t) dt} { \int_{t>4Gyr}SFH_{HZ09}(t) dt} \\
  f^{HZ09}_2 &= \frac{ \int^{t<200Myr} SFH_{HZ09}(t) dt} { \int_{t>4Gyr}SFH_{HZ09}(t) dt} 
  \label{eq:sfr_ratios}
\end{align}
which are calculated for each pixel of the HZ09 maps. The $f_1$ and $f_2$ quantities are then normalized, i.e.,
\begin{equation}
    r^{HZ09}_{1/2} = \frac{f^{HZ09}_{1/2}} {<f^{HZ09}_{1/2}>} \label{eq:sfr_norm}
\end{equation}
and assigned individually to the nearest sources in the OGLE photometric catalog. The normalizing values for $r_1$ and $r_2$ are 0.106 and 0.018, implying a 9\% and 2\% contribution to the inner LMC's total mass from intermediate-aged and young stellar populations, respectively. The values of these star formation metrics are tabulated for each spatial bin in \autoref{tab:fields}. Values greater than one indicate an above-average presence of young or intermediate-aged stellar populations. 

A correlation can be seen between lower quality TRGB detections and higher values of $E(V-I)$, $r_{1}$, and $r_{2}$. It is evident that in the case of the LMC, the largest contributions of systematic error to measurements of the TRGB magnitude are reddening uncertainties, e.g., differential reddening and potential variations in the reddening law (expected to increase in amplitude in regions with higher total dust content), as well as contamination from intermediate-aged and young stellar populations, which will affect both the RC color and TRGB magnitude. In \autoref{subsect:geofit} and \autoref{subsect:lmc_trgb}, it will be shown that masking regions in which these effects are significant (i.e., using only the Rank 1 and 2 fields) makes possible the first determination of the LMC geometry from the TRGB, as well as the most unambiguous and highest accuracy TRGB measurement to date.

In \autoref{fig:fields} and \autoref{fig:lmc_fields2}, the results of the cross-validation are shown for the $E(V-I)_{S21}$ and the $r_1^{HZ09}$ maps, respectively. The spatial correlation between all three quantities is striking, with the TRGB rank parameter tracing the structure in both the reddening map and the star formation quantity almost exactly. Indeed, there is a clear trend of decreasing TRGB quality with larger reddening values and higher fractions of intermediate-aged stellar mass, as shown in the right panels of \autoref{fig:fields} and \autoref{fig:lmc_fields2}.

It is interesting to also consider the relationship between TRGB ranking and known dynamical structures in the LMC. A majority of the Rank 1 and 2 fields coincide with a pristine ``pocket'' of thr inner LMC located just south of the star-forming bar as verified via H-I observations \citep{McGee_1966, Rohlfs_1984, Kim_2003}, RR Lyrae \citep{Pejcha_2009}, and the RC \citep[][among others]{Skowron_2021}.
By contrast, the Rank 3 fields traced closely the narrow, star-forming bar in which most of the DEBs are found, while the majority of the Rank 4 fields were occupied by the 30~Dor star-forming region and nearby structures, like the LMC2 supergiant shell and the associated ``X-ray spur'' \citep{Points_1999, Knies_2021}.
Of course, the Rank 3 and 4 calibration fields coincide generally star forming structures, as traced by $H\alpha$ observations \citep{Kennicutt_1995}, the locations of Supergiant shells \citep{Kim_1999}, and Cepheids \citep{Nikolaev_2004}.
It is also worth noting that the young and intermediate-aged populations, as inferred from the HZ09 maps, as well as the Rank 3 and 4 fields, trace structures identical to those traced by the S21 $N_{RC}/N$ parameter (defined in their Section 3.1), which is expected to increase as the fraction of young RC stars increases. This cross-consistency across probes confirms the reliability of the integrated HZ09 SFH maps, and in turn the accuracy of the TRGB Ranking procedure.

\begin{figure*}
    \centering
    \includegraphics[width=\textwidth]{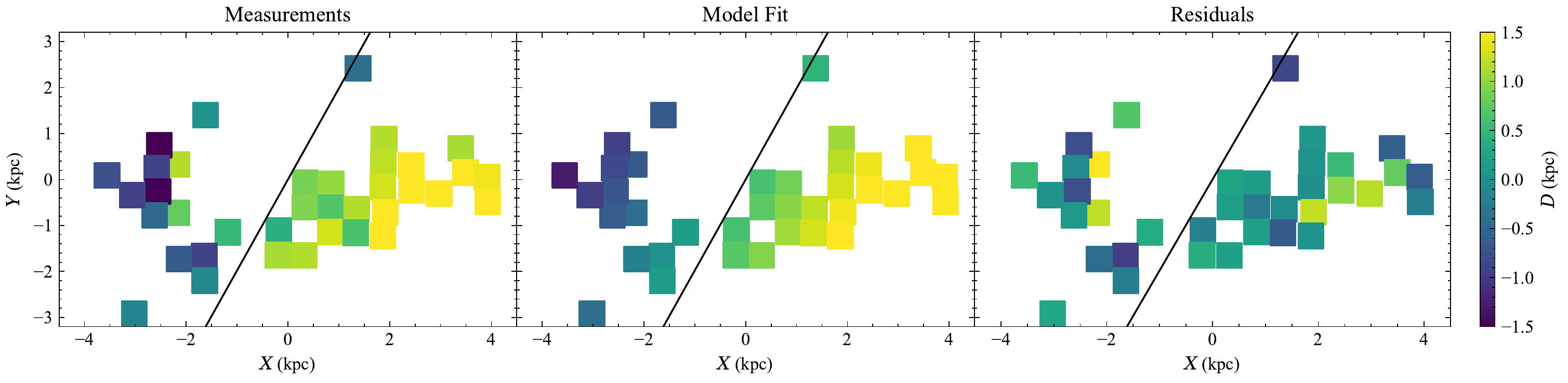}
    \caption{TRGB measurements made to the 41 Rank 1+2 OGLE-III fields used to determine the geometry of the LMC (left), best-fit model predictions (center), and residuals (right). The position of the line of nodes $\Theta = 153 \pm 12 \degree$ is plotted for reference.}
    \label{fig:trgb_geofit}
\end{figure*}

\begin{figure}
    \centering
    \includegraphics[width=\columnwidth]{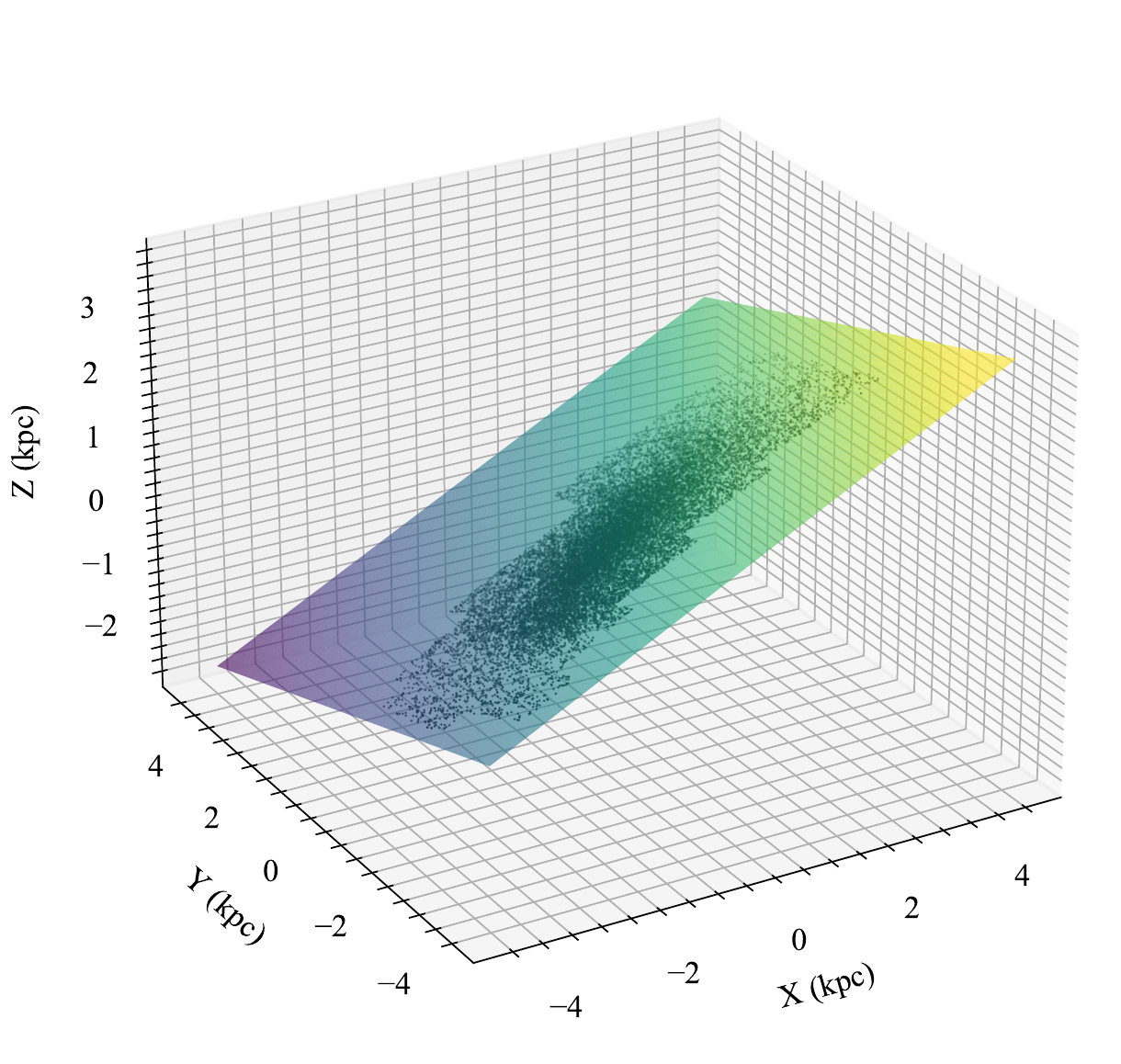}
    \caption{Three-dimensional visualization of the best-fit plane geometry (flat, hued surface) of the LMC using TRGB distances, for a position angle $\Theta = 153 \degree$ and inclination $i = 27 \degree$. For reference, a subset (20000 sources) of the total OGLE-III sample of RGB stars is overplotted (gray points).}
    \label{fig:lmc3d}
\end{figure}

In summary, the observed qualities of the CMDs, RGB LFs, and EDRs were used to rank the 141 spatial bins from 1 to 4 in order of decreasing TRGB measurement reliability. The TRGB Rankings were then compared with SFH maps from \citet{Harris_2009} and the reddening map from \citet{Skowron_2021}. The high-ranking fields were shown to be correlated with low levels of contamination from young and intermediate-aged populations, as well as low gas and dust content, which would otherwise reduce the accuracy of a TRGB calibration.
The fact that this was accomplished using only two-band $VI$ photometry of upper-RGB stars is encouraging for the use of the TRGB in lower S/N regimes (i.e., extragalactic applications) for which detailed maps of reddening and SFH cannot be acquired.

\subsubsection{Constraining the LMC Geometry using the TRGB} \label{subsect:geofit}

In this section, a direct determination of the LMC geometry is made using the TRGB magnitudes measured in the 41 OGLE-III fields classified as Rank 1 or 2. From these measurements, a strong signal of the geometric tilt is immediately apparent from the map of TRGB magnitude as a function of position in the galaxy (left panel of \autoref{fig:trgb_geofit}). Following the prescription of \citet{Nikolaev_2004}, an infinitely thin plane is fit to this distribution of TRGB distances (see middle panel of \autoref{fig:trgb_geofit}). The best fit model ($\Theta = 153 \pm 12 \degree$, $i=27\pm 3\degree$) is consistent with prior measurements, and the residuals are successfully de-trended (right panel of \autoref{fig:trgb_geofit}). In all panels of \autoref{fig:trgb_geofit}, the position of the best-fit line of nodes is shown.

The effect of adopted center on the fit results is tested by adopting any of the centers from \citet[][photometric center]{deVac_1972}, \citet[][Star count center]{Cioni_2000}, \citet[][Center from proper motions of young stars; adopted by P19]{VDM_2014}, and \citet[][center of RRL distribution]{Cusano_2021}. The maximal change to the best-fit parameters was only $1\%$, well within the $\sim 10\%$ uncertainties on the parameter estimates.

The best-fit LMC geometry is consistent with previous studies, including Cepheids \citep{Nikolaev_2004}, DEBs \citep{Pietrzynski_2019}, Carbon stars \citep{Weinberg_2001,VDM_2002}, and RR Lyrae \citep{Cusano_2021}. This provides further evidence that the high-resolution S21 reddening maps accurately trace sight lines shared with TRGB stars (in regions with relatively low levels of star formation and dust content), otherwise the geometric signal would not have been so clearly revealed and found to be consistent with prior studies.

In \autoref{fig:lmc3d}, the best-fit plane is plotted in three-dimensional space, along with a subsample of RGB stars in the OGLE-III photometry. This is the first time the TRGB has been used to measure quantitatively the geometry of the LMC and the results are convincing. This is made possible by the TRGB ranking system established in \autoref{subsect:trgb_measure}, which allowed only accurate TRGB measurements to enter the geometric fit. This consistent result also requires that the TRGB is measurably flat over most of the color range of the LMC, and that the S21 reddening map is accurate. This further supports the TRGB zero point calibration presented here as being the most accurate to date, and potentially the most precise measurement that can possibly be made in the LMC.

\subsubsection{LMC Geometric Corrections} \label{subsect:lmc_geo}
\begin{figure}
    \centering
    \includegraphics[width=\columnwidth]{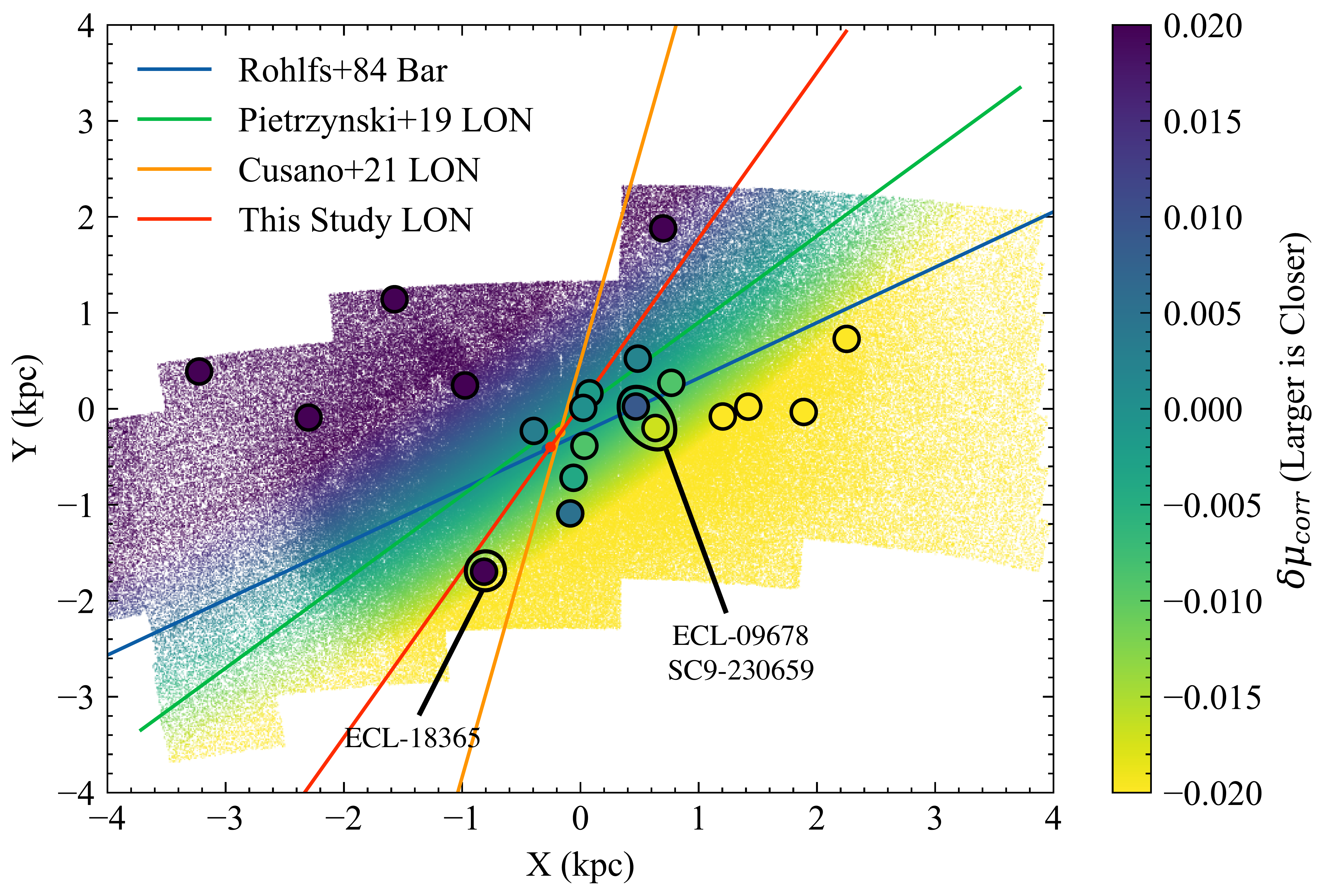}
    \caption{Upper RGB ($V-I > 0.5$~mag, $I < 17$~mag) stars from OGLE-III are plotted and colored according to their associated geometric distortions, as computed with \autoref{eq:weinberg_dist} and the fit parameters from \citet[][P19]{Pietrzynski_2019}. Also plotted are the 20 DEBs from P19, color-coded according to the ``corr'' values tabulated in their Table 1. The lines of nodes from P19 and \citet[][C21]{Cusano_2021} are plotted (blue and orange lines, respectively). For reference, the position angle of the Bar as determined from HI observations \citep[][]{Rohlfs_1984} is plotted (green line). DEBs with anomalous ``corr'' values in P19 Table 1 are annotated and discussed in the text. Equatorial coordinates are converted to physical ones according to the equations of \cite{Weinberg_2001} and $\mu=18.477$~mag. Note the inverted scale as compared to \autoref{fig:trgb_geofit} and \autoref{fig:lmc3d}. This is to be consistent with the P19 representation of the geometric distortions as \emph{corrections} to be made, rather than measured distances.
    }
    \label{fig:lmc_geo}
\end{figure}

In order to accurately calibrate the zero point of the TRGB in the LMC, the effect its three-dimensional structure has on the measured TRGB must be modeled and corrected for. To do so, the line-of-sight distance to a test particle embedded in the infinitely thin LMC plane is computed via Equation (A4) from \citet{Weinberg_2001}, which is reproduced below,
\begin{align} \label{eq:weinberg_dist}
    & r(\alpha, \delta; \{\theta, i\}) = \\
    & -R_{\mathrm{LMC}} \cos i \{ \cos\delta\sin(\alpha-\alpha_0)\sin\theta\sin(i) \nonumber \\
    & + [\sin\delta\cos\delta_0 - \cos\delta\sin\delta_0\cos(\alpha-\alpha_0)]\cos(\theta)\sin(i) \nonumber \\ 
    & + \cos(i)(\cos\delta \cos\delta_0\cos(\alpha-\alpha_0) + \sin\delta\sin\delta_0) \}^{-1}  \nonumber
\end{align}
Note that the angle $\theta$ in the above equation is measured North of West, while the Astronomical convention is East of North, defined as $\Theta = \theta - 90\degree$, following the nomenclature of \citet{VDM_2002}.

Three thin plane geometries are considered here: (1) that determined by P19 from their sample of 20 DEBs ($\Theta = 132\degree$, $i=25\degree$), (2) that determined by \citet[][C21]{Cusano_2021} from a sample of 30000 RR Lyrae observed by the VMC (VISTA Magellanic Clouds) Survey ($\Theta = 167\degree$, $i=22\degree$), and (3) the LMC geometry determined in the previous section ($\Theta = 153\degree$, $i = 27\degree$) using the TRGB directly.
These three set of parameters comfortably cover the typical range of values \citep[e.g.,][]{Weinberg_2001, VDM_2002, Nikolaev_2004} and will therefore successfully probe the effect that adoption of a geometric model has on the final TRGB calibration.
The C21 geometric corrections were found to best sharpen the final TRGB measurement, which is consistent with the RRLs sampling a distribution similar to that of old, metal-poor TRGB stars, though all three sets were consistent to 0.002~mag in terms of the final zero point calibration. See \autoref{app:fields} for a detailed evaluation and comparison of the geometric corrections.

\begin{figure*}
    \centering
    \includegraphics[width=\textwidth]{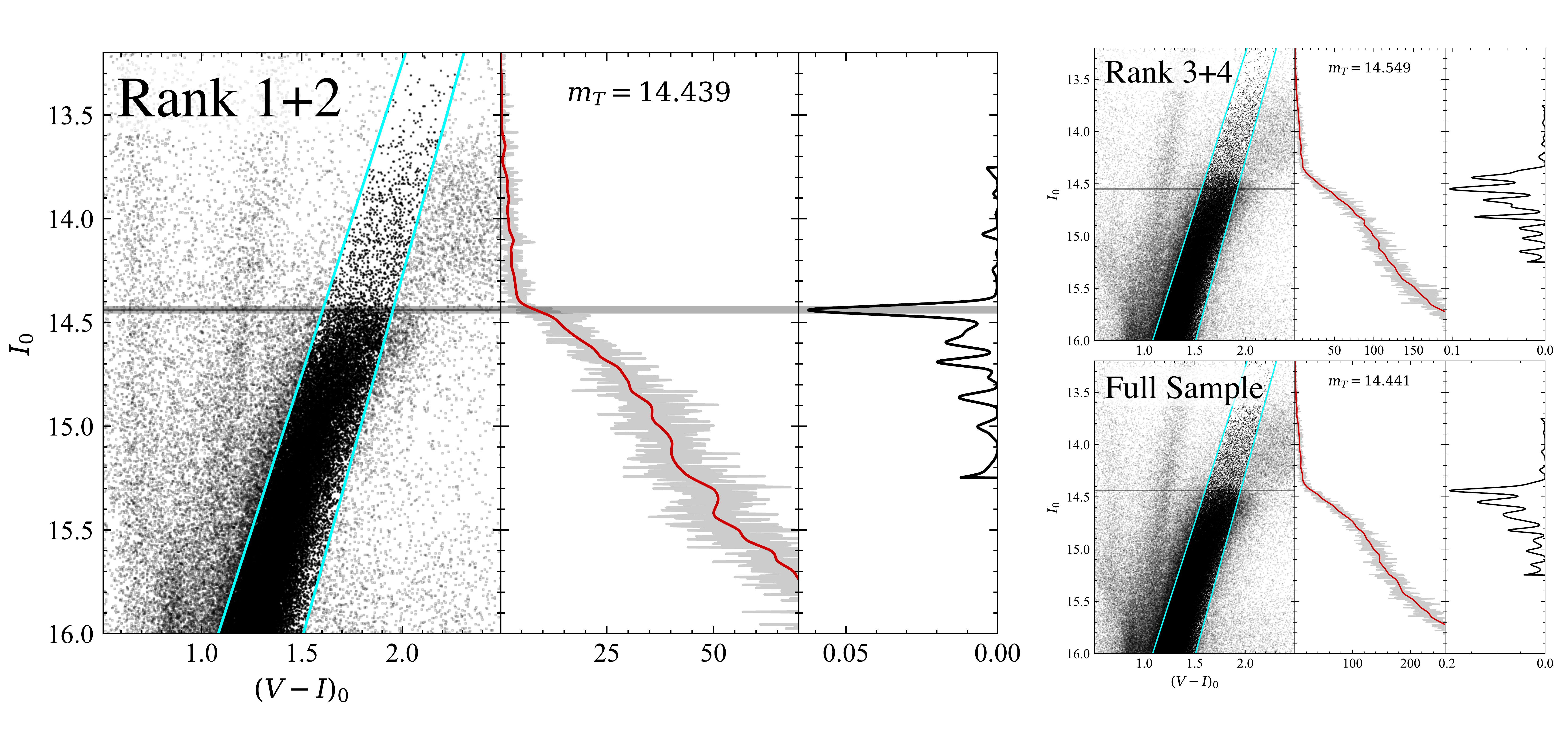}
    \caption{
    TRGB measurements in the LMC for the calibrating Rank 1+2 sample (left), the Rank 3+4 sample (top-right), and the full catalog (bottom-right). All photometry is dereddened using the S21 reddening maps.
    Note the color selection used here (cyan lines) is narrower than that used in the field-by-field measurements (\autoref{fig:lmc_example}), so as to remove metal-rich RGB stars that exhibit a strong dependence of Tip magnitude with metallicity/color (open circles), leaving only metal-poor RGB stars (filled dots) to define the zero point calibration.
    }
    \label{fig:lmc_trgb}
\end{figure*}

To confirm that the distortion corrections computed here are accurate, the predictions from \autoref{eq:weinberg_dist} were compared with the corrections presented by P19 in the ``corr'' column of their Table 1. However, a few significant anomalies (annotated in \autoref{fig:lmc_geo}) were found; these are discussed in detail in the Appendix.\footnote{It appears the depiction of the line of nodes in Figure 2 of P19 was not accurately projected onto the sky, being more skewed from the bar than it should be for the quoted position angle.} In summary, adopting the Corr. values exactly as tabulated in P19 Table 1 actually returned a different weighted average distance ($\mu = 18.4800 \pm 0.0061$~mag) than their own least-squares fit for the LMC geometry ($\mu = 18.477 \pm 0.004$~mag). However, adopting my own distortion corrections (computed with \autoref{eq:weinberg_dist} and the P19 determination of $\Theta$ and $i$) returned a perfectly consistent weighted average distance ($\mu = 18.4768 \pm 0.0046$~mag). For reference, the weighted average of the raw, uncorrected DEB distances is $\mu = 18.4771 \pm 0.0060$~mag. Thus, it is confirmed that the prescription adopted in this paper (\autoref{eq:weinberg_dist}, $\{\Theta, i\}$) for computing distortion corrections must be accurate.

\subsubsection{Composite TRGB Measurements} \label{subsect:lmc_trgb}

\begin{figure}
    \centering
    \includegraphics{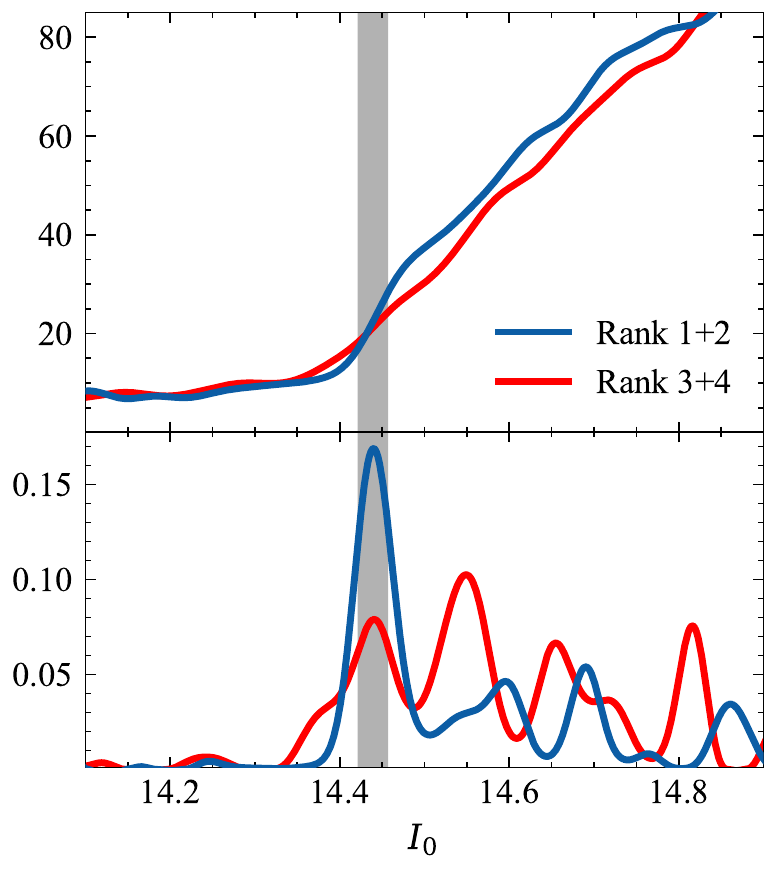}
    \caption{LMC RGB LFs and EDRs for the Rank 1+2 (blue) vs. Rank 3+4 (red) samples. The $2\sigma$ interval of the final TRGB measurement is also shown (gray band). Note the Rank 1+2 curves have been scaled up by a factor of 2.1 to match the number statistics of the Rank 3+4 curves, to enable better visual comparison.
    }
    \label{fig:overplot}
\end{figure}

For the final calibration, photometry of the Rank 1 and 2 fields is combined into a single CMD. This approach is preferred over, e.g., a field-by-field analysis since it is most likely to converge to a representation of the mean population of TRGB stars, making it more resilient to outliers in the distributions of metallicity, dust, and line-of-sight depth of the LMC. Additionally, the S21 reddening measurements made to the Rank 1 and 2 fields should be of higher accuracy than to the Rank 3 and 4 fields because the RC stars observed to these fields dominated by old stellar populations will better match the RC populations in the Outer LMC that defined their calibration. This is demonstrated explicitly in \autoref{fig:lmc_example} and \autoref{fig:example_fields}, where the S21 reddening corrections clearly improved the sharpness and precision of the TRGB feature in the Rank 1 and 2 fields, while the same sharpening was not observed for the Rank 3 and 4 fields.

From the Rank 1+2 CMD, a color-selection box is imposed to isolate metal-poor TRGB stars with magnitudes that displayed no measurable trend with photometric color. The color-selection is defined by a blue edge with slope $-3$~mag~mag$^{-1}$ and TRGB color $(V-I)_0 = 1.60$~mag, and a red edge with slope $-3.5$~mag~mag$^{-1}$ and TRGB color $(V-I)_0 = 1.95$~mag. This specific color range was adopted to capture the Tip stars with magnitudes measured to be flat as a function of $(V-I)_0$ color. Varying the blue edge of the color-cut between 1.45 and 1.70 had no measurable effect on the TRGB magnitude. The metal-rich ($(V-I)_0 > 1.95$~mag) TRGB stars are visibly fainter and the red edge was determined by reducing its location from $(V-I)_0=2.20$~mag until the measured TRGB magnitude no longer increased in brightness, which was at $(V-I)=1.95$~mag.

In \autoref{fig:lmc_trgb}, the CMDs, RGB LFs, and EDRs are plotted for the Calibrating sample (Ranks 1 and 2). For reference, TRGB detections are also shown for the Rank 3+4 fields and the full sample.
The Tip detection in the Rank 1+2 sample is definitive, with the entire ``transition'' region from AGB to RGB populations (i.e., the TRGB feature) spanning only 0.08~mag (a decent visual estimate of the $3$-sigma measurement interval). Indeed, the contrast of the TRGB in this sample is high enough that its magnitude can be identified to very high accuracy from visual inspection of the CMD alone.
By contrast, there is no clear Tip in the low-ranking sample, with the sharp step seen in the high-ranking sample's LF replaced by a comparatively featureless line that extends from 14.4 to 14.8~mag, before turning over to a slightly flatter RGB LF. This broadened ``Tip'' feature is also apparent as a blur of Tip stars in the CMD, and reflected by the forest of equally significant peaks seen in the EDR over the same magnitude range. This smearing of the TRGB transition feature is likely due to significant contamination from younger stellar populations as well as differential reddening.
To highlight these stark differences, in \autoref{fig:overplot} the RGB LFs and EDRs are overplotted for both the Rank 1+2 sample (blue curves) and for the Rank 3+4 sample (red curves).

\subsection{SMC} \label{subsect:smc}

As in the case of the LMC, the SMC OGLE photometry is dereddened using the S21 reddening map and a TRGB measurement is attempted to each of the 40 OGLE-III fields of the SMC, again with a ranking assigned to each measurement.

\begin{figure}
    \centering
    \includegraphics[width=\columnwidth]{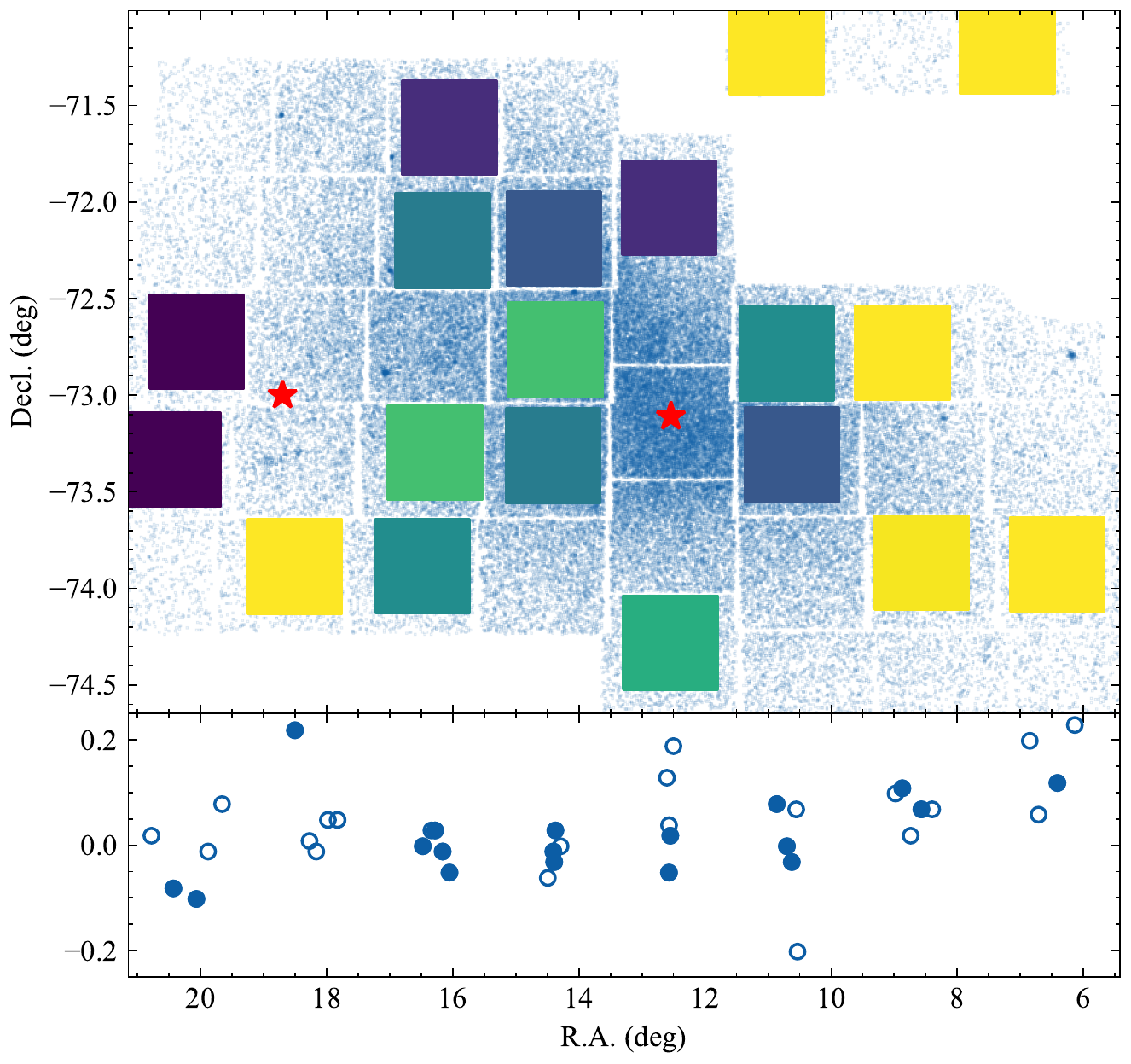}
    \caption{SMC TRGB magnitudes vs. Spatial Position. \textit{Top:} Map of TRGB magnitudes (colored squares) to OGLE-III SMC fields with TRGB features classified as Rank 1 or 2, zeroed to $I_0 = 14.927$~mag. Positions of catalog sources are plotted underneath (blue dots). The adopted positions for the center and the Wing are plotted (red stars). \textit{Bottom:} TRGB distance difference vs. R.A., after marginalizating over Decl. axis for high-ranking (filled circles) and low-ranking (open circles) fields.}
    \label{fig:smc_map}
\end{figure}

\subsubsection{Field-by-field TRGB} \label{subsect:smc_trgbs}

Similar to the case of the LMC, the TRGB is measured to the 40 individual OGLE-III fields that cover the main body of the SMC. The majority of the fields had clear, uncontaminated RGB features, except for a narrow strip of fields coincident with the narrow, star-forming bar, as traced by Cepheids \citep{Scowcroft_2016, Ripepi_2017} and DEBs \citep{Graczyk_2020}. However, the lower number statistics, irregular geometry, and back-to-front depth of the SMC reduced the precision and accuracy of individual field measurements as compared to the LMC measurements (adjacent, high-ranking fields were in some cases measured to be 0.15~mag separated in their TRGB magnitudes). These effects suppressed the likelihood that an accurate geometric model of the SMC could be quantified from the spatial distribution of TRGB magnitudes. However, the data do show a clear E/W gradient in measured TRGB distance, with a composite CMD measurement to the Eastern SMC found to be $\sim\!0.1$~mag more distant at $I_0 = 14.999 \pm 0.04$~mag than the Western side with a TRGB measured to be $I_0=14.907\pm 0.015$~mag. In \autoref{fig:smc_map}, a field-by-field map of the high-quality TRGB magnitudes is shown and the results are qualitatively consistent with prior findings.

\subsubsection{Composite TRGB Measurement} \label{subsect:smc_composite}

\begin{figure}
    \centering
    \includegraphics[width=\columnwidth]{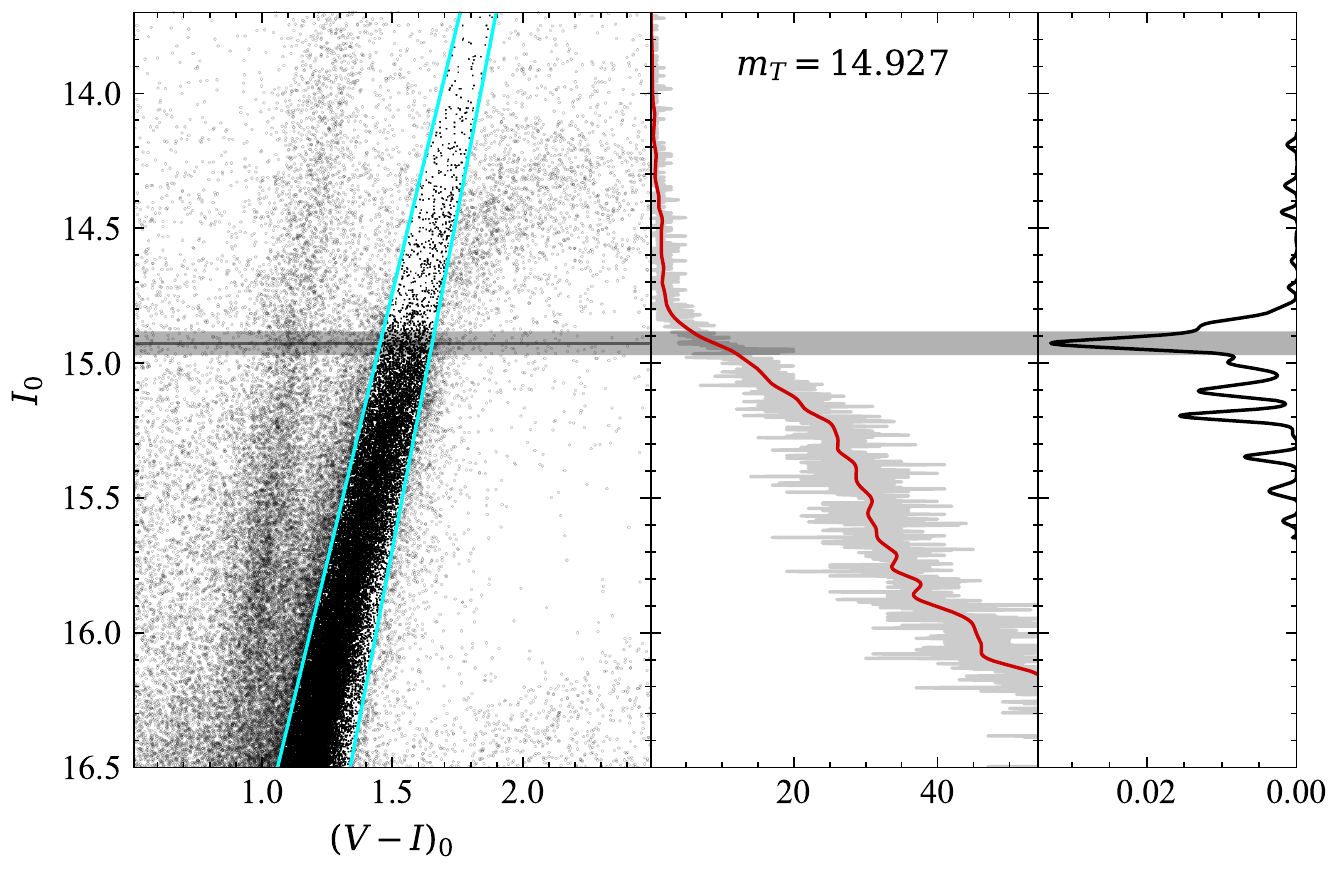} 
    \caption{TRGB detection for the SMC after masking the innermost region ($r_{GC} < 0.4$~deg) and the 0.7 deg centered on the Wing.
    }
    \label{fig:smc_trgb}
\end{figure}

In light of the significant fluctuations discussed in the previous section, the final SMC TRGB measurement will be approached differently from the case of the LMC. Star formation and internal dust reddening are less of a concern in the case of the SMC. The most troublesome source of uncertainty, however, is line-of-sight depth. The galaxy is significantly extended along the line of sight even at its central-most point \citep[4-5~kpc,][]{Muraveva_2018}, with debris and tidal tails exacerbating the problem in the dwarf galaxy's outer regions.

It is therefore decided to use as much area of the SMC main body as possible, since high-frequency variations in line-of-sight depth coupled with low number statistics could induce a selection effect when attempting to break the SMC into smaller regions. Masked from the calibrating sample is the SMC's central 0.4~deg at ($\alpha_0, \delta_0) = (12.54, -73.11)$~deg, shown to host relatively high levels of star formation \citep{Rubele_2018}, and a 0.7~deg radius region centered on the star-forming Wing at ($\alpha_0, \delta_0) = (18.75, -73.10)$~deg. Both regions were classified as Rank 4 and their exclusion from the calibrating sample considerably sharpens the measured TRGB feature, consistent with the findings in the LMC. The TRGB is then measured to be $I_0^{TRGB} = 14.927 \pm 0.022$~mag, with a detailed error budget presented in \autoref{subsect:budget}.

\section{TRGB Calibration} \label{sect:calib}

\subsection{On the Use of DEB distances to Calibrate the TRGB} \label{subsect:debs}

It was shown in \autoref{subsect:lmc_regions} that only roughly one-third of the OGLE-III footprint of the inner LMC is suitable for measurement of the TRGB, with the uncertainties spiking in regions of high recent star formation and observed dust content (the exact contribution of each to the total uncertainty is unclear). This notably includes the stellar neighborhoods within which the DEBs are embedded.
After all, the DEBs are predominantly found in high-reddening, actively star-forming regions of the LMC bar, where variations in the ratio of selective-to-total absorption are expected to be of higher amplitude \citep{Gordon_2003}, the mapping of reddening values from a discrete grid to individual TRGB stars is more uncertain (i.e., higher differential reddening), where there exists a significant population of more massive RGB stars (that reach fainter Tip luminosities), and where mixed populations could bias reddening determinations using the RC.

This fact is rather unsurprising given the young ages of the Araucaria Project's DEB sample \cite[$0.1-2.1$~Gyr,][]{Graczyk_2018}, but is still largely overlooked in recent works \citep{Groenewegen_2019,Yuan_2019} which specifically targeted the TRGB in DEB fields. Note that \citet{Jang_2017_color} also targeted the DEB fields in an effort to reduce the size of geometric corrections, but, different to the other two DEB-based TRGB calibrations, they included an explicit term in their error budget for intermediate-aged contamination (0.02~mag), as well as a conservative estimate of the uncertainty on their average TRGB magnitude (0.04~mag). In \autoref{sect:trgb_measure} and \autoref{app:fields}, the bias in TRGB measurements made in star-forming regions of the LMC was confirmed to be 0.05-0.10~mag.
Based on these findings, it is strongly advised that the LMC DEB fields not be used for calibration of the TRGB.

Indeed, it is not necessary to do so in order for the DEB geometric distance to be applicable to one's set of calibrating stars, so long as the stellar population in question traces the LMC's mean structures and that the accuracy of an adopted model of the geometric distortions can be confirmed (see \autoref{subsect:lmc_geo} and \autoref{app:geo_details}). Indeed, it is possible that the DEB-centric approach does not reduce line-of-sight depth uncertainties at all. After all, while likely, it is not required that the LMC's old stellar populations (typically observed in a more spheroidal distribution than younger stellar populations) are perturbed in exactly the same way as the young stellar component of which the DEBs are a part. On the other hand, it is very likely that the two stellar components share the same \emph{mean distance} to their centers to within the uncertainties. Thus, the optimal route to calibration of the TRGB in this galaxy is to cover as much area as possible in the inner LMC, while retaining only the high-quality TRGB features discussed in \autoref{sect:trgb_measure}. Doing so at once minimizes the systematic uncertainties due to assumed geometric model, dust extinction, and star formation.

A question naturally arises from this discussion: is the star-forming bar (where the majority of the DEBs are measured) distended/disjoint from the body of the LMC where the metal-poor TRGB stars are measured? In other words, is the mean DEB distance to the bar of the LMC located at the same line-of-sight distance as the center of the LMC's older more spheroidal stellar component(s)? Indeed, this possibility has been considered \citep{Zaritsky_2004, Nikolaev_2004}. However, it is encouraging that the TRGB calibration presented here agrees with the canonical value of $-4.05$~mag, and is in statistical agreement with the SMC calibration, suggesting that the amplitude of such an effect, if any, is below the uncertainties of the present calibration. Recently, \citet{Cusano_2021} showed that a sample of 30000 RRL in the LMC follow a smooth almost-spheroidal distribution, and this is likely the case for the calibrating TRGB stars used here.

\subsection{TRGB Zero Points} \label{subsect:zpcal}

Here, the composite TRGB measurements presented in \autoref{sect:trgb_measure} will be tied to the P19 and G20 DEB distances from the Araucaria Project, and high-accuracy zero point calibrations will be determined separately for the SMC and LMC.

For the LMC, the uncertainty in the P19 distance estimation is dominated by the remaining uncertainty ($0.018$~mag) in calibration of the surface brightness-color relation for the late-type DEBs \citep{Graczyk_2018}. Combining the P19 DEB distance with the Rank 1+2 TRGB magnitude, the TRGB zero point over the color range $1.60 < (V-I)_0 < 1.95$~mag is determined to be $M_I = \MtrgbLMC \pm \calSTATerrLMC_{stat} \pm \calSYSerrLMC_{sys}$~mag. 

\citet{Graczyk_2020} presented a distance to the SMC with a quoted accuracy to better than 2\%. They considered four separate models of the SMC geometry using their DEB distances, averaged the model fits, then took the standard deviation of the modeled distances as the statistical uncertainty on their final distance. Combining the measured TRGB magnitude $I_0^{TRGB} = 14.927 \pm 0.022$~mag  with the G20 DEB distance $ \mu = 18.977 \pm 0.016 (stat) \pm 0.028 (sys)$~mag, the TRGB zero point over the color range $1.45 < (V-I)_0 < 1.65$~mag is determined to be $\MtrgbSMC \pm \calSTATerrSMC_{\mathrm{stat}} \pm \calSYSerrSMC_{\mathrm{sys}}$.

The detailed error budgets for these zero point calibrations is presented in \autoref{subsect:budget}.

\subsection{Color Dependence of the $I$-band TRGB}\label{subsect:trgb_colors}

\begin{figure}
    \centering
    \includegraphics[width=\columnwidth]{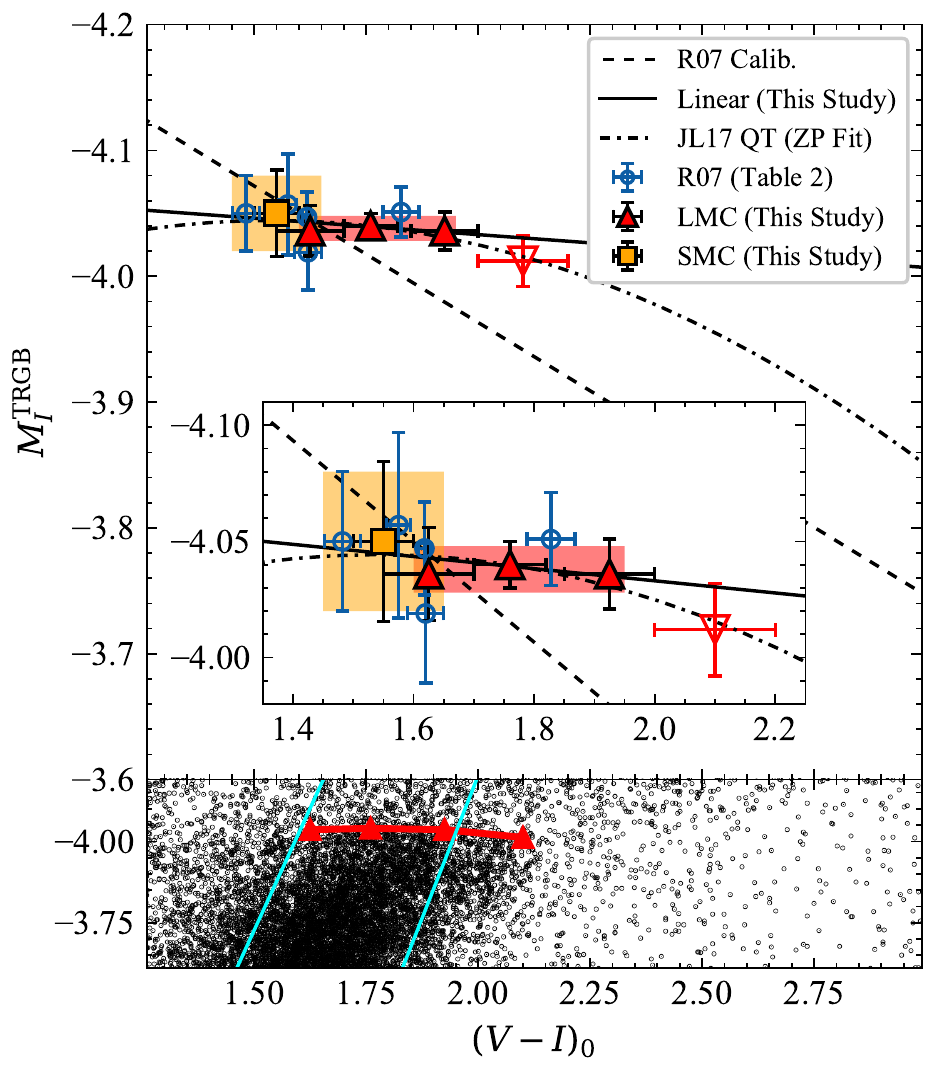}
    \caption{Color dependence of the $I$-band TRGB. Plotted are TRGB measurements from three color bins in the LMC (red triangles), the one measurement from the SMC (orange square), and TRGB zero point calibrators from Table 2 of \citet[][blue circles]{Rizzi_2007}. Filled points are used to constrain a linear dependence of the TRGB magnitude on color (solid black line). All points (except the R07 calibrators) are used to determine a zero point calibration of the QT color calibration (dot-dashed black line) from \citet{Jang_2017_color}. Also plotted are the new flat zero point calibrations for the SMC (orange band) and LMC (red band). For reference, the $VI$ slope of 0.217 mag mag$^{-1}$ from R07 is plotted (dashed black line). \textit{Bottom:} The LMC Rank 1+2 CMD is shown with the corresponding color bin measurements (red triangles) and the color cut used for the flat zero point calibration (cyan lines).
    }
    \label{fig:rizzi_comp}
\end{figure}

It is known that the metallicities and photometric colors of TRGB stars follow a well-defined relation \citep{Dacosta_1990, Bellazzini_2001, Valenti_2004}, and that in the $I$-band this dependence is flattened considerably, particularly for blue (metal-poor) Tip stars. 

There are two frequently-used methods to control for curvature in the I-band TRGB sequence. The first is the use of a color cut to exclude metal-rich TRGB stars that exhibit a much stronger steeper dependence than their metal-poor ($[Fe/H] < -0.5$~dex) counterparts. This has recently been used by the CCHP \citep{Beaton_2016, Freedman_2019} to measure $H_0$. In some cases \citep[e.g.,][]{Hatt_2018, Hoyt_2019}, the photometry in the $V$-equivalent band is sufficiently shallow that incompleteness and stringent data quality cuts replicate the effects of a red-edge color cut. This is the case for the bulk of the CCHP sample \citep[though see][for evidence of a 0.02~mag color-dependent effect in the case of M96]{Mager_2020}.

The other widely used method is application of an explicit color-correction \citep[e.g.,][]{Rizzi_2007, Jang_2017_color}. A drawback of this implementation is that high signal-to-noise imaging in the bluer band (typically V, or a similar effective wavelength) is required when attempting to measure a TRGB distance, otherwise the distance uncertainty is inflated significantly. This is why the CCHP uses the metallicity-independent, color cut approach to reduce the HST time needed to measure TRGB distances.

In terms of previous calibrations of the TRGB color dependence, \citet[][R07]{Rizzi_2007} used HST imaging of nearby galaxies to determine the slope of the $I$-band TRGB on the HST filter system. \citet[][JL17]{Jang_2017_color} performed an analysis similar to R07 but looked for an additional degree of freedom, i.e., a quadratic dependence of the TRGB magnitude with photometric color. Their quadratic fit to the TRGB was determined with HST photometry of TRGB stars in the outer regions of eight nearby, massive galaxies.

These two canonical color-corrections will now be discussed in the context of the new TRGB measurements in the MCs made here. Additionally, two new fits to the TRGB color dependence are presented: one including only TRGB stars with $(V-I)_0 < 1.95$~mag, and another using the full range of observed TRGB colors. To do so, the composite CMD in \autoref{fig:lmc_trgb} is divided into three color bins with a Tip magnitude measured in each bin (red points in \autoref{fig:rizzi_comp}). 
The color bins that define the LMC datapoints were varied in size and location. It was found that smaller bins than the $\sim 0.15$~mag ones adopted were too sensitive to small shifts to the bin edges, which would have introduced an additional uniqueness problem. It is thought that this stochasticity in small color bins of the composite CMD is due to residual white noise after application of the reddening and geometric corrections. Thus, the wider bins are adopted so the samples are large enough to ensure the TRGB magnitude measured in each color bin corresponds to the \emph{mean} population of TRGB stars located in these low SFR regions of the LMC, and is not skewed by any one spatial bin.

Added to the LMC points is the composite SMC measurement of \autoref{subsect:smc}, with the relative DEB distance uncertainty \citep[0.017~mag,][]{Graczyk_2020} added in quadrature to the measurement uncertainty (orange point in \autoref{fig:rizzi_comp}). This leaves four datapoints with uncertainties between 0.015~mag (for the central LMC points) to 0.035~mag (for the SMC). The R07 ground-based zero point calibrators (blue points) are plotted along with the flat LMC and SMC zero points (red and orange bands, respectively).

From the new MC data, insignificant evidence (a p-value of 0.48) is found for a color dependence for colors $(V-I)_0 < 1.95$~mag, tabulated as ``Linear (blue)'' in \autoref{tab:calibs}. Stronger evidence (p-value of 0.09) is found when including the reddest datapoint while the RMS scatter increases from 0.004~mag to 0.006~mag.

In either case, it is clear that these data on their own are nonideal for measuring the color dependence of the TRGB. Thus, the quadratic color dependence (QT) from JL17 is adopted, with a new zero point of their calibration determined here. The QT fit produces the smallest RMS scatter ($\sigma = 0.004$~mag) over the largest color range. Along with the flat LMC calibration, this is the calibration preferred by this author, with its zero point calibration valid over the full color range $1.45 < (V-I)_0 < 2.20$~mag, and is bolded in \autoref{tab:calibs}.

On the other hand, it becomes evident that the R07 TRGB slope in the J-C $VI$ filters (0.217 mag~mag$^{-1}$) is likely overestimated. The RMS scatter about the R07 line is nearly an order of magnitude larger than in the QT fit, ranging from 0.027 to 0.033~mag depending on the adopted color range. The cause of this is unclear, but given the consistency of their native HST filter slopes with the JL17 color calibration, the discrepancy is likely caused by inaccurate or outdated filter transformations from their native space-based measurements to ground-based J-C filters.

\begin{deluxetable*}{lllllc}
\tabletypesize{\small}
\tablecaption{Final TRGB Calibrations \label{tab:calibs}}

\tablehead{
\colhead{Type}        &
\colhead{Color Range} &
\colhead{Calibration} &
\colhead{P-value}   &
\colhead{Fit R.M.S.}  &
\colhead{Reference}
}
\startdata
Flat (SMC)   & $1.45 < (V-I)_0 < 1.65$ & $-4.050 ( \pm  0.022) $ & \nodata & \nodata & \ref{subsect:zpcal}\\
\textbf{Flat (LMC)}   & $1.60 < (V-I)_0 < 1.95$ & $-4.038 ( \pm  0.009) $ & \nodata & \nodata & \ref{subsect:zpcal}\\
Linear(Blue) & $1.45 < (V-I)_0 < 1.95$ & $-4.039 ( \pm  0.002) + 0.015(\pm0.018)(V_0 - I_0 -1.75)$ & 0.48 & 0.004 & \ref{subsect:trgb_colors} \\
Linear(All)  & $1.45 < (V-I)_0 < 2.20$ & $-4.036 ( \pm  0.003) + 0.051(\pm0.022)(V_0 - I_0 -1.80)$ & 0.09 & 0.006 & \ref{subsect:trgb_colors} \\
\textbf{QT (ZP fit)}  & $1.45 < (V-I)_0 < 2.20$ & $-4.044 ( \pm  0.002) + 0.091\tablenotemark{a}(V_0 - I_0 -1.5)^2 -0.007\tablenotemark{a}(V_0 - I_0 -1.5)$  & \nodata &  0.004 & \ref{subsect:trgb_colors} \& JL17
\enddata
\tablenotetext{a}{Adopted directly from \citet{Jang_2017_color}. The reader is referred to that study for uncertainties on the adopted curvature parameters and for transformations to HST bandpasses.}
\tablecomments{Preferred calibrations are bolded. Only statistical/formal uncertainties are quoted. For full uncertainties on the Flat (SMC) calibration, see the SMC column of \autoref{tab:budget}. For all other calibrations, the full zero-point uncertainty should be adopted from the LMC column of \autoref{tab:budget}.}
\end{deluxetable*}

One final piece of information to consider is the universality of the newly presented calibrations. A longstanding question of the TRGB concerns the extent to which age can shift the observed colors and magnitudes of TRGB stars, potentially breaking the assumption of universality in any single proposed calibration \citep{Salaris_2005_badtrgb}. Encouragingly, in this section it was shown that the JL17 QT color dependence -- based on observations in the stellar halos of $L_*$ galaxies -- describes very well the TRGB magnitude-color relation in the MCs (this study), Local Group dwarfs, and M33 \citep{Rizzi_2007}. This consistency indicates that for RGB stars found in these environments either: the age distributions are identical, or that age-dependent variations in the $I$-TRGB magnitude are minimal. In either case, the $I$-TRGB appears well-behaved and without a measurable bias across these host environments.

\subsection{Error Budget} \label{subsect:budget}

Described in this section is the error budget associated with the zero points in the newly presented TRGB calibrations. The uncertainties are tabulated in \autoref{tab:budget}.

\begin{deluxetable}{lllll}

\tabletypesize{\small}

\tablecaption{TRGB Zero-point Error Budget \label{tab:budget}}

\tablehead{
\colhead{Error Term} &
\multicolumn{2}{c}{LMC} &
\multicolumn{2}{c}{SMC} \\
\colhead{}      &
\colhead{stat.}  &
\colhead{sys.}   &
\colhead{stat.}  &
\colhead{sys.}  \\
\colhead{}      &
\colhead{(mag)} & 
\colhead{(mag)} & 
\colhead{(mag)} & 
\colhead{(mag)} } 
\startdata
Edge Detection    & 0.008   &  0.004  & 0.022  &  0.006   \\
Sample Selection  & 0.005   & 0.002 & 0.007 & 0.01 \\
Extinction        & 0.006   &  0.014  &  0.004  &  0.018  \\
DEB Distance\tablenotemark{a}      & 0.004   &  0.026  &  0.019  &  0.028  \\
Photometry        & 0.004 &  0.01   & 0.004 &  0.01   \\
\hline
Final Uncertainties & 0.012 &  0.032  & 0.030 & 0.039
\enddata
\tablenotetext{a}{LMC distance from \citet{Pietrzynski_2019} and SMC Distance from \citet{Graczyk_2020}.}
\end{deluxetable}

For the Edge Detection uncertainty, the error estimation methodology described in \autoref{subsect:lmc_trgb} is used. For the LMC, $\sigma_{min} = 0.007$~mag is found, and for the SMC 0.01~mag. Similarly, the quantity $\Delta m = 0.004$~mag for the LMC and 0.02~mag for the SMC, resulting in total statistical uncertainties 0.008~mag and 0.022~mag, respectively.
The TRGB magnitude is then measured with and without the Poisson weighting function, with the difference between the two adopted as an additional systematic error. In the LMC, this amounted to a 0.004~mag difference and in the SMC a 0.006~mag difference. Both values are adopted as systematic uncertainties on the edge detection.

To estimate the statistical uncertainty due to adoption of the S21 reddening map, their $\sigma_1$ and $\sigma_2$ quantities are used. These quantities are intended to characterize the amplitude of back-to-front differential reddening inherent to each S21 pixel, i.e., the statistical uncertainty associated with discretely sampling the distribution of dust along LMC sight lines. To quantify the effect this uncertainty has on the present calibration, the peak of the $\sigma_1$ distribution for the calibrating sample of TRGB stars (determined by selecting for stars within $0.02$~mag of the adopted TRGB magnitude) is divided by the square root of the sample size, resulting in $0.065/\sqrt{156} = 0.005$~mag for the LMC and $0.05/\sqrt{168}=0.004$~mag for the SMC. These quantities, after conversion to $A_I$, are included as statistical uncertainties in the Extinction row of \autoref{tab:budget}.

The dominant uncertainty in the reddening lies in its zero point. Systematic uncertainties on the S21 zero point of 0.014~mag and 0.018~mag are adopted for the LMC and SMC, respectively. These values are taken directly from the /emph{total} RMS scatter in the S21 calibration of the RC mean color, which used the SFD98/SF11 dust maps, as well as an empirically calibrated RC color vs. radius relation, to determine the RC intrinsic color throughout the MCs, ensuring that their reddenings are on the same system as the widely-used foreground dust maps (arguably the most powerful aspect of the S21 reddening map in terms of calibrating the TRGB), thereby making the present TRGB Calibration on the same extinction zero point as any TRGB measurement to which only integrated foreground reddening corrections are required (i.e. high galactic latitude targets).

The color-metallicity relation of the RC plays an important role in the S21 maps, especially since it is required to extrapolate the RC intrinsic color from the outer regions of both Clouds (where S21 could directly to the SFD98/SF11 maps) to the inner regions as used for the present calibration. To perform this extrapolation to the MC inner regions, S21 measured two empirical relations: a metallicity-radius relation for each Cloud and a color-metallicity relation for RC stars. Notably, both of these were determined empirically with measurements made in the MCs themselves (spectroscopic metallicities of red giants and star clusters paired with OGLE-IV photometry).
Also, the RC color-metallicity relation in S21 and the theoretical one presented in \citeauthor{Nataf_2021} differ by no more than 0.02~mag at either end of the LMC's metallicity range, suggesting the dependence of the RC color on metallicity is well-behaved. It is thusly concluded that the current adopted systematic reddening uncertainties (again, taken to be the \emph{full} scatter in the $(V-I)_{RC}$ vs. $E(V-I)_{SFD98/SF11}$ vs. radius relation for each Cloud) sufficiently account for uncertainties in the RC color-metallicity relation. Furthermore, any age effects on the RC intrinsic color are minimized by the use of only low-SFR regions of each Cloud, ensuring the population of RC stars that probe the reddening to the Calibration sample best match the population that defines the S21 calibration based in the Outer regions of each Cloud.

To estimate statistical uncertainties in the LMC calibration due to sample selection, the Rank 1+2 sample is jack-knife resampled to potentially identify anomalous detections. The final TRGB magnitude is found to be robust to resampling at the 0.005~mag level, and that value is adopted as a statistical uncertainty. In the case of the SMC, the masked inner region boundary is varied from $r=0.2$~deg to $r=1.0$~deg, with the full range of measured Tip values spanning 0.007~mag. That value is adopted as a statistical uncertainty under ``Sample Selection'' in \autoref{tab:budget}. Varying the boundary of the also-masked Wing region over the same range of values had no effect on the measured TRGB magnitude.

An additional systematic uncertainty in the SMC is determined by considering the effect that choice of adopted center has on the final TRGB magnitude. Considering a grid of centers that contain the values determined by \citet{Cioni_2000}, \citet{Ripepi_2017}, and \citet{Muraveva_2018}, the uncertainty due to adopted SMC center on the TRGB calibration is estimated to be 0.01~mag and this value is tabulated under the ``Sample Selection'' row in \autoref{tab:budget}.

In the case of the LMC, the effect of adopted plane geometry, from which distortion corrections were applied on a star-by-star basis, was also considered. Using any of the three sets of corrections result in TRGB magnitudes consistent to within 0.002~mag. Furthermore, it will be shown in \autoref{subsect:jl17} that using an \emph{empirical} set of distortion corrections is also consistent with the three adopted models to within 0.002~mag. This value is therefore adopted as a systematic uncertainty in the ``Sample Selection'' row of \autoref{tab:budget}.
In \autoref{subsect:geofit}, a direct fit was made to the LMC geometry using the TRGB distances. However, that determination is highly dependent on the S21 reddening map, which was essential in revealing the geometric tilt of the LMC. Therefore, to avoid circularity and keep the reddening and geometric calibrations independent from one another, the TRGB-determined geometry is not adopted for the calibration. It is simply noted that the newly-determined TRGB geometry is consistent with both the C21 and P19 geometries. The only remaining unknown uncertainty associated with the LMC geometry lies in the difficult-to-constrain possibility that the center of the distribution of metal-poor RGB stars in the LMC could be mis-aligned with the center of the young stellar component (traced by the DEBs) along the line of sight. Given the LMC's recent interaction with the SMC, this could be a possibility, but there is no justification that such a displacement, if any, would be significant. The uncertainties already folded into the P19 DEB distance are therefore considered more than sufficient to encompass any potential bias due to this phenomenon.

In the case of the SMC, its entire main body, after masking the regions of highest star formation, was used to calibrate the TRGB, thus any lateral distortions due to the galaxy's tilt should be symmetric and negligible in the average. However, there is still a substantial uncertainty due to the back-to-front depth, which should be symmetric, and thus a statistical error term. \citet{Muraveva_2018} determined the line-of-sight depth of the SMC's spheroidal component to be 4~kpc ($\pm 0.14$~mag) which places an absolute upper limit on our ability to converge on a central value for the mean distance to the SMC. If we assume one TRGB star provides an independent sampling of the mean SMC distance distribution, then the uncertainty due to back-to-front depth can be estimated via the standard error, i.e., $0.14/\sqrt(168) = 0.011$~mag where the number of TRGB stars was estimated an identical manner to the reddening uncertainty estimation earlier in this section. Of course, the TRGB magnitude cannot be defined by only one star, so this quantity is best seen as a lower limit. Therefore, the uncertainties on the Tip detection ($\sim 0.02$~mag) and DEB distance (also 0.02~mag) are considered a more robust constraint on this effect, estimated here to be 0.03~mag.

In the case of the LMC, there exist two OGLE photometric catalogs, the primary photometric maps \citep{Udalski_2008_maps} and the OGLE-Shallow photometry \citep{Ulaczyk_2012}. The two produced slightly different (0.004~mag) composite TRGB results and that quantity is adopted as an additional statistical uncertainty (since there is no clear choice of which set of photometry is the ``true'' baseline). The same quantity is also propagated to the SMC, despite the lack of an equivalent ``Shallow'' catalog to perform the same comparison. Lastly, for both calibrations a systematic uncertainty equal to 0.01~mag is adopted based on the OGLE collaboration's adopted uncertainty on their photometric zero points. 

\subsection{Prescription for Using this Calibration}
\label{subsect:guide}
If the reader is interested in using one of the calibrations in \autoref{tab:calibs} to measure a TRGB distance (preferred calibrations are bolded), it is strongly recommended to first ensure that any potential mismatch with the OGLE $I$ filter is taken into account (differences in $V$ that would affect the color dependence are negligible compared to potential zero point offsets between differently-defined $I$-band filter systems).

Then, if using a non-flat calibration, either use the color correction to flatten the TRGB feature in the CMD before measuring it \citep[a la][]{Madore_2009} or first measure the TRGB, then compute a color-dependent correction based on the mean color of the TRGB stars in question. If the color range of observed TRGB stars is larger than 0.2~mag, it is highly recommended that the former method is used, because in the latter case the Tip measurement to a sloped TRGB sequence can be biased. Furthermore, the use of a single mean color to represent the TRGB color distribution neglects non-linearities in the TRGB magnitude-color relation and implicitly assumes a normal distribution of TRGB colors, which is usually not the case due to the ``fanning out'' behavior of the TRGB feature (this problem will be discussed in detail in \autoref{subsect:g18}).

Once the color rectification (or mean correction, but again only if considering a narrow range in TRGB colors) has been made and an apparent magnitude has been determined, the zero point of the respective calibration can be used to determine a distance. Of course, the flat zero point calibrations can instead be directly used, as long as the color range of the calibration (SMC or LMC) matches the target galaxy's TRGB color range. 

No matter the calibration adopted, the uncertainties in the zero point will come from \autoref{tab:budget}. For example, if using the QT calibration, the uncertainty on its zero point would come from the LMC column of \autoref{tab:budget}, with the 0.008~mag Edge Detection statistical uncertainty replaced with the 0.002~mag standard error on the zero point fit parameter tabulated in \autoref{tab:calibs}. 

\section{Discussion} \label{sect:discussion}

\subsection{Optimal Measurement of the TRGB in the LMC} \label{subsect:sfr_trgb}

It was shown in \autoref{subsect:trgb_measure} and discussed in \autoref{subsect:debs} that the Araucaria Project's late-type DEBs are located in regions of the LMC associated with low quality TRGB features, caused by high rates of recent star-formation and dust content (see \autoref{fig:fields} and \autoref{fig:lmc_fields2}), and that a TRGB calibration in those fields should be avoided. In star-forming regions such as these it is suspected that the following systematic effects are at play: (1) the intermediate-aged stellar populations in these regions include higher mass RGB stars that reach \emph{lower} luminosities at the Tip of the RGB and are no longer standard candles, and (2) decreased precision in assigning a local reddening value to individual RGB stars, i.e., larger absolute reddening values signal concomitantly larger differential/internal reddening, as well as larger variations in $R_V$ along the same sightline. In \autoref{fig:lmc_trgb} and \autoref{fig:overplot}, it can be seen from the Rank 3+4 sample that these astrophysical systematics introduce multiple false edges in the RGB LF, effectively blurring the TRGB discontinuity by at least a factor of three or four when compared to the ideal case.

Measuring the TRGB in star-forming regions can lead to inconsistencies in defining the actual location of the TRGB. \citet{Groenewegen_2019} tested two different edge detection implementations for TRGB measurements made to fields centered on Cepheids and DEBs of the inner LMC. They found 0.02-0.04~mag offsets when using either a first- or second- derivative kernel to locate the strongest edge feature in their luminosity functions. A similar effect was seen in the Rank 3 and 4 fields here, where slight changes to the edge detection such as smoothing window size and choices of edge detection kernel introduced identical shifts in the measured location of the TRGB edge. This is also likely an explanation for why \citet{Gorski_2018} found that a commonly used maximum likelihood approach (which assumes a simple single-population RGB) to measuring the TRGB did not converge in their fields.

This problem is exacerbated when a field-by-field analysis is undertaken, as opposed to combining all photometry of a single galaxy into a composite CMD. Unless TRGB measurements to star-forming regions are properly controlled for, then field-by-field TRGB calibrations like those undertaken by \citet{Jang_2017_color} and \citet{Gorski_2018} will be systematically biased and be of poor precision, because low-quality Tip features (shown to be systematically biased to faint magnitudes) from low-ranking fields skew the average to biased values.

On the other hand, if the fields are combined into a composite CMD, then statistical power is aggregated into the discontinuity defined by the true, standard candle TRGB, while the nonstandard false edges from the low-ranking fields will simply add baseline noise to the RGB LF and EDR. This is demonstrated in \autoref{fig:lmc_trgb}, where the full sample (bottom-right) returns a TRGB magnitude that is fully consistent with the Rank 1 and 2 detection, albeit with much more structure in its LF and EDR. In the full sample measurement, the true TRGB magnitude (verified through consistency with the Rank 1+2 measurement) rises above the noise (sourced by the Rank 3+4 fields) despite the Rank 3 and 4 fields contributing roughly three times the number of stars as the Rank 1+2 sample.
This resilience of the composite approach to false TRGB features will be demonstrated again in \autoref{sect:lmc_discuss}, where the field-by-field calibrations of \citet{Jang_2017_color} and \citet{Gorski_2018} will be redetermined using a composite CMD, with the reddening and/or SFR biases of both studies subsequently corrected.

\subsection{An External Test of the LMC Calibration and Implications for the TRGB Distance Scale} \label{subsect:h0}

In this section, the new flat LMC calibration will be used to perform a fully independent check on the majority of the CCHP TRGB distances, and thus their measurement of $H_0$ \citep{Freedman_2019}. To do so, the CCHP distances (after shifting onto the new LMC zero point presented here) are compared with measurements made by the CosmicFlows collaboration \citep{Tully_2016} as tabulated in the Extragalactic Distance Database \citep[EDD,][]{Anand_2021}.

This comparison is based on the same imaging datasets, but the independent analyses differ in all other aspects, namely the EDD pipeline uses a photometric reduction program (DOLPHOT) different from the CCHP's (which uses DAOPHOT). Also, as already mentioned, the EDD employs an explicit color-correction for TRGB magnitudes \citep{Rizzi_2007}, as opposed to the CCHP's blue color-cut approach.\footnote{Note the CCHP does not employ an explicit color-selection. Instead, stringent signal-to-noise cuts made in the F606W-band replicate the effect, leading to a bluer TRGB sample than that measured by, e.g., the CosmicFlows team.} Furthermore, the EDD determines the TRGB magnitude via maximum likelihood fit, while the CCHP employs a nonparametric edge detection (identical to that used in this study). The details of the EDD methodology are summarized in \citet{Makarov_2006,Rizzi_2007,Jacobs_2009} and the details of the CCHP methodology are summarized in \citet{Beaton_2016, Hatt_2017, Freedman_2019}.

Of the 17 galaxies in the CCHP TRGB distance sample \citep{Freedman_2019, Jang_2021, Hoyt_2021}, 12 overlap with the current EDD database. These galaxies are, in order of distance: M101, NGC~4258, M66 (NGC~3627), M96 (NGC~3368), NGC~5643, NGC~4536, NGC~4526, NGC~4424, NGC~1448, NGC~1365, NGC~1316, and NGC~4038.\footnote{A more detailed comparison between the EDD/CosmicFlows and CCHP TRGB distances is forthcoming from the CosmicFlows team (G. Anand, priv. comm) and will likely incorporate the remaining CCHP galaxies that do not currently overlap with the EDD sample.} A direct comparison between the re-scaled CCHP distances and the exact quoted EDD distances returns a 0.024~mag offset, already a promising level of agreement.
\begin{figure}
    \centering
    \includegraphics[width=\columnwidth]{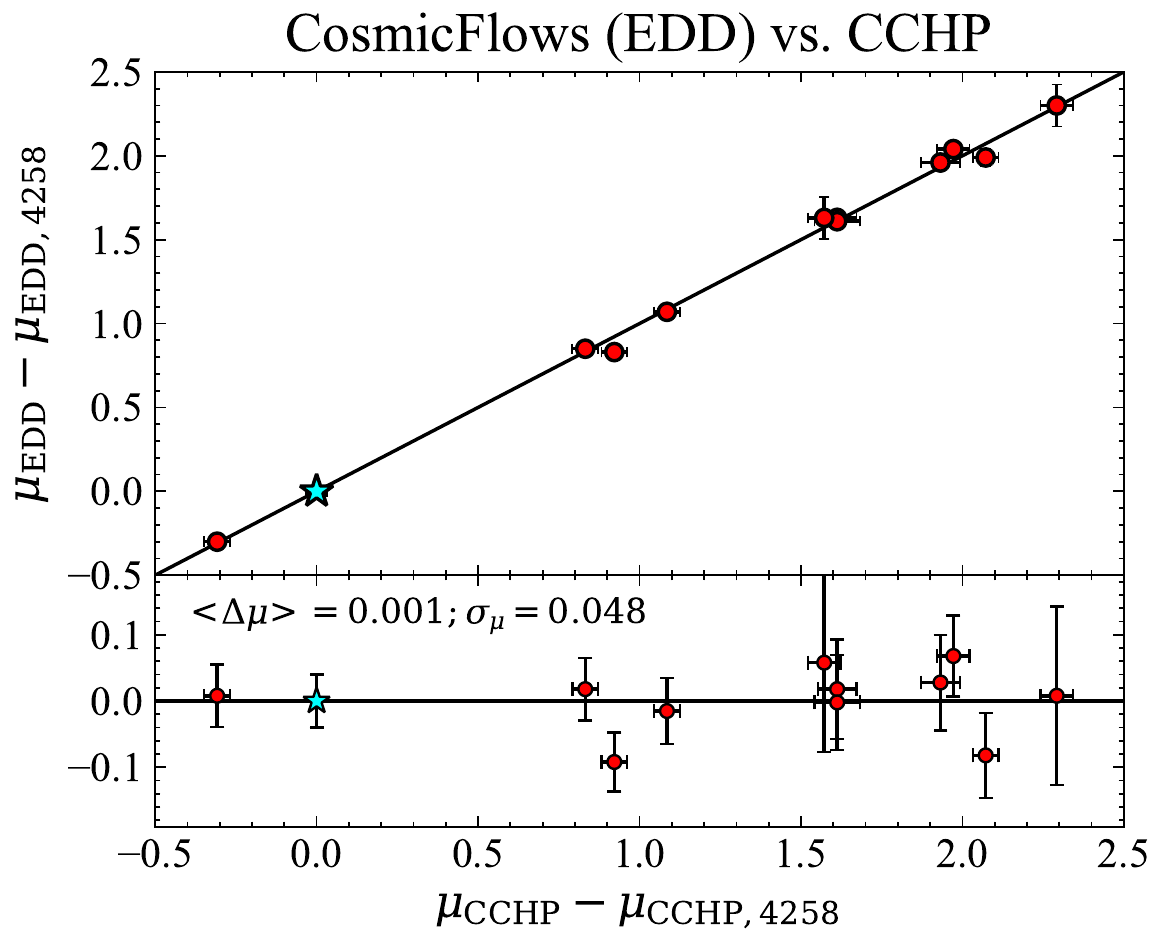}
    \caption{
    Comparison of CosmicFlows \citep[EDD,][]{Anand_2021} TRGB distances and CCHP \citep{Jang_2017_h0, Freedman_2019, Jang_2021, Hoyt_2021} distances made from the same HST imaging sets. CCHP distances were rescaled by 0.009~mag to match the independent TRGB zero point determined here from the LMC. Each set of distances is zeroed by each group's TRGB measurement to the geometric anchor galaxy NGC~4258 (blue star). In the top panel, a one-to-one line is shown for reference. In the bottom panel, the median offset value is shown (dashed line) along with the line of zero (solid line). The EDD entry for the foreground extinction to the galaxy NGC~5643 has been corrected from $A_I=0.278$~mag to $0.257$~mag, according to the SFD98/SF11 maps as compiled by IPAC/NED. The consistency between the CCHP TRGB zero point and the new one determined in this study, coupled with the excellent agreement with the independent EDD distances, provides strong evidence in support of the CCHP's TRGB-anchored measurement of $H_0$.}
    \label{fig:tully_comp}
\end{figure}

To test for a real systematic difference between the two sets of measurements, i.e., what would each set of measurements have to say about $H_0$, each group's distance scale is zeroed to their respective distance measured to NGC~4258, for which an independent geometric distance exists. Doing so will remove systematic differences between each group's set of TRGB distances (down to differences in measurement methodology and zero points), and allow for a homogeneous comparison. After zero-ing to the respective distances to NGC~4258, a remarkable 0.001~mag agreement is found between the two fully independent teams and methodologies, with a dispersion $\sigma = 0.048$~mag. The comparison is shown in \autoref{fig:tully_comp}.
The RMS scatter seen in this comparison is equivalent to the typical TRGB distance error quoted in Table 3 of \citet{Freedman_2019}, suggesting also that the TRGB distance uncertainties have been accurately estimated.

The re-analysis of the CCHP data undertaken by the CosmicFlows team is a fully independent check of the CCHP's TRGB distances, and thus their measurement of $H_0$. A similar stress test has yet to be applied to the SH0ES \citep{Riess_2016} Cepheid datasets (and thus their measured $H_0$). Currently, only a single galaxy from the SH0ES sample has had its Cepheid data independently reduced and analyzed at the image level \citep{Javanmardi_2021}, and that result agrees with both the original SH0ES distance and the TRGB \citep{Jang_2015, Freedman_2019} distances.

Of course, with this impressive level of cross-consistency established, the new TRGB calibration presented could be used to shift the CCHP value of $H_0$. However, I do not perform that simple exercise here and simply note the consistency of the new flat LMC calibration with that adopted by the CCHP (0.009~mag, or 0.4\% in $H_0$), and the agreement of the updated CCHP distances with the independent measurements made by the CosmicFlows/EDD team (0.001~mag, or 0.05\% in $H_0$).\footnote{Indeed, if a ``new'' $H_0$ value -- determined by merely shifting the central value of the CCHP measurement -- were quoted along with every proposed re-calibration of the TRGB zero point, then numerous, strongly covariant $H_0$ values would muddy the literature.}

\section{Comparison with Literature zero point Calibrations} \label{sect:lmc_discuss}

In this section, a brief summary of recently proposed calibrations of the TRGB will be presented. After, an in-depth discussion of recent independent, LMC-based calibrations of the TRGB will be presented. Each previous study will be updated to match the same reddening and distances assumptions made in this study. Then each study will be updated in light of this study's findings, at once revealing significant biases in some of these recent calibrations. Correcting for these biases will reveal a convergence in zero point calibration of the TRGB in the LMC.

\subsection{Summary of Recent Calibrations} \label{subsect:trgb_summary}

Summarized here are recent calibrations of the TRGB magnitude, and the reader is referred to \cite{Beaton_2018} for a review of earlier calibrations in the literature.

\citet[][JL17]{Jang_2017_color} used a sample of massive galaxies observed by HST to calibrate the TRGB's color dependence on the HST on-flight filter system (already introduced and applied in \autoref{subsect:trgb_colors}). One of their geometric zero points was set by TRGB stars in the DEB fields of the LMC, the same star-forming regions which have been shown to produce biased TRGB magnitudes. Their analysis and the source of the bias in their measurements is covered in \autoref{app:jl17}. An improved version of the JL17 analysis that corrects for the biases due to low-ranking TRGB features is presented in \autoref{subsect:jl17}.

\citet[][G18]{Gorski_2018} used a small sample of central fields observed by the OGLE-III survey to calibrate the TRGB's color dependence in $VIJHK$ wavelengths. They measured the line-of-sight reddening values to their chosen fields using the RC color, results that were later presented in \citet{Gorski_2020}. Similar to the JL17 LMC measurements, it will be shown that this study's TRGB measurements were biased to fainter magnitudes by $\sim 0.05$~mag due to inclusion of low-ranking TRGB features. A detailed discussion and attempted reproduction of their result is presented in \autoref{app:g18}. Two updates to remove the biases in the original G18 analysis are presented in \autoref{subsect:g18}.

\citet[][F19]{Freedman_2019} \citep[later updated in ][F20]{Freedman_2020} used $VIJHK$ photometry of the LMC, masked its innermost metal-rich regions, and used TRGB stars belonging to two galaxies (the SMC and IC~1613) as reddening zero points. They performed a simultaneous fit to the TRGB's wavelength-dependent zero points (adopting an external set of near-infrared slopes as a constraint) in order to determine the reddening directly to LMC TRGB stars. An updated TRGB calibration using the S21 reddening map (instead of solving for the extinction directly) is presented in \autoref{subsect:f20} and is in agreement with their original calibration.

\citet{Yuan_2019} presented a post-processing of the JL17 and F19/F20 TRGB calibrations, as opposed to making new TRGB measurements. The modifications to F19 proposed in \citeauthor{Yuan_2019} have since been pointed out \citep{Freedman_2020} as a misinterpretation of the method. Those authors also overlooked the conservative error budget considerations from JL17, designed to account for the inclusion of low-quality TRGB fields in star-forming regions (see \autoref{app:jl17}).
It will be seen that the \citeauthor{Yuan_2019} underestimation of uncertainties has inflated the significance of the current debate over the TRGB's zero point calibration.

\citet{Reid_2019} adopted a TRGB measurement made from a Cepheid disk field observed by HST in the galaxy NGC~4258 \citep{Macri_2006}, as well as a partial-disk/inner-halo field \citep{Mager_2008}. Similar to \citet{Yuan_2019}, they did not present new TRGB measurements, instead adopting them directly from \citet{Macri_2006} and \citet{Jang_2017_color}. The authors claimed a high-precision measurement (0.02~mag) to the Cepheid disk field comparable to that measured here from the Rank 1+2 fields in the LMC, despite that field exhibiting an ill-defined TRGB feature that would see it classified as a Rank 3 or 4 detection, as expected considering the disk environment in which that TRGB measurement was made. The reader is referred to Section 4.2 of \citet{Jang_2021} for a detailed explanation for how \citeauthor{Reid_2019} did not in their error budget account for the systematic uncertainties of metallicity, reddening, or age, with the cumulative TRGB uncertainty estimated to be at the 0.1~mag level in that particular disk field.

\citet{Soltis_2021} quoted a high-accuracy calibration of the TRGB and a proposed shift to the CCHP $H_0$ using Gaia EDR3 parallaxes to $\omega$~Cen. Similar to \citeauthor{Reid_2019} and \citeauthor{Yuan_2019}, they did not make a new TRGB measurement, and adopted a very old result \citep{Bellazzini_2004}, to which an optimistic reduction in the uncertainties was proposed.
In that result, small number statistics ($<200$ stars within 1~mag of the TRGB) in $\omega$Cen limit the inherent accuracy of any TRGB measurement. 
Additionally, the photometry \citep{Pancino_2000} was acquired through a nonstandard narrow-band filter ($I_{853}$ with $\Delta \lambda = 14$~nm on the WFI at the 2.2m at La Silla) for which there does not exist a robust transformation to the on-flight ACS/F814W filter system (the filter used to measure the TRGB distances that calibrate the Hubble constant), thus breaking the link between their calibration and the CCHP's TRGB distances. It has also been found that \citeauthor{Soltis_2021} likely underestimated their parallax uncertainties by at least a factor of two, a result of $\sim 5-10~\mu$as residuals in the Gaia EDR3 zero point solution \citep{Maiz_2021, Vasiliev_2021, Ren_2021}. Furthermore, the authors underestimated by a factor of two the uncertainty in their adopted foreground extinction correction; they adopted a 5\% uncertainty associated with the SFD98/SF11 maps, as opposed to the canonically adopted value of at least 10\% (see Appendix C, item 6 of SFD98).

\citeauthor{Soltis_2021} also proposed an LMC calibration via re-interpretation of the \citet{Gorski_2018} study, but it will be shown in \autoref{subsect:g18} and \autoref{app:g18} that their result is biased to even fainter TRGB luminosities than the already-biased \citeauthor{Gorski_2018} measurements because they did not account for metallicity effects.

\citet{Jang_2021} presented a new TRGB measurement and calibration based in the stellar halo of NGC~4258. All prior TRGB measurements made in that galaxy were either placed totally \citep{Macri_2006, Rizzi_2007, Reid_2019} or partially \citep{Mager_2008, Jang_2017_color} in the disk of that galaxy. The hard-to-constrain systematic uncertainties from mixed-populations and reddening effects (a similar situation to that addressed in the present paper) were at once resolved by their excursion into the galaxy's stellar halo. Though the astrophysical systematics were consequently minimized, the halo archival imaging they were forced to use (the only available in the stellar halo) is older (2003-04) and at relatively low S/N for an absolute calibration ($\sigma_{F814W} \sim 0.1$~mag for a typical Tip star). As a result, their error budget is dominated by photometric uncertainties (including the difficult task of bridging pre- and post-SM4 ACS/WFC), which could be immediately reduced to a fraction of their original size via a modern HST imaging program in the halo of NGC~4258.

\begin{figure}
    \centering
    \includegraphics[width=0.92\columnwidth]{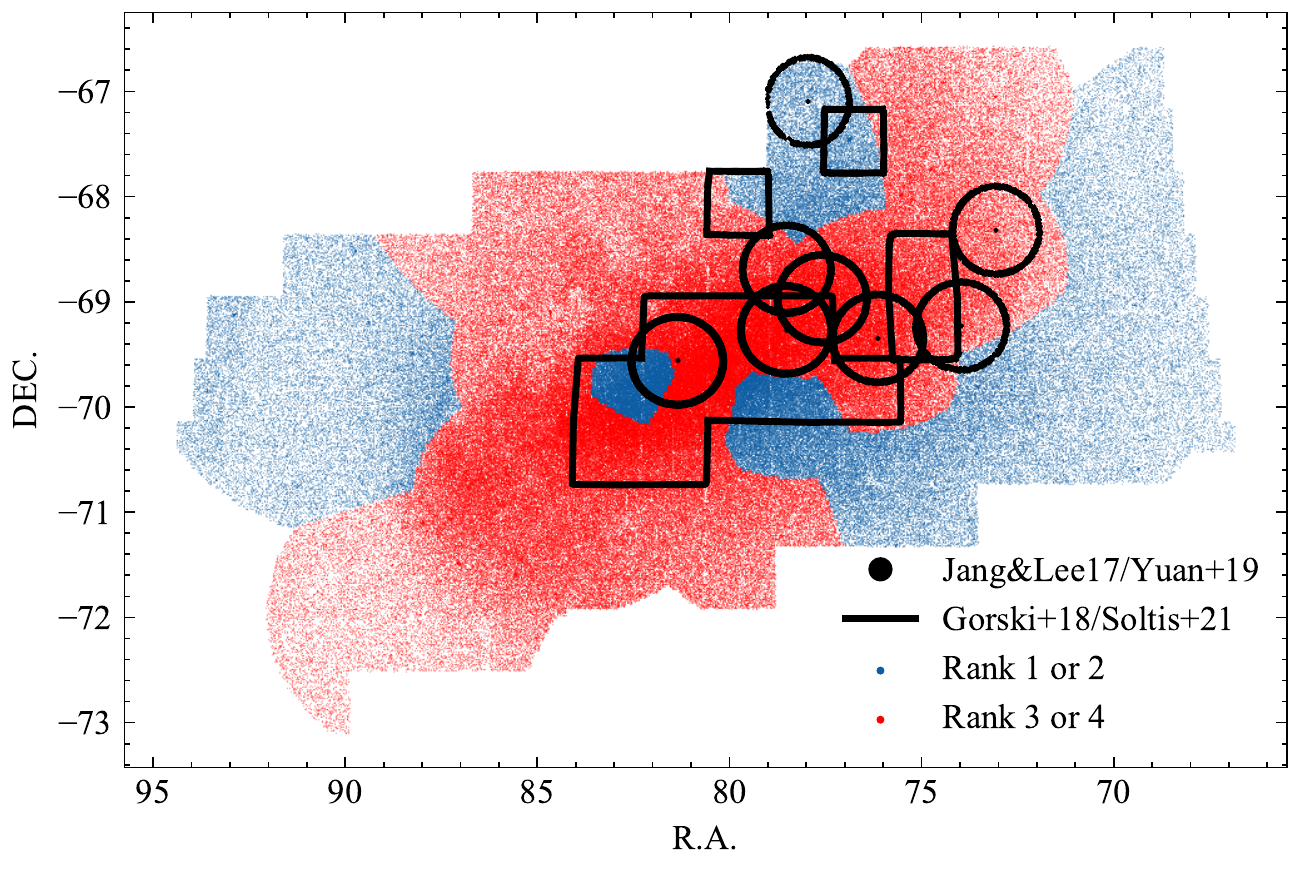}
    \caption{Locations of recent literature calibrations based in the LMC plotted over the field-by-field Rankings shown in \autoref{subsect:lmc_regions}. For clarity of illustration, the JL17 fields are decreased in radius by a factor of two from their actual size. Note the majority of these fields used in prior TRGB calibrations are associated with a low TRGB quality Rank and high rates of star formation.}
    \label{fig:rank_map}
\end{figure}

\subsection{Discussion and Renalysis of Recent LMC calibrations}

In this section, three recent studies that calibrated the $I$-band TRGB in the Clouds using OGLE-III photometry are revisited and updated. Two of these previous studies \citep{Jang_2017_color, Gorski_2018} included TRGB measurements classified as Rank 3 or 4 that biased their calibrations (see \autoref{fig:rank_map}). These biases were subsequently underestimated by \citet{Yuan_2019} and \citet{Soltis_2021} in their recent proposed re-calibrations of the CCHP's TRGB distance scale.

\begin{figure}
    \centering
    \includegraphics[width=\columnwidth]{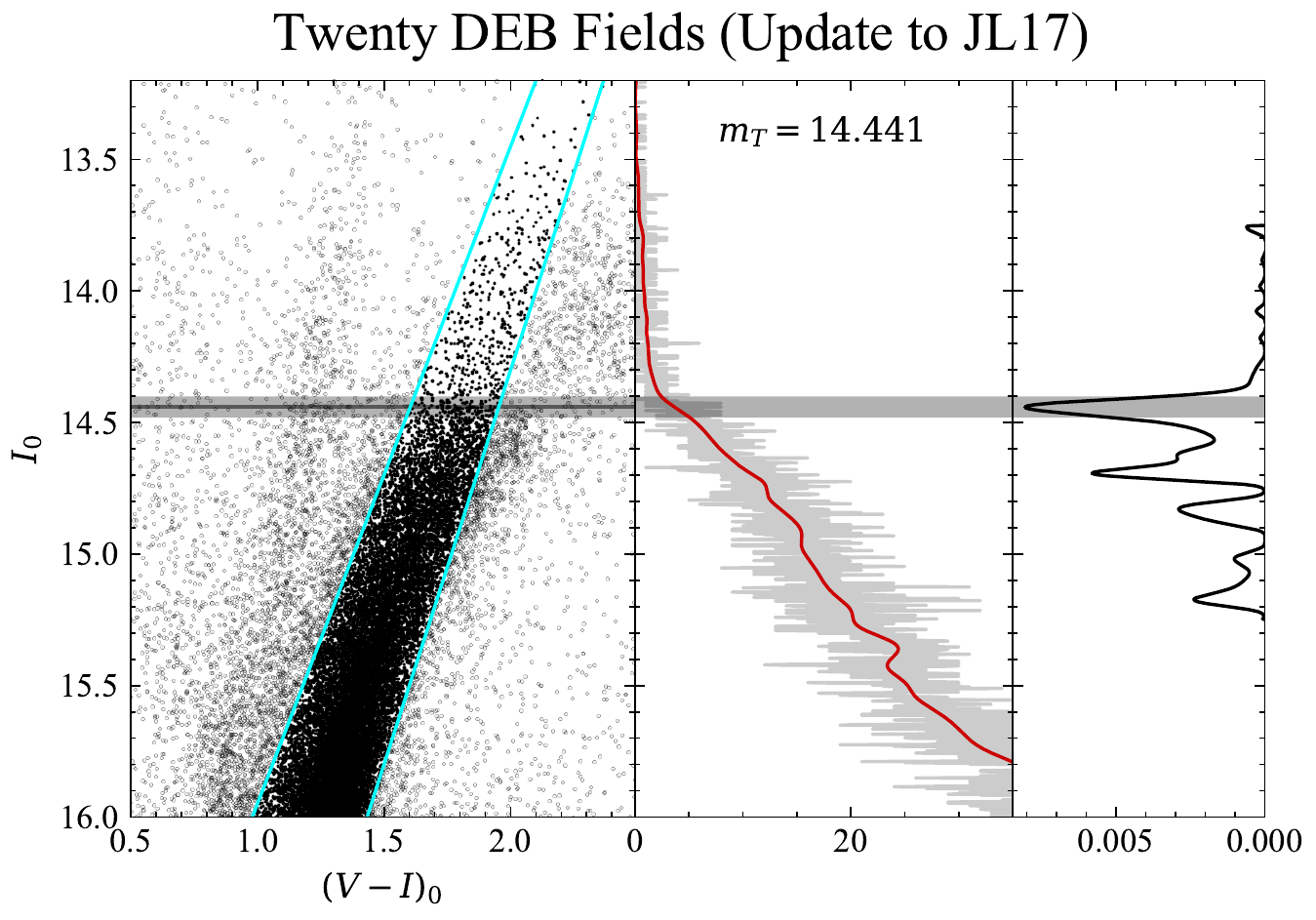}
    \caption{
    Update to the JL17 analysis. Eight $r<50'$ DEB-centered fields are increased to twenty $r < 35'$ fields, \emph{empirical} geometric distortion corrections are used as opposed to model-dependent ones, and only high-ranking (1 or 2) regions are considered.
    }
    \label{fig:jl17_update}
\end{figure}

\subsubsection{JL17} \label{subsect:jl17}
The JL17 analysis is updated and improved in the following ways: (1) an increase from eight to 20 DEB fields, (2) a decrease in individual field area from $50'$ to $35'$, (3) removal of Rank 3+4 regions, (4) use of empirical geometric corrections, as opposed to model predictions (e.g., \autoref{subsect:lmc_geo}), and (5) a composite CMD approach, as opposed to a mean of individual field measurements. 

The model-independent geometric distortion corrections are computed by finding, for every source, the weighted average distance to all P19 DEBs located within $35'$. This updated measurement $I_0 = 14.441 \pm 0.015$ is shown in \autoref{fig:jl17_update} and is perfectly consistent with this study's main result of $I_0 = 14.439 \pm 0.008$~mag. This novel TRGB measurement was made without any assumption of the underlying geometry of the LMC by using only the DEB distance measurements to create empirical geometric corrections. This provides an independent check on the accuracy of the modeled geometric corrections used in the main analysis. This update to the JL17 measurement is labeled ``JL17 Blue'' under the Updates block in \autoref{tab:lit_summ}.

\begin{figure*}
    \centering
    \includegraphics[width=\columnwidth]{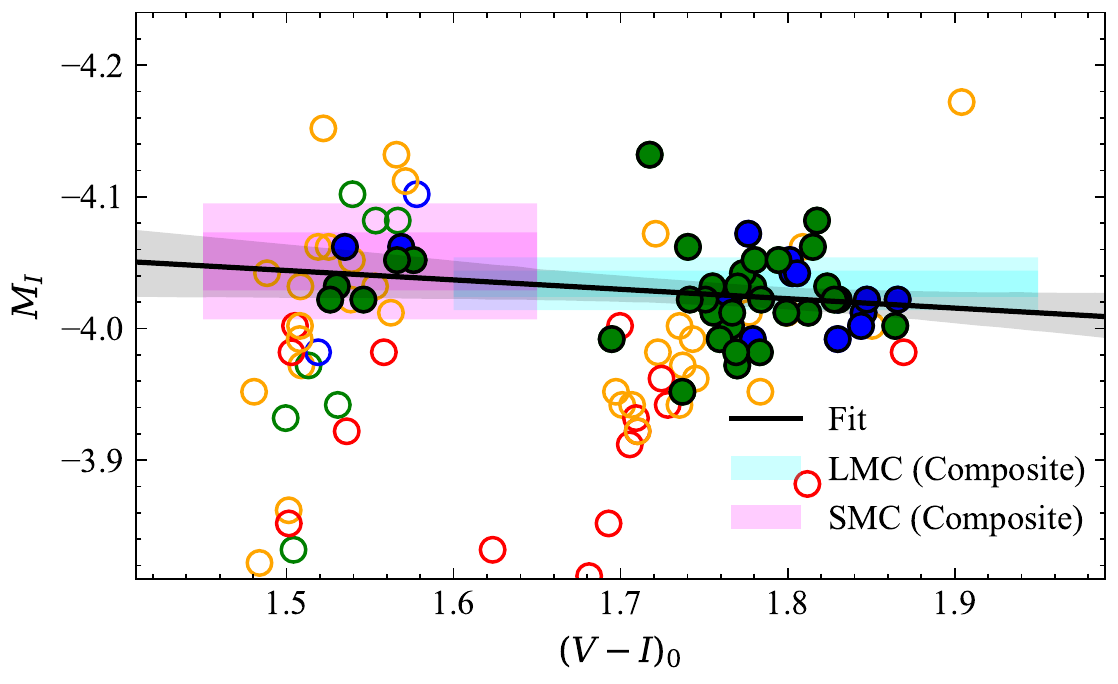}
    \includegraphics[width=\columnwidth]{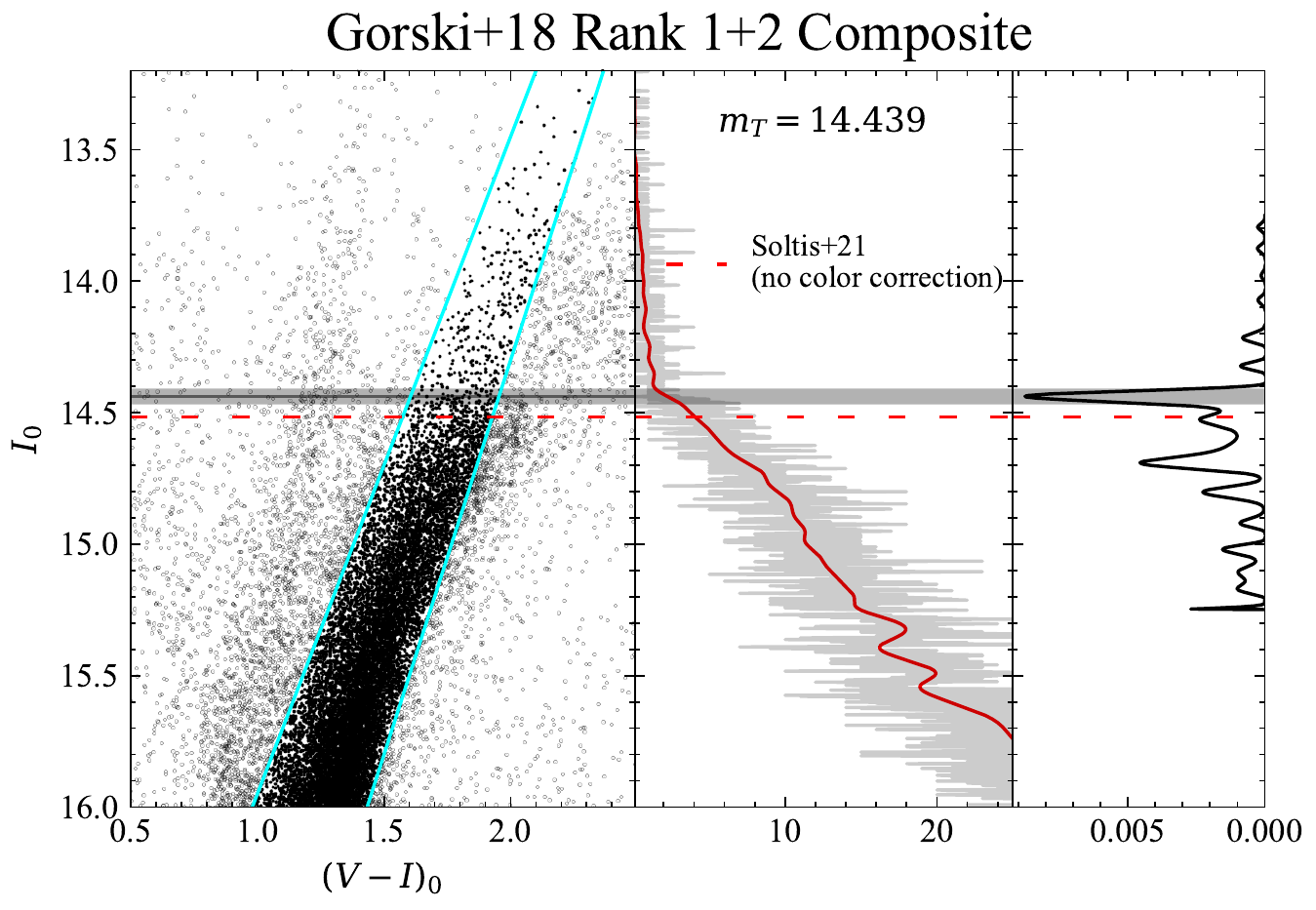}
    \caption{Two alternative updates to the G18 analysis.
    \textit{left:} Field-by-field TRGB calibration identical to the original G18 approach. The starting number of fields considered was increased from 33 to all 156 across the MCs, and only those classified as Rank 1 or 2 were included in the fit (closed, green and blue circles). Significant geometric outliers in the SMC were rejected (open, blue and green circles). The 95\% CI of the best fit line (black line and gray bands) is plotted. $2\sigma$ bands are shown for the flat SMC (magenta) and the LMC (cyan) calibrations from the main analysis. Due to the relative LMC-SMC DEB distance, there is an additional \emph{systematic} uncertainty of 0.017~mag associated with all SMC points seen in the left-hand side of the figure. Taking that systematic into account makes the best-fit line statistically consistent with zero.
    \textit{right}: Composite TRGB detection for the same fields used by G18, de-reddened using the S21 empirical reddening maps. Also shown is a TRGB calibration presented from this same dataset in \citet{Soltis_2021} where the authors did not control for metallicity/color (dashed red line).}
    \label{fig:g18_update}
\end{figure*}

\subsubsection{G18} \label{subsect:g18}

Two alternate updates to the G18 analysis are presented here: a similar field-by-field analysis to theirs, and a composite CMD approach. 

The updated field-by-field analysis will increase the starting sample of calibrating fields from 39 to all 156 OGLE-III fields across the MCs.\footnote{See \autoref{app:g18} for confirmation that the Tip measurement methodology used here produces results consistent with theirs.} This results in a total of 47 Rank 1+2 datapoints (6 SMC, 41 LMC) used for fitting a slope to the TRGB, to be compared to the seven (three SMC, three LMC) Rank 1+2 datapoints contained in the original G18 sample. The results are presented in the left panel \autoref{fig:g18_update}, along with the results from the primary composite analysis (magenta and cyan bands). The zero point at the mean LMC color of $(V-I)=1.80$ is determined to be $-4.026 \pm 0.008$~mag (formal fit uncertainty only), which is in very good agreement with the value in \autoref{subsect:zpcal}.

The slope of $0.071 \pm 0.049$ determined from this method is in mild disagreement with the finding in \autoref{subsect:trgb_colors} that the TRGB magnitude is only significantly sloped for colors $(V-I)_0 > 1.95$~mag. This points to a bias inherent to measuring the run of TRGB magnitude with color in this way and why the analysis in \autoref{subsect:trgb_colors} is preferred. By compressing the colors of TRGB stars from an individual field/bin into a single mean value, the wings are clipped from the underlying TRGB color distribution, and asymmetries in the distribution of colors, as well as nonlinearities in the TRGB magnitude-color relation, are necessarily missed, thus biasing the eventual fit.
For example, the reddest bins in the field-by-field analysis have typical mean colors $(V-I)_0 \simeq 1.85$~mag with $\sigma \sim 0.2$~mag. Thus, if the TRGB magnitude is flat for colors $(V-I)_0 < 1.95$~mag (as confirmed in the composite LMC CMD and \autoref{subsect:trgb_colors}), then Tip magnitudes measured to these spatial bins are being skewed to fainter magnitudes by the reddest Tip stars, which are all redder than the mean color they are being represented by. This compresses the measured TRGB magnitude-color relation along the color axis, which artificially steepens the measured slope.

The color bins adopted in \autoref{subsect:trgb_colors} do not suffer from this bias because they were constructed from the composite CMD, i.e., they were derived from the fully populated distribution of Tip stars in the LMC. Put another way, the problem boils down to order of operations. To properly sample the underlying population of TRGB stars in color-magnitude space, the photometry of individual fields would have to first be marginalized over spatial position in order build up number statistics in the wings of the metallicity distribution. Then, the TRGB sample can be split into individual color bins (as was done in \autoref{subsect:trgb_colors}), so that the wings of the magnitude-color distribution are well-sampled and can be accurately modeled. Note how not one point in the left panel of \autoref{fig:g18_update} has a color $(V-I)_0 > 1.90$~mag, despite there being a sizeable population of metal-rich Tip stars at those colors (as confirmed in \autoref{sect:trgb_measure} and \autoref{subsect:trgb_colors}). The red Tip stars pull the average TRGB magnitude to fainter values and redder colors, while the blue tip stars do not change the TRGB magnitude, but still pull the mean TRGB color bluewards, introducing a bias in a slope measured from this method; that is, a steeper slope is measured at bluer TRGB colors.

This is a situation that is fundamentally impossible to overcome when attempting to compress the full photometric information of Tip stars with a large, asymmetric color spread into a scalar quantity. \citet{Durbin_2020}, however, have presented a novel method aimed at addressing this exact problem by using data-driven 2-D distributions to represent Tip star photometry from single spatial bins. In their case each bin was an individual galaxy, as opposed to individual LMC fields as considered here, but the situations are analogous.

Now the alternate G18 calibration is determined from a composite CMD. This approach is preferred to a field-by-field analysis due to its resilience to the biases induced by high-frequency variations in star formation, differential reddening, and line-of-sight depth\footnote{Inversely, should the systematic uncertainties induced by high local SFR and differential reddening be properly controlled for, as done by the TRGB ranking procedure in \autoref{sect:trgb_measure}, then accurate field-by-field TRGB magnitudes can be used to determine the orientation of the LMC plane, as was done in \autoref{subsect:lmc_geo}.}, as well as sample selection bias.
Therefore, a composite CMD is built from the 14 fields of G18 and is shown in the right panel of \autoref{fig:g18_update} along with the RGB LF, EDR, and measured Tip magnitude. After a selection is made to isolate blue (metal-poor) TRGB stars,
it is found that the TRGB as determined from the G18 composite CMD agrees with that measured from the Rank 1+2 calibrating sample of this study over the same color range.
The G18 composite is then convolved with the Rank 1+2 calibrating sample, and an $I_0 = 14.439 \pm 0.015$~mag is determined. This value is tabulated in \autoref{tab:lit_summ} as G18 Blue under the Updates block.

Also plotted in the right panel of \autoref{fig:g18_update} is the TRGB calibration proposed by \citeauthor{Soltis_2021} (red, horizontal dashed line), which they made from a simple average of the G18 measurements and the S21 reddening maps (the exact same datasets used to produce \autoref{fig:g18_update}). It is found that their proposed TRGB magnitude is not supported by the data. This is because their approach -- the simple mean of the raw TRGB magnitudes presented in G18 -- did not account for the metallicity/color dependence of the $I$-band TRGB, and was skewed towards anomalously faint values in the G18 measurements.

\subsubsection{F20} \label{subsect:f20}

\begin{figure}
    \centering
    \includegraphics[width=\columnwidth]{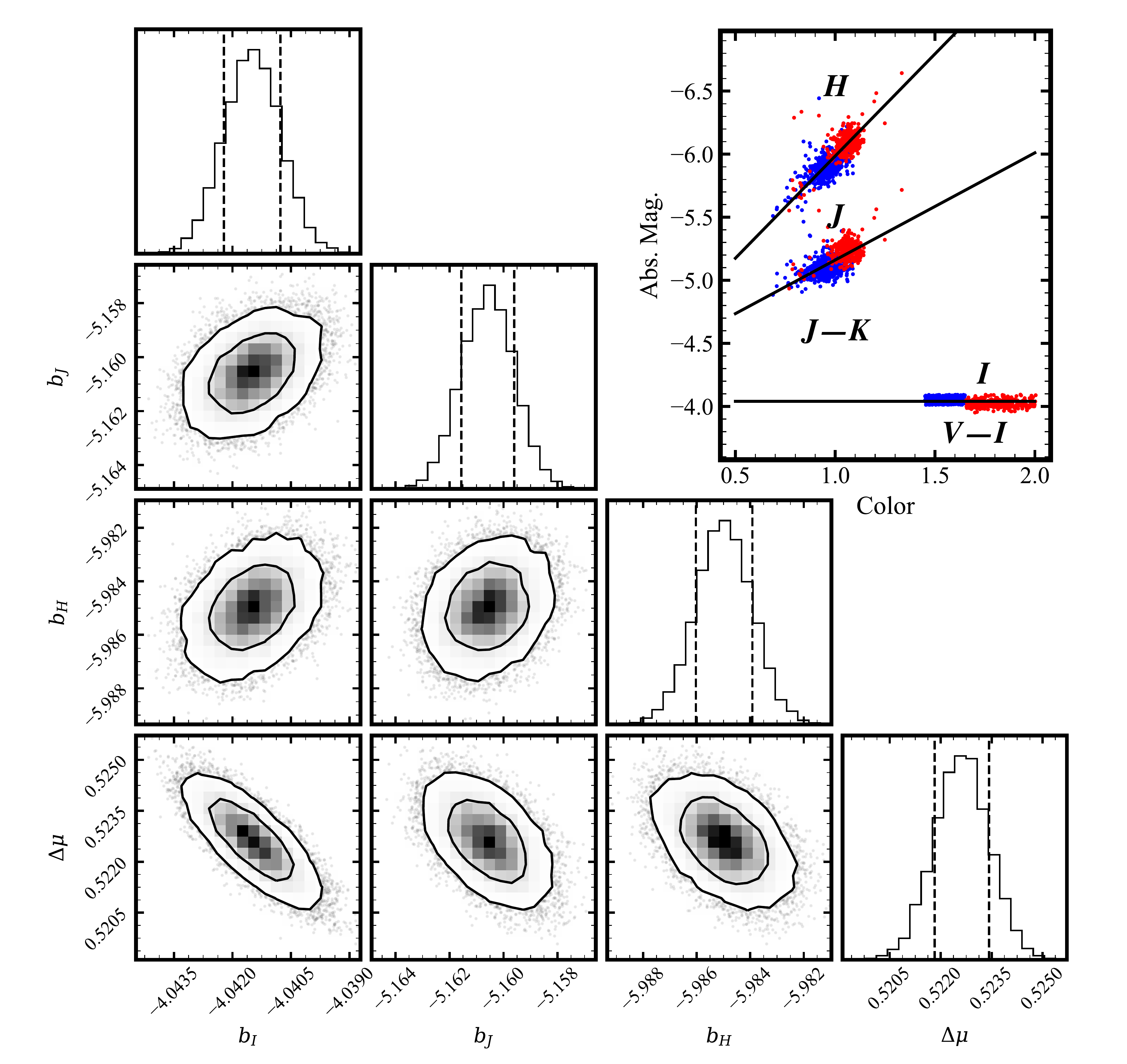}
    \caption{A reformulation of the F20 multi-wavelength TRGB technique. S21 reddening corrections are applied, as opposed to directly fitting for the reddening to Tip stars as in the original study. MCMC parameter estimation of TRGB zero points and the distance between the SMC (blue points) and LMC (red points). Best-fit zero points using slopes from \citet{Madore_2018} are plotted (black lines). The zero points are absolutely calibrated by the P19 DEB distance ($\mu =18.477$~mag).}
    \label{fig:mcmc}
\end{figure}

An independent test of the F20 calibration can be made by using their same technique, but first de-reddening the OGLE-III photometry with the S21 maps, then seeing if the TRGB zero points agree with those determined by F20, who directly fit for the average reddening to their TRGB stars.
Near-infrared (NIR) TRGB slopes from \citet{Madore_2018} and $VIJHK$ OGLE-2MASS Tip star photometry from the SMC and LMC are input to the \textit{emcee} package \citep{Foreman-Mackey_2013} to determine the best-fit parameter values for the $IJH$ zero points and the distance between the SMC and LMC, $\Delta \mu$. Note $M_V$ and $M_K$ are not included in the fit because they are, by definition, forced to be equal to $M_I$ and $M_J$ as per their $(V-I)$ and $(J-K)$ colors, respectively. A Gaussian likelihood and uniform, positive priors are assumed for all parameters. 

The corner plot and resulting fits are shown in \autoref{fig:mcmc}. The fit returns $\Delta \mu = 0.523$~mag, $M_I = -4.041$,  $M_J = -5.159$, $M_H = -5.893$. The formal uncertainties of the fit are negligibly small because of the large number statistics ($N\sim300$) and small photometric uncertainties ($< \! 0.01$~mag). Also, the TRGB slopes were adopted as external constraints, with their uncertainties not propagated, or e.g., being included in the fit as priors. The true uncertainty in the zero points is estimated to be closer to 0.02~mag in $VI$ and 0.03~mag in $JHK$, after accounting for uncertainties in the Tip star identification process and in the NIR slopes. The $I$-band TRGB result $-4.041 \pm 0.02$~mag is included in the ``Updates'' block of \autoref{tab:lit_summ} and is in very good agreement with the main analysis.

Agreement is found with the original F20 zero points, despite the use of an entirely independent probe of the reddening to TRGB stars. Furthermore, the LMC-SMC differential modulus ($\Delta \mu = 0.523 \pm 0.03$~mag), determined using only the $VIJHK$ TRGB magnitude-color relations and the S21 reddening map, is in good agreement with the fully independent, geometric DEB value $\Delta \mu = 0.50 \pm 0.017$~mag \citep{Graczyk_2020}. There is a clear cross-consistency at the 0.02~mag (1\%) level between the F20 zero point calibration (and their adopted NIR slopes), the S21 maps, as well as the P19 and G20 DEB distances.

\subsection{Convergence to a single TRGB zero point} \label{subsect:trgb_zp}

\begin{deluxetable}{lll}
\tablecaption{Literature Zero points based in the LMC. \label{tab:lit_summ}}

\tablehead{
\colhead{} &
\colhead{Method} &
\colhead{$M_I^{TRGB}$}
}
\startdata
Original \\
\hline
JL17 & Blue     & $-3.970 \pm 0.090 $  \\
JL17 & QT       & $-3.998 \pm 0.082 $   \\
Yuan+19\tablenotemark{a} & Blue & $-3.952 \pm 0.046 $\\
G18\tablenotemark{b}  & Linear & $-4.088 \pm 0.011 $ \\
F20 & Blue   & $-4.047 \pm 0.034 $  \\
\hline
Renormalized \\
\hline
JL17 & Blue    & $-3.993 \pm 0.059 $  \\
JL17 & QT      & $-4.020 \pm 0.046 $  \\
Yuan+19\tablenotemark{a} & Blue & $-3.991 \pm 0.027 $ \\
G18\tablenotemark{b} & Linear & $-3.997 \pm 0.011 $ \\
Soltis+21\tablenotemark{c} & Flat & $-3.960  \pm 0.011 $\\
\hline
Updates \\
\hline
JL17 & Blue   & $-4.036  \pm 0.015 $  \\
G18  & Linear & $-4.026  \pm 0.008 $  \\
G18  & Blue   & $-4.030  \pm 0.02  $ \\
F20  & Blue   & $-4.041  \pm 0.02  $ \\
\hline 
\textbf{This Study} \\
\hline
& Blue  & $\MtrgbLMC \pm 0.008 $ \\
& Linear & $-4.036 \pm 0.003 $ \\
& QT (ZP Fit)\tablenotemark{d} & $-4.036  \pm 0.002 $ 
\enddata
\tablenotetext{a}{Adopted Blue-TRGB measurements from JL17, but not their uncertainties.}
\tablenotetext{b}{G18 did not present an additional error budget beyond the standard errors on their fit parameters.}
\tablenotetext{c}{Adopted simple mean of raw G18 TRGB measurements and did not control for metallicity/color.}
\tablenotetext{d}{Quadratic color dependence adopted from JL17 with updated zero point fit from \autoref{subsect:trgb_colors}}
\tablecomments{For congruous comparison between methods (flat vs. color correction), zero points of sloped calibrations were scaled to the average LMC TRGB color. Original values are quoted without DEB distance uncertainties. Renormalized quantities are quoted without DEB distance or the S21 reddening uncertainties. Updates and This Study's values are quoted with only their formal fit/measurement uncertainties, since all other uncertainties are equal.}
\end{deluxetable}

\begin{figure}
    \centering
    \includegraphics{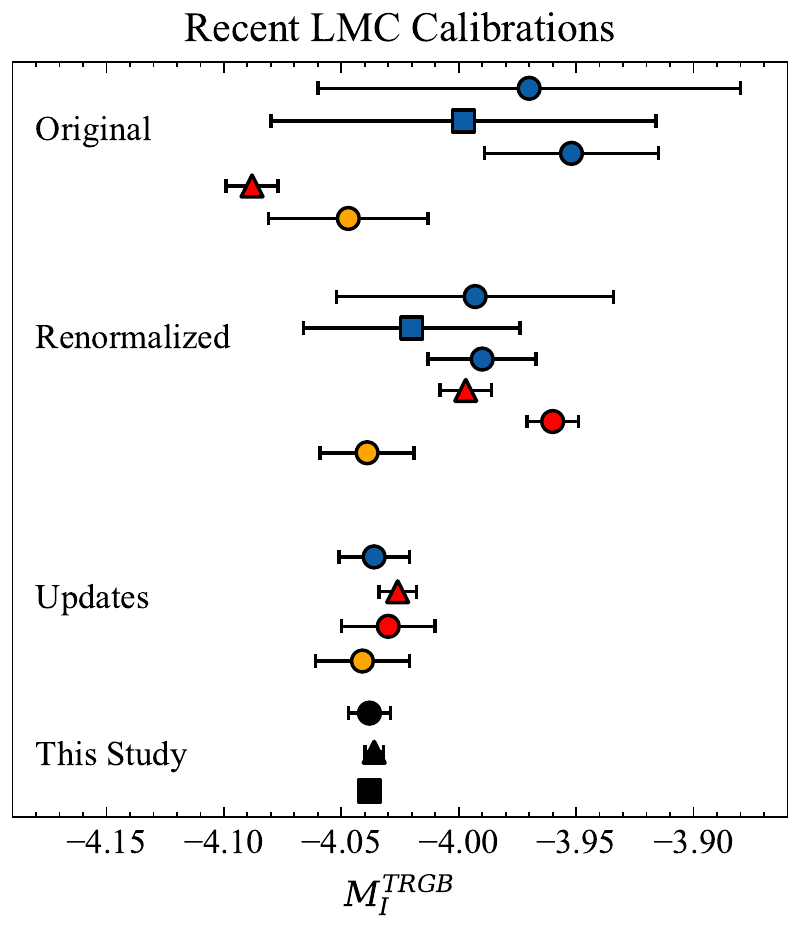}
    \caption{Exact graphical representation of \autoref{tab:lit_summ}, with flat (circles), linear (triangles), and quadratic (squares) calibrations color-coded according to the parent study of JL17 (blue), G18 (red), and F20 (orange). This Study's results from \autoref{tab:calibs} are also shown (black markers). Once the reddening and distance zero points are normalized to the latest state-of-the-art measurements, and the low-quality TRGB features and subsequent biases are controlled for, a clear convergence to a single TRGB zero point is found.}
    \label{fig:trgb_cals}
\end{figure}

The three independent LMC studies, and the pair of derivative studies, are included in \autoref{tab:lit_summ} and plotted \autoref{fig:trgb_cals}. The original values are tabulated in the ``Original'' segment, with the values after renormalization to the P19 and G20 DEB distances and the S21 reddening map in the ``Renormalized'' section. Both the original and renormalized values are derived in \autoref{app:revisits}.

Included in the ``Updates'' blocks are the results from the updated analyses that corrected for biases due to TRGB measurements made in star-forming regions. Each update in the previous section provided a useful check on the primary results of this study.

The updated JL17 result was made using \emph{empirical} geometric distortion corrections. This result was in perfect agreement when performing the same analysis with the same modeled corrections adopted for the main analysis. The uncertainties adopted in the original JL17 study were confirmed to be accurate and not overestimated. \citet{Yuan_2019} subsequently deflated the JL17 uncertainties, thereby forcing the biases in the JL17 LMC measurements to become statistically significant.

The G18 analysis was updated to include an order of magnitude more high-ranking TRGB fields than the original study. It was found that the steep TRGB slope originally determined by G18 was due to their very small sample size, the biases in their low-ranking TRGB measurements, and the inherent limitation when attempting to use scalar quantities to represent the continuous color distributions of TRGB stars. \citet{Soltis_2021} adopted the raw TRGB magnitudes from G18, ignoring the original study's determination of a sloped TRGB. This assumption is particularly inaccurate for the G18 fields which were chosen to be in the most central, and thus most metal-rich, regions of the LMC.

The F20 analysis was reformulated to no longer fit directly for the reddening to TRGB stars, instead adopting the S21 reddening corrections as an external constraint. Also, the minimization was performed in absolute magnitude vs. color space, as opposed to extinction vs. central wavelength space as done in the original study. A TRGB zero point within 0.005~mag of the original F20 result was found. Furthermore, the relative SMC-LMC distance determined using just the S21 reddening map and the TRGB magnitudes and colors was in very good agreement with the DEB distance. This established a 0.02~mag cross-consistency between the F20 result \citep[and their adopted slopes from][]{Madore_2018}, the P19 and G20 DEB distances to the SMC and LMC, and the S21 reddening map of both Clouds.

Finally, this study's calibrations are displayed in the bottom segments of both table and figure. The final zero points (Updates and This Study) all landed between $-4.026$~mag and $-4.041$~mag, indicating a strong convergence to a single value for the TRGB zero point in the LMC. It is thus concluded that a recent debate over zero point calibration of the TRGB in the LMC is settled, and was caused by systematic biases in literature measurements \citep{Jang_2017_color, Gorski_2018}, and a subsequent underestimation of the uncertainties in those measurements \citep{Yuan_2019, Soltis_2021}.

\section{Summary and Conclusions} \label{sect:conclusion}
The new high-accuracy calibration presented in this study is determined from three fundamental sets of measurements: (1) the OGLE-III photometry (calibrated to OGLE-IV), (2) the S21 reddening map based on OGLE-IV observations, and (3) the P19 and G20 DEB distances to the MCs. Barring a catastrophic failure underlying any one of those pillars (see \autoref{subsect:f20} for evidence against such a failure), it is likely that this new zero point calibration -- and its associated error budget -- is comprehensive and definitive.

For calibration of the TRGB in the LMC, the OGLE photometry was divided into spatial bins which were then ranked based on the quality of their TRGB features.
The locations of high-quality TRGB regions were found to correlate with regions host to negligible rates of star formation as well as minimal levels of internal gas and dust. The highly-ranked TRGB regions were then used to measure $\Theta_{LON} = 153 \pm 12 \degree$ and $i = 27 \pm 3 \degree$ for the 3D orientation of the LMC plane. In the SMC, no significant differential effects due to dust or star formation were expected nor observed, though a blurring of the Tip feature due to line-of-sight depth effects was apparent, and a significant E to W tilt, with the Eastern side closer, was confirmed.

The resulting CMDs, luminosity functions and TRGB edge detector response functions as measured from both MCs were unambiguous and of high precision, particularly in the case of the LMC, with sharp edge features setting the TRGB zero point in each galaxy to unprecedented accuracy.
The TRGB zero points determined separately from each Cloud, as well as the cross-galaxy fits to the color dependence of the TRGB (all of which are tabulated in \autoref{tab:calibs}), are in excellent agreement with the canonical value of $-4.05$~mag for the old, metal-poor TRGB magnitude.

The new measurements favor a shallow, quadratic TRGB color dependence and not a steep, linear one for TRGB stars with $(V-I)_0 < 2.2$~mag. The quadratic functionality is also predicted by theory and appears to be the most accurate representation of the intrinsic $I$-band TRGB magnitude-color relation. Furthermore, minimal deviation from the best-fit TRGB magnitude relation was seen ($0.01$ RMS scatter) for datapoints including the Clouds, Local dwarf dSphs, and M33, that were fit with an external \citep{Jang_2017_color} color dependence determined in the halos of $L_*$ galaxies. This consistency indicates that potential biases in the TRGB magnitude-color relation due to age/SFH are minimal across these host environments.

The dominant uncertainties in the LMC zero point calibration are both the P19 calibration of the NIR surface brightness-color relation (0.018~mag) and the S21 reddening zero point (adopted to be 0.014~mag). The dominant uncertainty in the SMC is the effect its line-of-sight depth has on accurate triangulation of its mean distance, which is conservatively captured by the combined statistical uncertainty in the edge detection and DEB distance entries of \autoref{tab:budget} ($\simeq 0.03$~mag).

In \autoref{sect:lmc_discuss}, all recent independent $I$-band TRGB calibrations based in the MCs were revisited and reanalyzed.
After normalizing over reddening and distance assumptions (and excluding one study that did not account for metallicity effects), the \emph{full range} of independent LMC-based TRGB zero point calibrations was found to be $0.05$~mag, already a promising level of agreement.
During the revisit, some ambiguities and inconsistencies were uncovered and are documented in the Appendix. It was found that many of these inconsistencies, and the original studies' biases, were caused largely by the inclusion of regions of the LMC with high recent star-formation and high dust content -- both of which were shown in the present study to significantly degrade TRGB measurement accuracy.

Once these systematics are controlled for, a more consistent picture of the TRGB zero point is revealed. After masking star-forming regions of the LMC from calibration, all three recent literature calibrations converged towards the highest accuracy one presented here, becoming consistent with each other to within 0.015~mag (see \autoref{subsect:trgb_zp} and \autoref{tab:lit_summ}). It is therefore concluded that a ``discrepancy'' that has arisen in the literature in regards to zero point calibration of the $I$-band TRGB is not real, and is an artefact borne of underestimated uncertainties and biased measurements. In fact, the situation is quite the opposite, with the old, metal-poor $I$-band TRGB proving to be a remarkably consistent distance indicator when measured in the appropriate environments.

I see three paths forward for further improvement of the TRGB distance scale, in particular in its role as an anchor of the Hubble Diagram:
\begin{enumerate}
    \item Calibration of the slope(s) and zero point(s) of the TRGB magnitude-color relations in IR space-based bandpasses in the halo of NGC~4258, as well as an improved F814W-band calibration.
    \item Two-fold increase in the number of SN~Ia Host galaxies with HST observations designed to measure TRGB stars in their stellar halos.
    \item Thorough cross-examination of the zero point offsets between various $I$-band photometric systems and $I$-equivalent bands on board HST.
\end{enumerate}

Item 1 is arguably the most efficient path to decreasing systematic uncertainties in current TRGB-anchored measurements of $H_0$. Because variations of the TRGB with age and metallicity are difficult, if not impossible, to disentangle empirically \citep{Salaris_2005_badtrgb}, calibrating the TRGB in an environment most similar to the CCHP SN~Ia Host sample (stellar halos of $\sim \! L_{*}$ galaxies) is ideal, and NGC~4258 provides just that. Furthermore, calibration of the TRGB directly with HST (and later Webb) imaging bypasses the additional uncertainties associated with transforming from ground-to-space filter systems, which appear to remain uncertain at the 0.02~mag level \citep{Riess_2019, Freedman_2019}.

Item 2 is immediately attainable provided enough Hubble Space Telescope time and is currently the largest contribution to the uncertainty in $H_0$. As of now, there exist only twenty SNe~Ia anchored by the TRGB distance scale, and it is imperative that number be increased to a statistically significant size to enable more rigorous statistical analyses of distance ladder $H_0$ measurements. In light of an internal 2-3$\sigma$ discrepancy in the SH0ES team calibration of the Cepheid Leavitt Law \citep{Efstathiou_2020}, and the overwhelming self-consistency of modern and historical TRGB calibrations \citep[][this study]{Lee_1993, Bellazzini_2004, Rizzi_2007, Jang_2017_color, Gorski_2018, Freedman_2020, Jang_2021}, it would appear that the TRGB is currently the most efficient route to accomplish this goal.

Item 3 relates to the transformations between filter systems, and the uncertainty in those systems' own photometric zero points. Indeed, \citet{Riess_2016} presented a set of synthetic ground-to-HST filter transformations thought to be accurate to 4~mmag. However, \citet{Riess_2019} later empirically measured departures from those synthetic zero points ranging from 0.016 to 0.04~mag, reflecting $4-5\sigma$ shifts. Direct HST imaging of RGB stars in the Clouds could be a powerful option to link this work's ground-based calibration in the MCs to TRGB distance measurements made with HST, though careful attention must be paid to both saturation effects and charge-transfer leaks.

\acknowledgments
I am indebted to Prof. Wendy L. Freedman for her support and tutelage throughout my graduate studies, as well as very helpful comments on, and discussions regarding, this manuscript.
I thank Prof. Andrzej Udalski and the OGLE Collaboration for their sustained efforts on their revelatory survey of the Magellanic Clouds, and for making the photometry publicly available.
I thank Dr. Jan Skowron for making publicly accessible their reddening maps of the MCs.
I am thankful for frequent, fundamental insights from Dr. Barry Madore.
I am grateful to Prof. Hsiao-Wen Chen for encouraging discussions.
I thank Dr. Rachael Beaton for stimulating discussions regarding the Clouds.
I am grateful to Dr. Mark Seibert for his key contributions to the distance scale.
I thank Dr. In Sung Jang for helpful comments and clarifications.
I thank Dr. Brandon Hensley for suggesting a dive into the rich literature of observations acquired of the Clouds.
I am grateful to past and current members of the Carnegie Chicago Hubble Program from whom I have had the privilege to learn.
I thank the many staff and researchers at Carnegie Observatories that have provided a welcoming and highly productive environment throughout my Ph.D.
\vspace{5mm}

\software{matplotlib \citep{Hunter_2007},
          emcee \citep{Foreman-Mackey_2013},
          astropy \citep{2013A&A...558A..33A}   
          numpy \citep{Harris_2020}
          }

\clearpage

\appendix

\section{Gaia EDR3 Proper Motion Cleaning} \label{app:pmclean}
As demonstrated by \citet{Luri_2021}, the proper motions of stars to the MCs have been determined to remarkable precision in the Gaia EDR3 release. In this section, a procedure is detailed for removing sources in the Clouds with a very high probability of being located in the foreground. A more sophisticated procedure, e.g., the one described in \citet{Lindegren_2021}, is not necessary here since the goal is simply a mostly-cleaned TRGB ($G \sim 15.8$~mag for both Clouds) and not an accurate parallax (offset) determination from either Cloud.

First, all bright Gaia EDR3 sources within 5.5~deg of the LMC center ($\alpha=80.9^d, \delta=-69.75^d$) and 3~deg of the SMC center ($\alpha=12.5^d, \delta=-73.0^d$) are selected to contain upper RGB stars in the OGLE-III footprint. The SQL query for the LMC was,
\begin{verbatim}
SELECT edr3.source_id, edr3.ra, edr3.dec, 
edr3.parallax, edr3.parallax_error,
edr3.pm, edr3.pmra, edr3.pmra_error, 
edr3.pmdec, edr3.pmdec_error,
edr3.ruwe, edr3.phot_g_mean_mag, 
edr3.phot_rp_mean_mag, edr3.bp_rp,
FROM gaiaedr3.gaia_source as edr3
WHERE 1=CONTAINS(POINT('ICRS',edr3.ra,edr3.dec),
CIRCLE('ICRS',80.9,-69.75,5.5))
AND edr3.phot_g_mean_mag < 17
\end{verbatim}
with the same for the SMC but for its respective center and a limit $G < 17.5$~mag. This returned 570119 and 200718 sources for the LMC and SMC, respectively. To ensure that only sources with well-measured proper motions are considered for removal from the science catalog, a cut on $RUWE < 1.4$ is made, the same threshold adopted by \citet{Lindegren_2021} for sources with good fits to their astrometry.

\begin{figure}
    \centering
    \includegraphics[width=0.95\columnwidth]{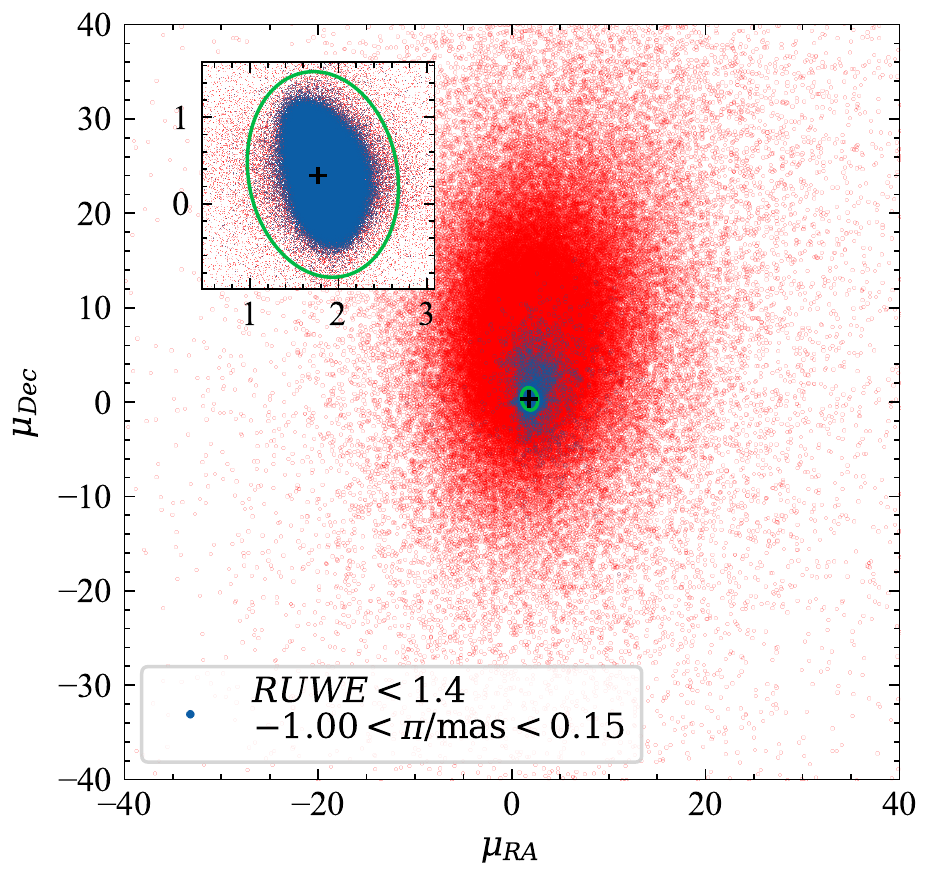}
    \caption{Distributions of EDR3 proper motions for the LMC. Sources used to define an elliptical proper-motion profile (green ellipse) are plotted (blue points) along with those outside the parallax cut (red points). The LMC proper motion determined by \citet{Luri_2021} is plotted for reference (black plus sign). All sources outside the green profile are excluded as foreground. To avoid clipping real LMC sources, only sources with $RUWE < 1.4$ are considered for cleaning, i.e., all OGLE sources without a Gaia counterpart or with $RUWE > 1.4$ are retained.}
    \label{fig:pmdist}
\end{figure}

\begin{figure*}
    \centering
    \includegraphics[width=0.45 \textwidth]{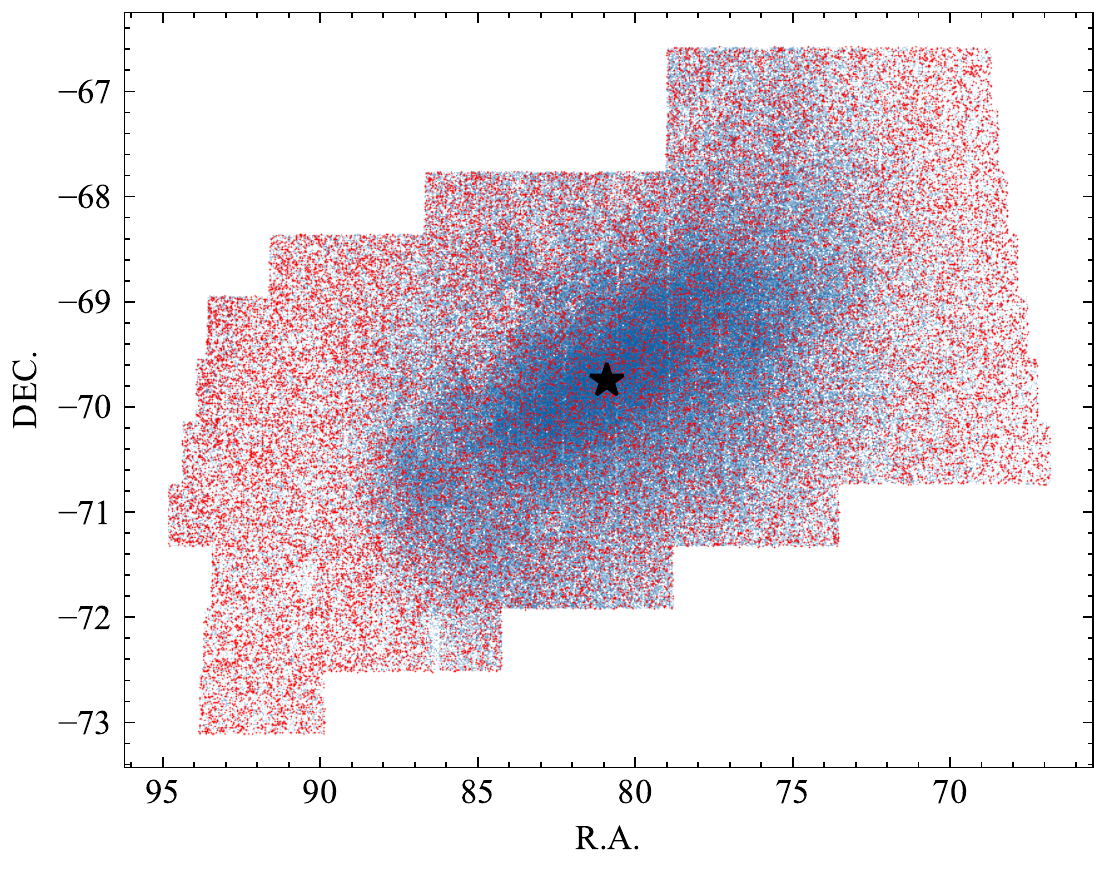}
    \includegraphics[width=0.45 \textwidth]{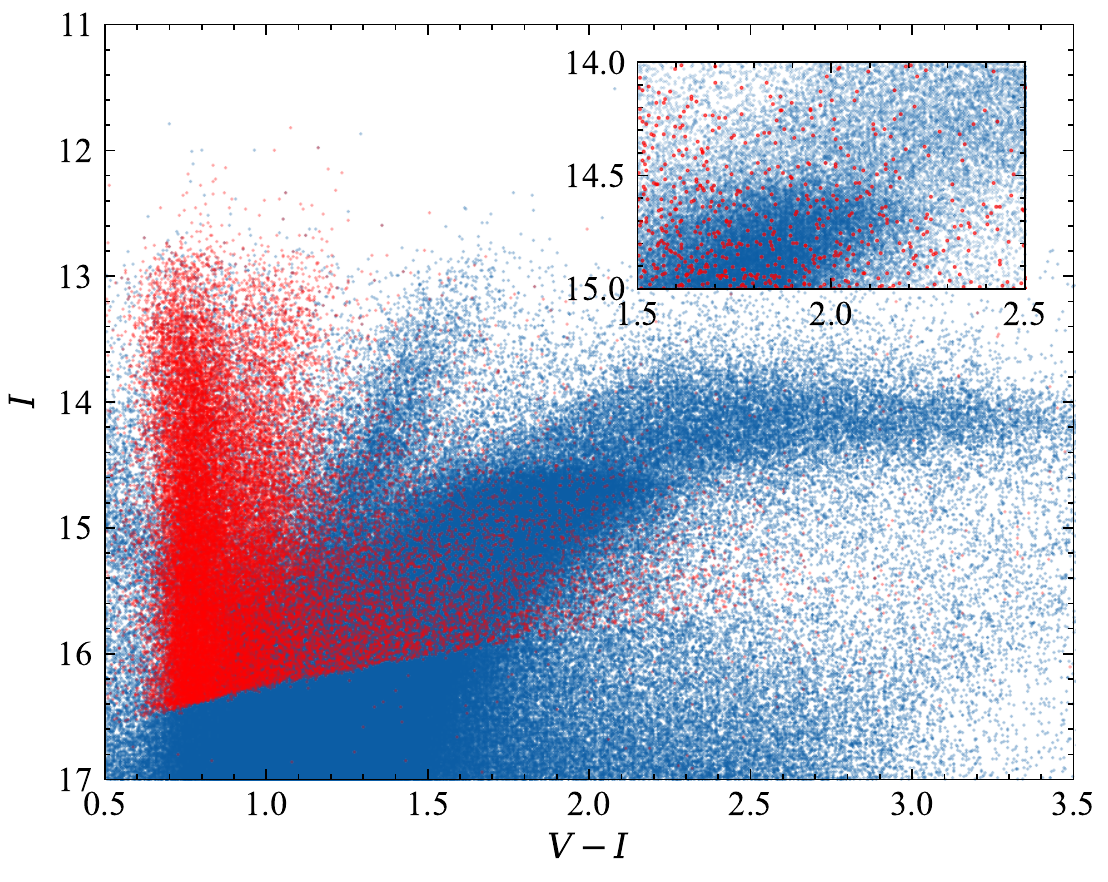}
    \caption{Successful results of proper motion cleaning in the LMC as demonstrated by the distribution of sources on the sky (left) and in the color-magnitude diagram (right). Sources identified as foreground (red dots) are plotted over LMC member stars (blue dots). A zoomed view of the TRGB is shown in the inset.}
    \label{fig:pm_results}
\end{figure*}
To isolate the underlying distribution of LMC proper motions, an initial cut of $-1 < \pi < 0.15 $~mas is made. The resulting sources are plotted in \autoref{fig:pmdist} as blue dots, with the cut sources in red. Note this does \emph{not} define the final selection and is only used temporarily to filter out most of the contamination from foreground sources when determining the 2D proper motion profile. In the inset of \autoref{fig:pmdist}, an extreme overdensity of sources is visible near the expected value for the proper motion \citep{Luri_2021}, which is plotted as a black crosshair.
The green curve represents the elliptical profile adopted to represent the distribution of LMC proper motions; everything outside the profile is thus excluded as a foreground source. The results of the proper motion cleaning for the LMC are shown in \autoref{fig:pm_results}, where a uniform distribution in position and in the color magnitude diagram at the TRGB confirms that the clipped sources are located in the foreground. In the CMD panel, the TRGB is shown up close in the inset. PM cleaning removes 35,000 definite foreground sources from the LMC OGLE-III catalog, and 15,000 from the SMC OGLE-III catalog.

\section{Discussion of Anomalous Geometric Distortion Corrections in P19} \label{app:p19_geos}

During computation of the geometric distortion corrections described in \autoref{subsect:lmc_geo}, it was noticed that the ``corr'' values contained in P19 Table 1 do not appear consistent with the predictions from a smooth plane model. Earlier in \autoref{fig:lmc_geo} the DEBs ECL-09678 and SC9-230659 were highlighted to illustrate the point. The two are separated by only $19'$ but are tabulated as having \emph{predicted} line-of-sight corrections
that differ by 0.026~mag. 
Another illustrative outlier is ECL-18365 which, according to its listed coordinates, is located on the more distant, SW side of the line of nodes, while the P19 ``corr'' value states that it is predicted to be 0.033~mag 
\emph{nearer}. It is not possible for a smooth planar model to produce these high-frequency variations in predicted line-of-sight distance.

Summarizing this situation, it was found that the standard deviation of the DEB distances \emph{increases} from $\sigma_{\mu} = 0.026$~mag (without any geometric correction) to $\sigma_{\mu} = 0.028$~mag, after adding the correction values quoted in P19 Table 2. Conversely, if the corrections are instead re-computed here using \autoref{eq:weinberg_dist} and the P19 planar fit ($\Theta=132\degree$, $i=25\degree$), the dispersion in the distortion-corrected DEB distances decreases to $\sigma_{\mu} = 0.022$~mag, which is closer to the typical uncertainty ($0.019$~mag) in each DEB distance measurement. It would appear that the parameters quoted by P19 for their LMC planar fit were reported accurately, but some minor inconsistencies were introduced in the presentation of the data (their Table 1 and Figure 2). While slightly puzzling, the actual effect on computing the mean distance to the LMC is trivial. When using any of the three sets of DEB distances -- (1) with no geometric correction, (2) after applying the correctiong tabulated in the ``corr'' column of P19 Table 2, or (3) using the updated geometric corrections computed here -- the maximal difference in the mean LMC distance is only 3~mmag.

\section{Field-by-field TRGB Measurements} \label{app:fields}
In this section, more details are provided regarding the TRGB measurement and ranking procedure introduced in \autoref{subsect:trgb_measure}. 

In \autoref{tab:detect_summary}, summary statistics for the individual TRGB field measurements are presented for each set of spatial bins (25 Voronoi and 140 OGLE), for each Rank, and for the four different sets of geometric corrections: None, P19, C21, and the geometry determined in \autoref{subsect:geofit}. The lower-ranked fields exhibited larger dispersions and systematically fainter TRGB magnitudes. As concluded in \autoref{subsect:lmc_regions}, this is due to increased levels of recent star formation as well as higher dust and gas content in these regions. Furthermore, it can be seen that all three sets of geometric corrections decreased the dispersion in TRGB magnitudes measured to Rank 1 and 2 fields, while having no effect for the Rank 3 and 4 fields, indicating the TRGB measurement uncertainties in the latter are dominant over the effect of the LMC's three-dimensional orientation.

\begin{figure*}
    \centering
    \includegraphics[width=0.9\textwidth]{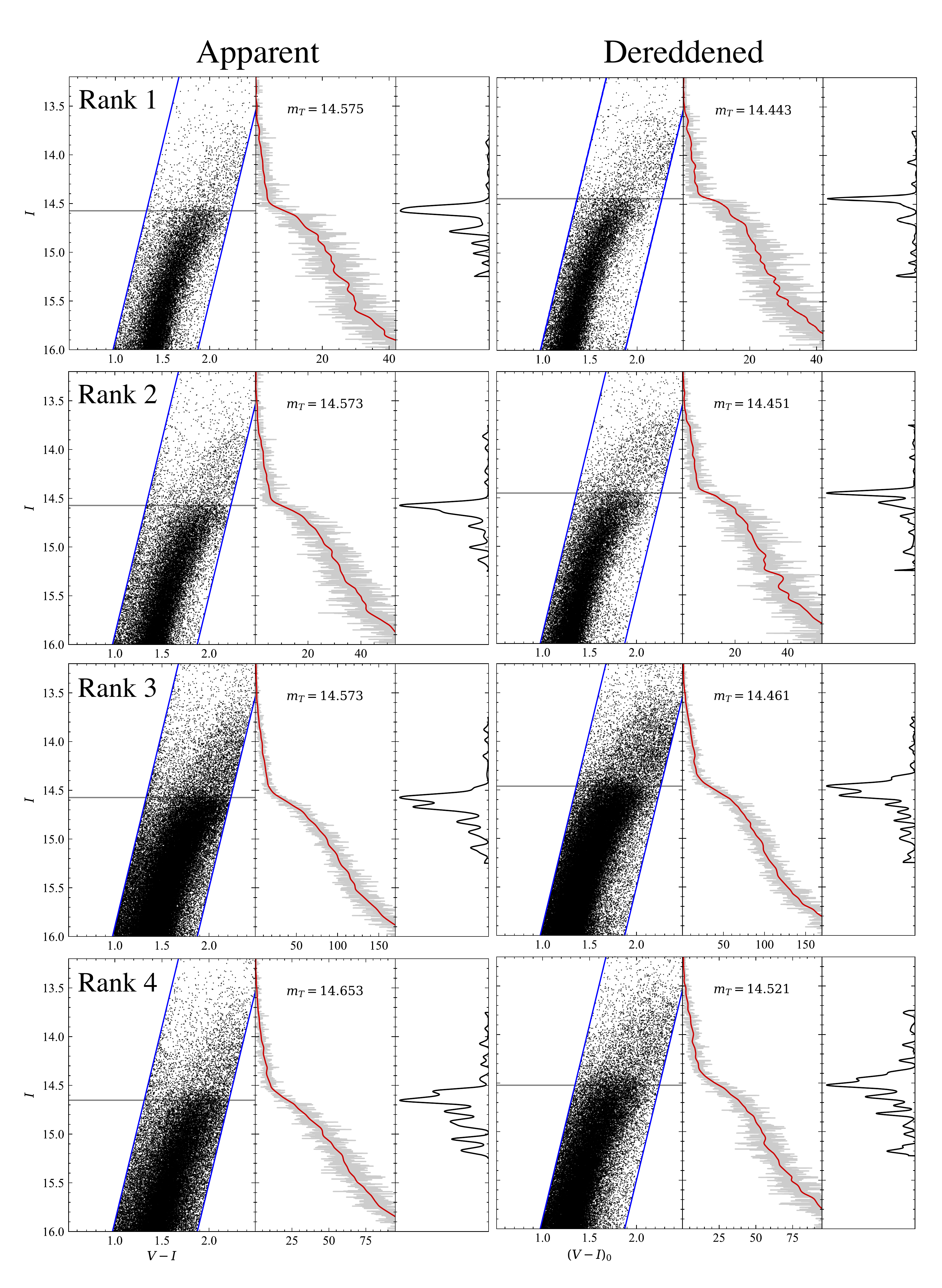}
    \caption{Apparent and S21-dereddened TRGB Detections for the Rank 1 through 4 samples in the LMC. Note that the color selection of $1.35$~mag $< (V-I) < 2.25$~mag shown here includes a much wider range of colors than that used in the zero point calibration ($1.60$~mag $< (V-I)_0 < 1.95$~mag). Therefore, the TRGB magnitudes printed here should not be considered equivalent TRGB calibrations (see \autoref{subsect:zpcal} and \autoref{fig:lmc_trgb} for the final adopted LMC zero point calibration).}
    \label{fig:example_fields}
\end{figure*}

In \autoref{fig:example_fields}, composite CMDs (with C21 geometric corrections) are shown for the Rank 1, 2, 3, and 4 samples for both apparent and S21-reddening-corrected photometry. The trend already illustrated in \autoref{fig:lmc_example} is seen again, with the S21 reddening corrections improving the sharpness and quality of the Rank 1 and 2 fields, while no noticeable improvement is observed in the Rank 3 and 4 fields, suggesting in these regions that either the assumptions in the reddening corrections break down and/or that the problem in these fields is an age effect on the TRGB magnitude or the RC color, and thus not rectifiable via RC-determined reddening corrections.

\begin{deluxetable*}{lcccr|ccr}

\tabletypesize{\small}
\tablecaption{Summary of LMC TRGB Measurements for Different Adopted Geometries. \label{tab:detect_summary}}

\tablehead{
\colhead{Geo.}           &
\colhead{Rank}           &
\colhead{$<\!I_0^{TRGB}\!>$\tablenotemark{a}}   &
\colhead{$\sigma$\tablenotemark{a}}       &
\colhead{N\tablenotemark{a}}              &
\colhead{$<\!I_0^{TRGB}\!>$\tablenotemark{b}}   &
\colhead{$\sigma$\tablenotemark{b}}       &
\colhead{N\tablenotemark{b}} \\ [-0.2cm]
\colhead{Corr.} &
\colhead{}      &
\colhead{(mag)} & 
\colhead{(mag)} & 
\colhead{}      &
\colhead{(mag)} & 
\colhead{(mag)} &
\colhead{}      }
\startdata
None  \\
\hline
& 1    & 14.462 & 0.035 & 3  & 14.475 & 0.028 & 11 \\
& 2    & 14.482 & 0.031 & 5  & 14.469 & 0.056 & 28 \\
& 3    & 14.512 & 0.039 & 10 & 14.522 & 0.066 & 19 \\
& 4    & 14.524 & 0.045 & 7  & 14.612 & 0.075 & 16 \\
& All  & 14.498 & 0.052 & 25 & 14.515 & 0.082 & 74 \\
\hline
P19 & ($\theta=132\degree$, & $i=25\degree$) \\
\hline
& 1   & 14.451 & 0.005 & 3  & 14.453 & 0.014 & 11 \\
& 2   & 14.458 & 0.030 & 5  & 14.457 & 0.038 & 28 \\
& 3   & 14.503 & 0.043 & 10 & 14.520 & 0.063 & 19 \\
& 4   & 14.532 & 0.055 & 7  & 14.557 & 0.183 & 16 \\
& All & 14.497 & 0.056 & 25 & 14.494 & 0.103 & 74 \\
\hline
C21 & ($\theta=167\degree$, & $i=22\degree$) \\
\hline
& 1   & 14.459 & 0.014 & 3  & 14.458 & 0.020 & 11 \\
& 2   & 14.442 & 0.022 & 5  & 14.455 & 0.031 & 28 \\
& 3   & 14.501 & 0.039 & 10 & 14.514 & 0.070 & 19 \\
& 4   & 14.538 & 0.058 & 7  & 14.556 & 0.194 & 16 \\
& All & 14.492 & 0.056 & 25 & 14.496 & 0.106 & 74 \\
\hline
TRGB & ($\theta=150\degree$, & $i=30\degree$) \\
\hline
& 1   & 14.444 & 0.027 & 3  & 14.446 & 0.016 & 11 \\
& 2   & 14.447 & 0.025 & 5  & 14.458 & 0.034 & 28 \\
& 3   & 14.503 & 0.041 & 10 & 14.517 & 0.068 & 19 \\
& 4   & 14.541 & 0.060 & 7  & 14.525 & 0.183 & 16 \\
& All & 14.492 & 0.060 & 25 & 14.486 & 0.100 & 74
\enddata
\tablecomments{
The field-by-field detections that contributed to this table contained TRGB stars that spanned the color range $1.35 < (V-I)_0 < 2.25$~mag, while the color range for the final composite calibration was restricted to $1.65< (V-I)_0 < 1.95$~mag. Therefore, these raw averages should not be taken as equivalent zero-point calibrations, because they will be systematically biased due to metallicity, age, and mixed-population effects.}
\tablenotetext{a}{Statistics from the Voronoi Tip Measurements.}
\tablenotetext{b}{Statistics from Tip measurements to the OGLE-defined fields.}
\end{deluxetable*}

\section{Performance Evaluation of LMC Geometric Corrections} \label{app:geo_details}

\begin{figure}
    \centering
    \includegraphics[width=\columnwidth]{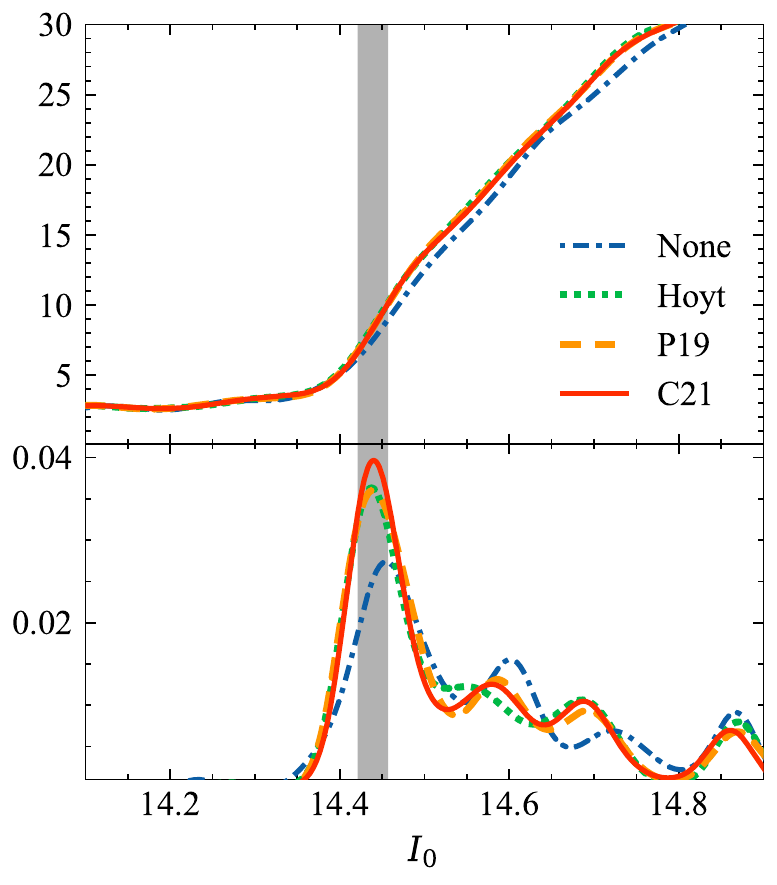}
    \caption{Comparison of RGB LFs and EDRs for the four different sets of geometric corrections: None (Dot-dashed blue), P19 (dashed orange), this study (dotted green), and C21 (red solid). The three sets of corrections result in consistent TRGB edge locations which are sharper than the measurement without geometric corrections. The C21 corrections produce the sharpest edge feature and are adopted for the final calibration. The $2\sigma$ band representing the C21 statistical uncertainty $\sigma = 0.008$~mag is plotted for reference.}
    \label{fig:geocor_compare}
\end{figure}
In \autoref{fig:geocor_compare}, the Rank 1+2 TRGB detections for all four sets of geometric corrections (no correction vs. P19 vs. C21 vs. this study) are compared. The C21 corrections are confirmed to produce the sharpest composite TRGB detection (with an estimated uncertainty $\sigma = 0.008$~mag) and are adopted for the final calibration.

\section{Notes on Revisited Analyses} \label{app:revisits}

In \autoref{sect:lmc_discuss}, a trio of recent calibrations based in the Clouds were revisited and updated in order to correct for or identify limitations in the original studies. In this section, the original measurements will be examined closely and inconsistencies will be identified. A reproduction of each study starting at the catalog level is then attempted and the sources of these biases are confirmed to be a result of some of original studies' TRGB measurements having been made in low-ranking fields. Then, to enable congruous comparison with the TRGB calibration determined here, the original studies are rescaled onto the S21 reddening map and the P19 and G20 DEB distances.

\subsection{JL17} \label{app:jl17}

\subsubsection{Original Study}

After externally calibrating the run of the $I$-band TRGB magnitude with $VI$ color, JL17 measured the TRGB in $r \leq 50'$ fields centered on the eight DEBs of \citet{Pietrzynski_2013}. They used the \citet{Haschke_2011} RC reddening maps and the \citet{Pietrzynski_2013} DEB distance (and accompanying geometric corrections) to bring their TRGB measurements onto an absolute system and determine a TRGB zero point. They provided two calibrations: a ``blue-TRGB'' which selected for only metal-poor TRGB stars, $M_I = -3.970 \pm 0.102$~mag
and a calibration using their externally-determined quadratic color correction 
($M_I = -3.998\pm 0.096$~mag for $(V-I)_0 = 1.80$~mag). These calibrations are tabulated in \autoref{tab:lit_summ} under the ``Original Studies'' block as ``JL17 (flat)'' and ``JL17 (QT)'', respectively, with the DEB distance uncertainties subtracted out. It is found that the JL17 TRGB measurement is biased by 0.05~mag on account of including measuring the TRGB in the star-forming DEB fields. The authors noticed this and adopted accurately large uncertainties (0.04~mag in the measurement and an additional 0.02~mag due to intermediate-aged contamination).

\citet{Yuan_2019}, however, in their adoption of the JL17 blue-TRGB calibration, did not also propagate the JL17 uncertainties, thereby increasing the statistical significance of the bias in their proposed calibration. Their calibration $M_{F814W} = -3.97 \pm 0.046$~mag is shifted back to $M_I$ by their measured offset $I - F814W = 0.018$~mag and entered under the ``Original'' block of \autoref{tab:lit_summ}.

Looking at the measurements made by JL17 (their Figure 10 and Table 5), there is an apparent internal discrepancy between the individual QT and ``blue TRGB'' measurements for their eight EB fields. As an introduction, the blue edge of their RGB selection box for the blue-TRGB measurements was positioned at $(V-I)_0 = 1.5$~mag, exactly at the pivot color of their QT correction. Thus, all of their QT corrections in the LMC should have monotonically brightened the TRGB magnitudes by a predictable amount based on the mean color of TRGB stars in those fields. This however was not always the case. For example, their field EB8 TRGB measurement became 0.05~mag \emph{fainter} after the QT color correction, which should not be possible. Similarly, fields EB6 and EB7 both became 0.08~mag brighter when measured in the QT-rectified CMD. However, if one directly computes the expected color-correction based on their QT function (their Table 5),
the exact correction values are 0.027~mag and 0.017~mag, for mean colors $(V-I)_0= 2.00$~mag, and $(V-I)_0 = 1.90$~mag, respectively. This inconsistency between the $QT$ and blue TRGB detections is likely real and a reflection of the ambiguous nature of the TRGB detections in those fields. Put another way, even small modifications to the underlying photometry (via color-rectification) and sample selection (color selection box) can induce discrepancies in the measured TRGB magnitude up to a factor of four larger than expected. This will now be confirmed.

\begin{figure}
    \centering
    \includegraphics[width=\columnwidth]{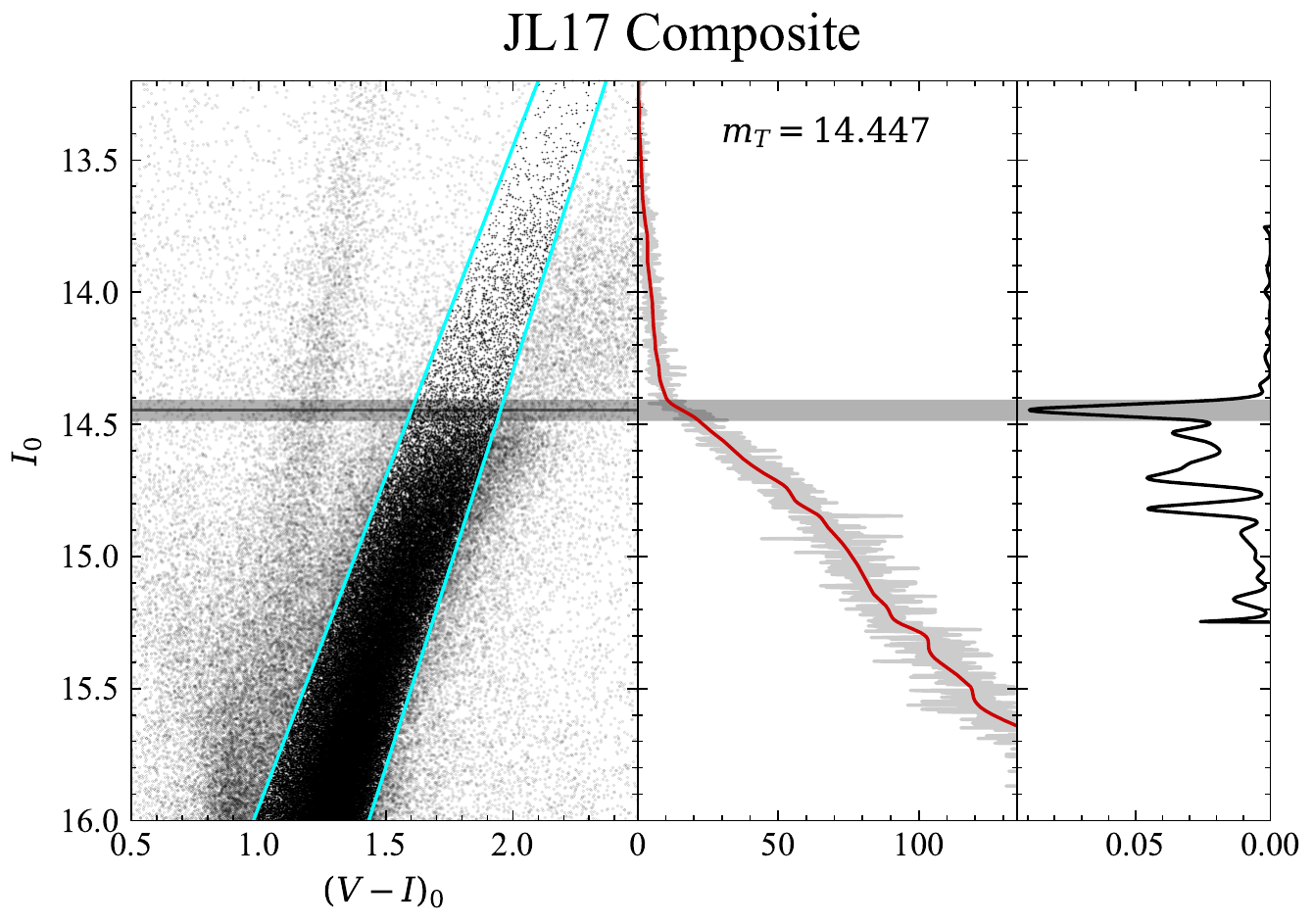}
    \includegraphics[width=\columnwidth]{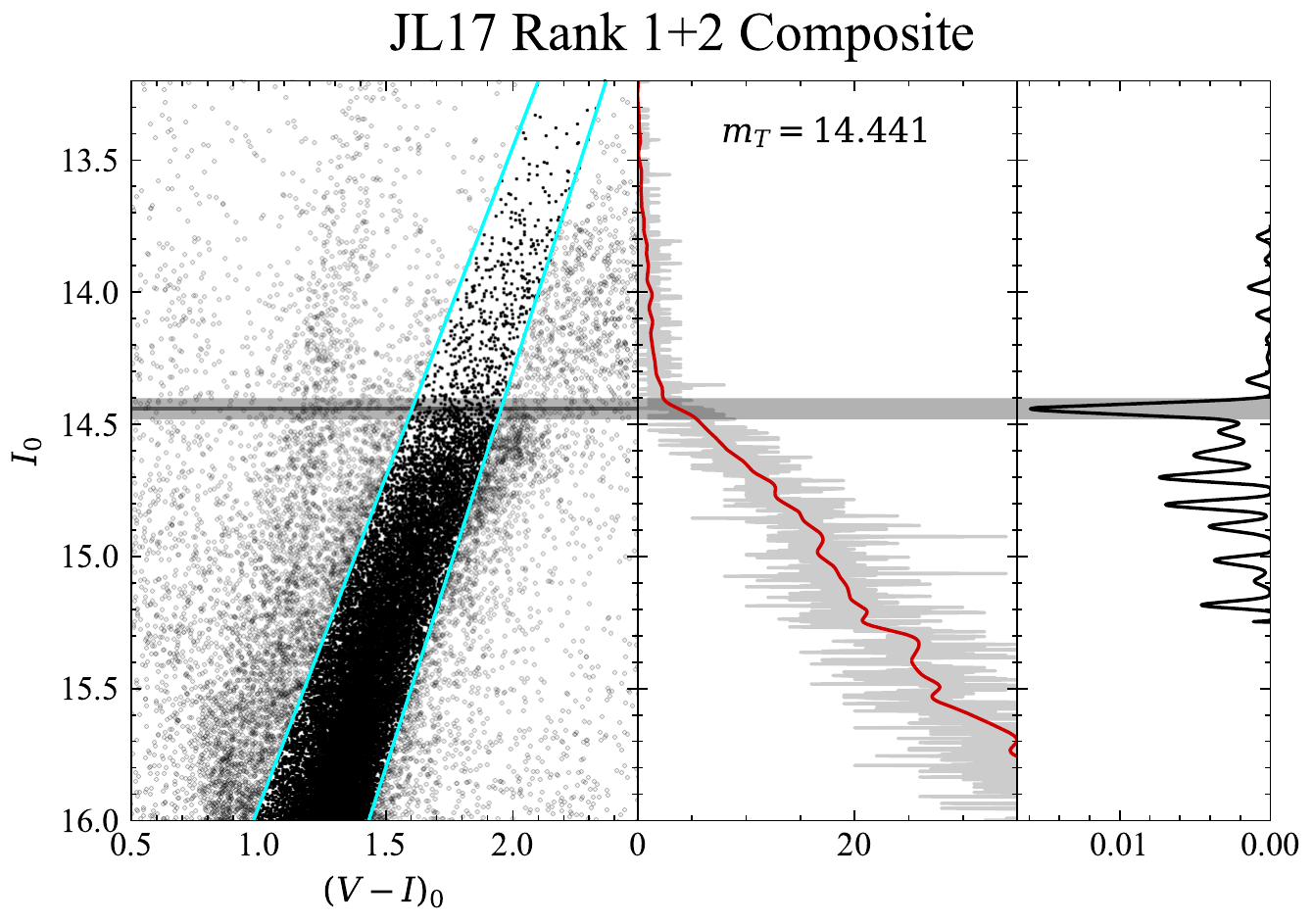}
    \includegraphics[width=\columnwidth]{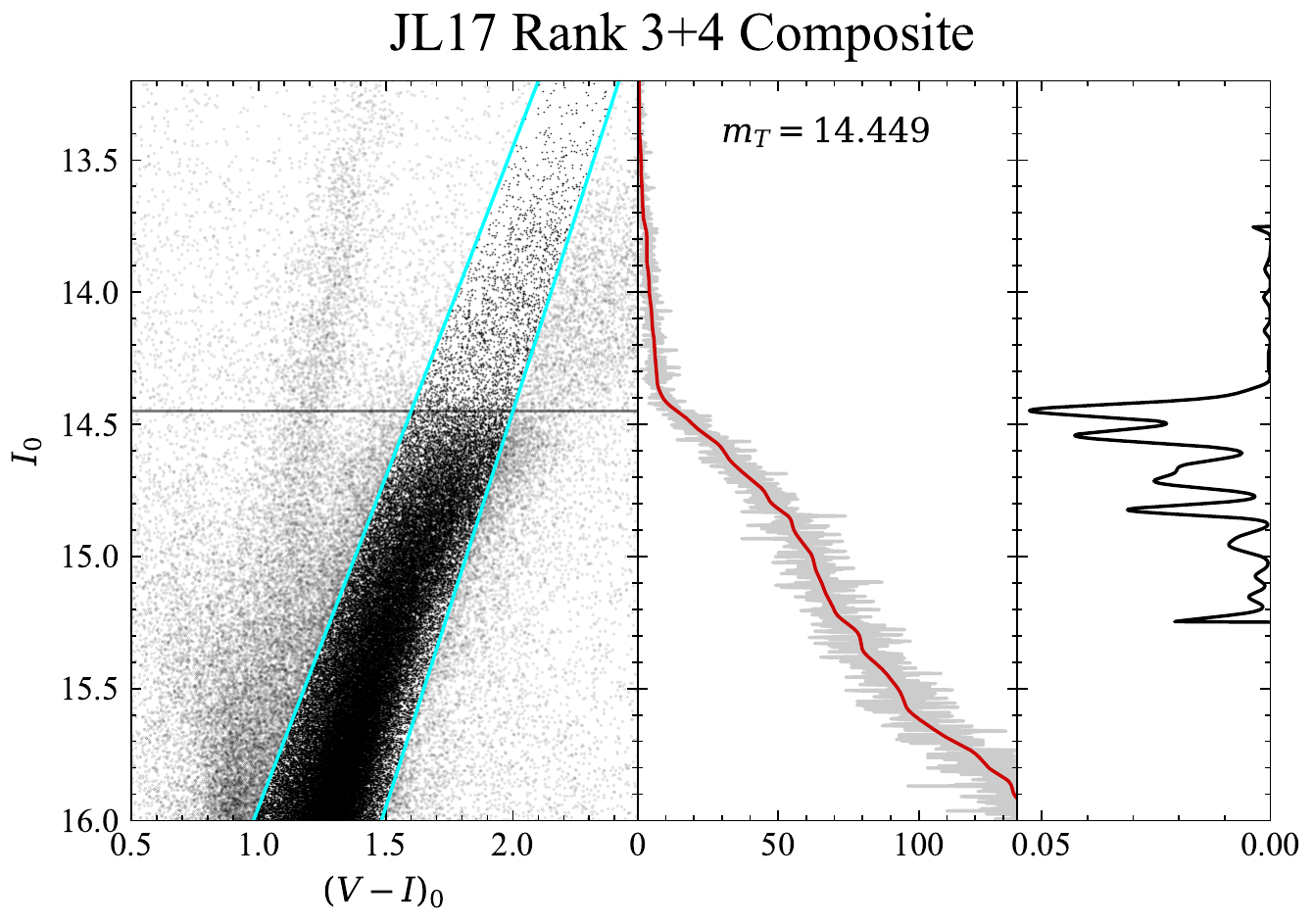}
    \caption{Composite TRGB detections for the JL17 fields. \textit{Top:} The full sample. Despite a long tail and false edges, a TRGB measurement consistent with this study is found. \textit{Middle:} The intersection of the JL17 sample with the Rank 1+2 fields. A clear peak is seen at the same magnitude as the full sample. \textit{Bottom:} Same with the rank 3+4 sample. The bimodality at $I_0\simeq 14.45$ and $I_0\simeq 14.55$ is visible. The approach by JL17 essentially took the mean of these two peaks and split the difference as the uncertainty. The composite CMD approach corrects for the bias in the JL17 calibration which was skewed to fainter magnitudes by fields with anomalously faint TRGB features.}
    \label{fig:jl17_composites}
\end{figure}

The TRGB is re-measured to each of the JL17 fields and it is found that the EDRs for five of the eight DEB-centered fields exhibit strongly bimodal, low-ranking responses. Only three of eight JL17 fields were found to exhibit clear, unambiguous TRGB measurements: EB4, EB7, and EB8. My measurements to these fields are in agreement with those of JL17 (after correcting for differences in reddening), but for the TRGB measurement to EB7 for which I find an unambiguous bright peak 0.10 mag brighter than that measured by JL17. The five low-ranking fields overlap considerably (seen near $\alpha = -77\degree, \delta = -69 \degree$ in \autoref{fig:rank_map}), so this feature is likely sourced by the same local star-forming dust clump. Indeed, there are 640000 sources contained in all eight fields (including duplicates). That number decreases to 340000 after exclusion of duplicates.

To illustrate the bimodality in the low-ranking fields, the TRGB was measured to the eight JL17 fields in two ways: allowing the edge detector to trigger anywhere, and where it was forced to trigger on the brighter of the two observed peaks. From the former unrestricted analysis, the mean TRGB was found to be $14.50\pm0.040$~mag, while the latter set the mean TRGB magnitude is $14.46 \pm 0.011$~mag. The former is identical to what JL17 originally measured to their eight fields (after taking into account the updated DEB distances and reddening map). In the latter case, the smaller scatter and brighter mean magnitude would suggest that the brighter of the two peaks represents the true TRGB magnitude in each field (i.e., a mixed population). 

This is confirmed in the top panel of \autoref{fig:jl17_composites}, where a composite CMD and TRGB detection for the JL17 fields. By simply removing the duplicate sources (which are disproportionately located in regions of low TRGB quality) and measuring the TRGB magnitude via edge detection in a composite CMD+LF, the value $I_0 = 14.449 \pm 0.03$~mag is determined, which is perfectly consistent with the high accuracy result of this study. Then, if the TRGB is measured from the set intersection of the JL17 fields and the Rank 1+2 fields, the Tip detection is sharpened and measured to be $I_0 = 14.443 \pm 0.015$~mag, as shown in the middle panel of \autoref{fig:jl17_composites}.

The source of the bimodality is confirmed clearly in the bottom panel of \autoref{fig:jl17_composites}, which shows an attempted TRGB measurement made to the JL17 fields that overlap only with the low-quality Rank 3+4 sample of the present study. The smeared Tip feature characteristic of these star-forming regions is clearly visible, as well as the bimodality at $I_0\simeq14.45$~mag and $I_0\simeq14.55$~mag. The large scatter and systematic bias seen in the JL17 TRGB measurements -- which those authors took into consideration in their error budget -- are thus immediately explained. 

\subsubsection{Renormalization to Homogeneous Reddening and Distance Measurements}

To accurately compare the original JL17 results with this study's results, the JL17 zero points are shifted by the average offset between the \citet{Haschke_2011} and S21 reddening maps \citep[$\Delta E(V-I) = +0.032$~mag][]{Skowron_2021} and updated from the \citet{Pietrzynski_2013} DEB distances to the P19 distance \citep[$\Delta \mu = +0.016$~mag][]{Pietrzynski_2019}. Their blue-TRGB calibration becomes $M_I = -3.993 \pm 0.059$~mag and the QT zero point becomes $M_I = -4.020 \pm 0.046$~mag, where the extinction and distance uncertainties have been subtracted out. These renormalized zero points are tabulated in \autoref{tab:lit_summ} under the ``Renormalized'' block and are consistent with this study's measurements to within the uncertainties.

The \citet{Yuan_2019} zero point is also shifted by the average offset between the \citet{Haschke_2011} and S21 reddening maps, giving $M_I = -3.991 \pm 0.027$~mag.

\subsection{G18} \label{app:g18}

\subsubsection{Original Study}

\citet{Gorski_2018} presented a multi-wavelength calibration of the TRGB magnitude using a selection of 14 (out of 116 available) centrally located, metal-rich fields. It is unclear why that specific subset of fields was chosen, though it is stated they employed a TRGB quality control procedure similar to this study's.
On account of the high frequency variations in star formation and dust content in the LMC, and in line-of-sight depth in the SMC, this very small sample is likely not representative of the underlying mean population in either galaxy. They determined a TRGB zero point of $-4.119 \pm 0.008$~mag at $(V-K)_0 = 3.8$~mag. After extrapolating to the mean color $(V-K)_0 = 4.0$~mag of their LMC fields, the LMC zero point is $-4.088 \pm 0.011$~mag, which is entered into the Original block of \autoref{tab:lit_summ}. An attempted reproduction of their results, and subsequent inconsistencies found, are documented in \autoref{app:g18}.

As in the previous section, the G18 analysis is revisited at the photometry level to verify their TRGB measurements can be reproduced. To each of their adopted fields (5/40 in the SMC and 14/116 in the LMC), the apparent TRGB magnitude is determined. Good agreement is found with the G18 TRGB measurements presented in their Table 1, with an offset of $I^{TRGB} - I_{G18}^{TRGB} = +0.006 \pm 0.03 $~mag when using an un-weighted Sobel kernel vs. $-0.017 \pm 0.028 $~mag when using the Poisson-weighted Sobel (the ``PN'' filter, following the nomenclature defined in G18).
Thus, it is confirmed that both sets of methodologies are consistent at the catalog level. 

There is also a slight disagreement seen between the TRGB as measured via weighted and unweighted EDRs, consistent with G18's findings, and those of \citet{Groenewegen_2019}, which is explained by the measuring the TRGB to regions of mixed stellar populations (see \autoref{subsect:sfr_trgb} for a detailed discussion).

\begin{figure}
    \centering
    \includegraphics[width=\columnwidth]{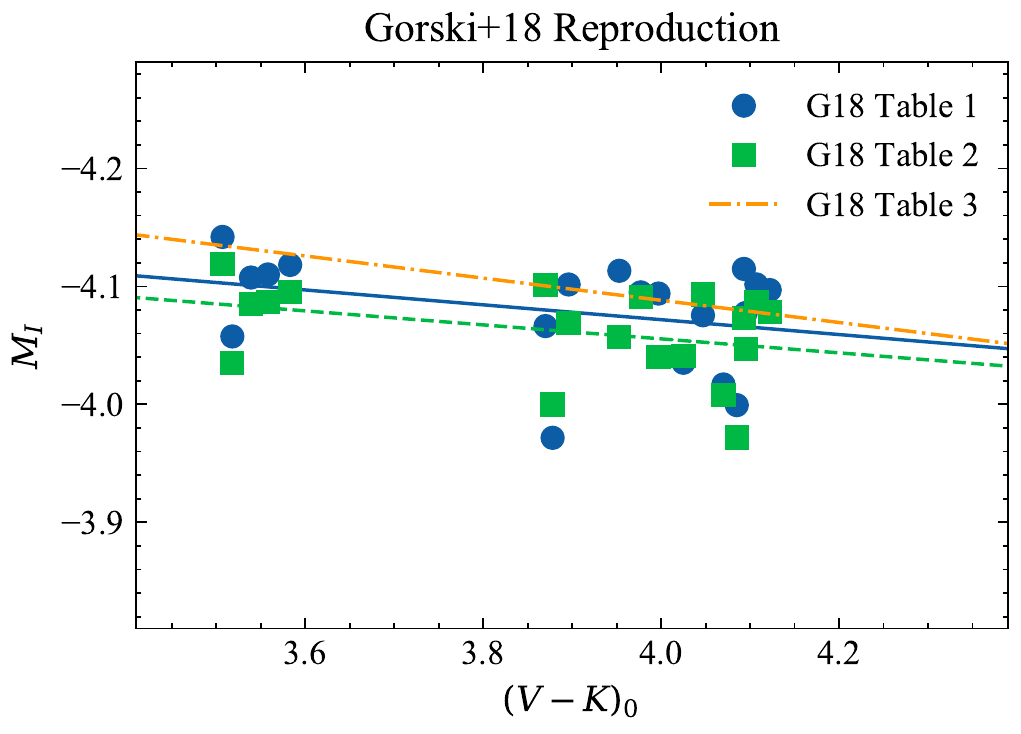}
    \includegraphics[width=\columnwidth]{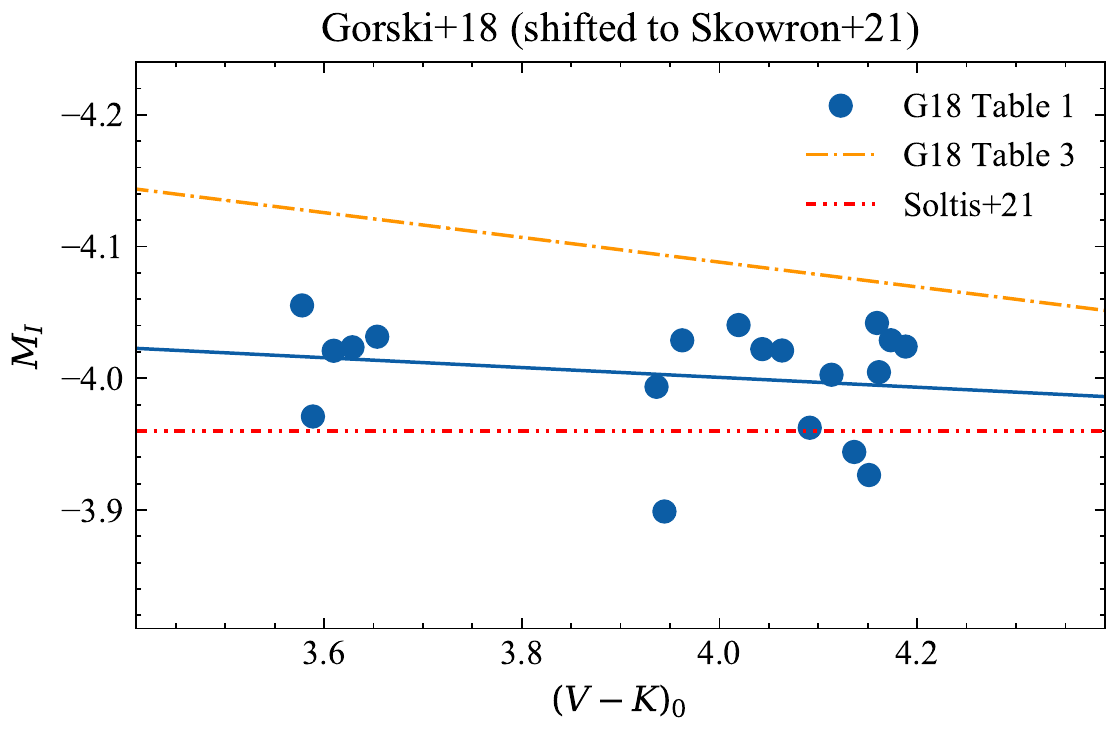}
    \caption{
    \textit{Top:} Direct reproduction of Tables 1, 2, and 3 from G18. TRGB magnitudes from their Table 1, after applying their Table 2 corrections onto an absolute system, are plotted (blue circles) with the best-fit line (solid blue line). Absolute magnitudes from Table 2 (green squares) are plotted with a best-fit (dashed green line). The calibration presented in G18 Table 3 is also plotted (dot-dashed orange line). \textit{Bottom:} Same but after shifting the colors and magnitudes onto the S21 reddening zero point (a shift of $\Delta E(V-I) = 0.048$ and 0.045~mag for the SMC and LMC, respectively). The \citet{Soltis_2021} result, which used the G18 measurements but ignored color/metallicity effects is shown (red dot-dot dashed line) and is not supported by the data.}
    \label{fig:g18_reprod}
\end{figure}

After confirming consistency with the G18 raw measurements in Table 1, it is sensible to now check for cross consistency between their Table 1 (apparent measured magnitudes), Table 2 (absolute magnitudes), and Table 3 (final calibration). However, I was unable to establish consistency between any of the three tables. The absolute magnitudes in Table 2 could not be reproduced from the Table 1 apparent magnitudes, the tabulated reddenings (and values of $A_{\lambda}$, the tabulated geometric corrections and the \citet{Pietrzynski_2013} distance. Furthermore, the calibration of Table 3 could not be reproduced from a fit to either the data in Table 1 or Table 2. Most clearly, the absolute magnitudes and dereddened colors in Table 2 are not the same values plotted in their Figures 6, 7, and 8 which were used to determine their calibration. The source of disagreement between the values quoted in their Table 2 and those plotted in Figures 6, 7, and 8 is unclear. For example, the brightest $M_I$ value in their Table 2 is $-4.119$~mag, while the brightest $M_I$ value in Figures 6, 7, and 8 is close to $-4.18$~mag.

It was also unclear from where G18 derived their adopted ratios of selective to total absorption. In their paper it is stated that they used a ``\citet{SFD98} reddening law.''\footnote{Note that the reddening law adopted by SFD98 is a combination of the O'Donnell (1994) and the Cardelli, Clayton, and Mathis (1989) curves.} However, after direct inspection of SFD98 Table 6, the G18 adopted value of $A_V = 1.08 R_V$ for $R_V=3.1$ could not be confirmed, and the same is true for the rest of their adopted ratios of selective-to-total absorption.

This internal inconsistency is summarized in \autoref{fig:g18_reprod}, where Tables 1, 2, and 3 are compared directly in a reproduction of their Figure 6. The first data are the $I_{trgb}$ values presented in their Table 1, de-reddened using the $E(B-V)$ values in their Table 2, and brought onto an absolute scale using $\mu_{\mathrm{LMC}} = 18.493$ and $\mu_{\mathrm{SMC}} = 19.003$, as well as the geometric corrections tabulated exactly as in Table 2. These values are plotted as filled blue circles. Secondly, the $M_I$ and $(V-K)_0$ values directly from G18 Table 2 are plotted as green squares. For both sets of data a least-squares fit was made and the resulting lines are plotted as solid blue, and dashed green, respectively. Neither fit agreed with the final calibration quoted by G18, plotted as a dot-dashed orange line. For comparison, the three fit equations are, in order of Table 1 to 3,

\begin{align}
\label{eq:g18_reprod}
    M_I^{G18} &= -4.084(\pm 0.011) + 0.063(\pm0.045)(V_0-K_0 - 3.8) \nonumber \\
        &= -4.067(\pm 0.009) + 0.059(\pm0.039)(V_0-K_0 - 3.8) \nonumber \\
        &= -4.107(\pm 0.008) + 0.094(\pm0.034)(V_0-K_0 - 3.8) \nonumber
\end{align}

As can be seen, consistency was not found between any of Tables 1, 2, or 3 in G18.

As mentioned in the main text and shown in \autoref{fig:rank_map}, the majority of the G18 fields are coincident with regions with a low TRGB ranking. The fields used by \citeauthor{Gorski_2018} are listed here, and included in parentheses is the fraction of sources that are coincident with a Rank 1 or 2 (Good TRGB quality) region as defined in this study: LMC-100 (0\%), LMC-102 (0\%), LMC-103 (55\%), LMC-111 (0\%), LMC-112 (70\%), LMC-116 (0\%), LMC-120 (100\%), LMC-126 (0\%), LMC-127 (55\%), LMC-161 (0\%), LMC-162 (46\%), LMC-163 (23\%), LMC-169 (0\%), LMC-170 (0\%).

It is also useful to consider the nine low-ranking G18 fields in the context of their local astrophysical environments. Fields LMC-100, LMC-102, LMC-111, and LMC-161 are contained in the young bar's highest star-forming regions. LMC-120, LMC-126 and LMC-127 are dominated by sources in the densest regions of star-formation in the Western side of the LMC, located at the ``root'' of the Western arm. And LMC-163 and LMC-170 are coincident with the trailing plume of star formation and dust that extends southwards from 30 Dor (see \autoref{fig:fields} and \autoref{fig:lmc_fields2} for reference).

\subsubsection{Renormalization to Homogeneous Reddening and Distance Measurements}

To enable accurate comparison with this study's calibrations, the magnitudes from their Table 1 and colors from G18 Table 2 are shifted by the differences between the \citet{Gorski_2020} and S21 maps ($(E(V-I) = 0.048$~mag and 0.045~mag for the SMC and LMC, respectively). The magnitudes are then shifted by the difference between the SMC DEB distance adopted by G18 and the updated G20 measurement ($\delta \mu_{SMC} = 18.977 - 19.003 = 0.026$~mag) and between the \citet{Pietrzynski_2013} and P19 distance to the LMC ($\delta \mu_{LMC} = 18.477 - 18.493 = 0.016$~mag). In the bottom panel of \autoref{fig:g18_reprod} the renormalized datapoints are plotted.

The best-fit line to the renormalized data is $M_I^{G18'} = -4.008(\pm 0.012) + 0.037(\pm0.045)(V_0-K_0 - 3.8) $. Scaling the zero point to the mean LMC color of $(V-K)_0 = 4.1$~mag returns a value $M_I = -3.997 \pm 0.018$~mag, which is entered into \autoref{tab:lit_summ} under the Renormalized block. By simply updating to modern DEB distances and to the S21 reddening map, the slope is considerably flattened as compared to that originally measured by G18 (orange dot-dashed line).

\citet{Soltis_2021} quoted an $M_I = -3.96 \pm 0.011$~mag using these same data from G18. However, they took a simple average of the raw G18 magnitudes and did not account for metallicity effects. As a result, their quoted calibration is biased by a further $\sim \! 0.04$~mag, for a total bias of 0.08~mag to a fainter TRGB magnitude.

\subsection{F20} \label{app:f20}

\citet{Freedman_2020} presented a calibration of the TRGB zero point using a differential, multi-wavelength technique that simultaneously constrained the distances and reddenings to TRGB stars. Their original result $M_I = -4.047 \pm 0.046$~mag is tabulated as F20 Blue under the Original block of \autoref{tab:lit_summ} with the DEB uncertainty subtracted out.

In their discussion of literature distances measured to the LMC, F20 made the argument that \citet{Haschke_2011} and \citet{Yuan_2019} had probably underestimated the reddening to the inner LMC, a claim which was later disputed by \citet{Nataf_2021}. F20's argument was simple (and forms the foundation of the S21 reddening maps) and based on the fact that \citet{Haschke_2011}, and subsequently \citet{Yuan_2019}, found reddening values to the \emph{inner} LMC that were comparable to those measured in the SFD98/SF11 maps to the \emph{outer} regions of the LMC, where the foreground dust maps become reliable due to the decreased density of LMC member stars. The only way to reconcile the low (mean) reddening values posited by \citet{Yuan_2019} would be to require that: (1) there is no gradient in internal dust content between the inner and outer LMC, or (2) the entirety of the SFD98/SF11 Galactic dust maps are systematically overestimated by $\sim 0.025$~mag in $E(B-V)$. Both scenarios are highly unlikely; if true, they would upend decades of keystone astrophysical measurements and knowledge. Instead, this problem is immediately resolved by the S21 analysis, which directly calibrated the RC mean color by using the SFD98/SF11 dust map paired with empirical, homogeneous measurements made in the MCs themselves.

There is no renormalization to be made to the F20 result. That calibration is already on the P19 DEB distance scale. They also constrained the reddening to LMC TRGB stars simultaneously with the distance between the LMC and SMC.
However a quick comparison of their measured reddening value with the S21 map can be made. If TRGB stars are selected from the OGLE-III photometry in an identical way as in F20, the mean reddening as measured by the S21 reddening maps is $E(B-V)_{S21} = 0.091$~mag, while the 68\%-ile is $0.085^{+0.029}_{-0.020}$~mag, both in good agreement with the F20 estimate of $E(B-V) = 0.097 \pm 0.012$~mag.

F20 also presented a separate TRGB zero point calibration using the SMC (similar to that done in \autoref{subsect:smc}) with the SFD98/SF11 foreground estimate of $A_I=0.056$ and the \citet{Graczyk_2014} DEB distance and an apparent TRGB measurement $I=14.93$~mag (equal to the extinction-corrected value measured in \autoref{subsect:smc}) to measure an $M_I = -4.09 \pm 0.03_{stat} \pm 0.05_{sys}$~mag. However, given the agreement of their \emph{apparent} TRGB measurement with the extinction-corrected measurement in \autoref{subsect:smc_composite}, it is likely that the apparent TRGB magnitude quoted by F20 was already corrected for foreground extinction. Adopting instead 14.93 as the extinction-corrected TRGB, their SMC determination becomes $-4.035 \pm 0.03_{stat} \pm 0.05_{sys}$~mag, which is in very good agreement with the measurements of this work.

As a check, the apparent TRGB magnitude is measured from the SMC using the same error estimation methodology of \autoref{subsect:trgb_measure} and determined to be $I = 14.99 \pm0.03$~mag. Adopting the updated G20 DEB distance, and again the SFD98/SF11 foreground estimate, gives $M_I = -4.040 \pm 0.04_{stat} \pm 0.03_{sys}$~mag. This value is in very good agreement with that determined in this study.

\newpage
\bibliographystyle{aasjournal}

\end{document}